\documentclass[preprint,10pt,authoryear]{elsarticle}
\usepackage[utf8]{inputenc}
\usepackage[T1]{fontenc}

\usepackage[a4paper,top=3.6cm,bottom=2.2cm,left=2.3cm,right=2.3cm]{geometry}

\usepackage{amsmath,amssymb,amsfonts}
\usepackage{bm}

\usepackage{graphicx}
\usepackage{booktabs}
\usepackage{threeparttable}
\usepackage{longtable}
\usepackage{array}
\usepackage{multirow}
\usepackage{caption}
\usepackage{subcaption}
\usepackage{float}
\usepackage{placeins}
\usepackage{tabularx}
\usepackage{pdflscape}
\usepackage{xcolor}

\usepackage{microtype}

\usepackage{enumitem}
\setlist[itemize]{topsep=3pt,itemsep=2pt}
\setlist[enumerate]{topsep=3pt,itemsep=2pt}

\usepackage{ragged2e}
\newcolumntype{Y}{>{\RaggedRight\arraybackslash}X}
\newcolumntype{P}[1]{>{\RaggedRight\arraybackslash}p{#1}}
\newcolumntype{Q}[1]{>{\Centering\arraybackslash}p{#1}}

\usepackage{hyperref}
\hypersetup{
    hidelinks,
    pdftitle={Machine Learning-Based Bitcoin Trading under Transaction Costs},
    pdfauthor={Andrei Bysik and Robert Ślepaczuk}
}

\journal{arXiv}

\begin{document}

\begin{frontmatter}

% ==================================================
% TITLE
% ==================================================
\title{Machine Learning-Based Bitcoin Trading Under Transaction Costs: Evidence From Walk-Forward Forecasting}

% ==================================================
% AUTHORS
% ==================================================
\author[a]{Andrei Bysik\fnref{fn1}}

\author[b]{Robert Ślepaczuk\corref{cor1}\fnref{fn2}}

\address[a]{Quantitative Finance Research Group, Faculty of Economic Sciences, University of Warsaw, ul. Długa 44/50, 00-241 Warsaw, Poland}

\address[b]{Quantitative Finance Research Group, Department of Quantitative Finance and Machine Learning, Faculty of Economic Sciences, University of Warsaw, ul. Długa 44/50, 00-241 Warsaw, Poland}

\cortext[cor1]{Corresponding author. Email address: rslepaczuk@wne.uw.edu.pl}

\fntext[fn1]{ORCID: 0009-0006-7250-7670}

\fntext[fn2]{ORCID: 0000-0001-5227-2014}

% ==================================================
% ABSTRACT
% ==================================================
\begin{abstract}
This paper investigates whether machine-learning forecasts of hourly BTC/USDT returns can be converted into economically meaningful trading performance after transaction costs. Using approximately 70,000 hourly observations from 2018--2026, XGBoost, LSTM, and iTransformer are evaluated in a 27-fold walk-forward protocol. All three models produce positive gross trading performance in selected configurations, but naive sign-based strategies fail once transaction costs of ten basis points are imposed. A cost-aware execution filter, which permits trades only when the forecast magnitude exceeds a transaction-cost-based threshold, sharply reduces turnover and restores profitability in selected configurations. The strongest long-only XGBoost strategy produces annualised returns above 65\% with a Sharpe ratio above one. Additional tests show that technical indicators improve performance in selected cases, EGARCH-derived features do not provide uniformly robust gains, and XGBoost is descriptively stronger than the neural alternatives, although bootstrap evidence does not support formal statistical dominance. Loss-function and model-selection effects are secondary and statistically fragile. The results show that the main obstacle in hourly cryptocurrency trading is not only weak predictability, but also the way forecasts are converted into trades.
\end{abstract}

% ==================================================
% KEYWORDS
% ==================================================

\end{frontmatter}

% ==================================================
% BODY TEXT SETTINGS
% ==================================================
\setlength\parindent{1cm}

% ====================================================================
% 1. INTRODUCTION
% ====================================================================
\section{Introduction}
\label{sec:introduction}

Whether asset prices can be forecast, and whether such forecasts can survive the journey from a statistical model to a traded strategy, remains a central question in financial economics. The Efficient Market Hypothesis implies that publicly available information should already be reflected in prices, leaving little room for systematic abnormal profits \citep{fama1970efficient}. At the same time, empirical evidence on momentum, value premia, and related anomalies suggests that some predictable structure may exist, even if it is weak, unstable, and difficult to exploit in practice \citep{jegadeesh1993returns,asness2013value}. Machine learning has renewed this debate by offering flexible tools for modelling nonlinear interactions in financial data, but better statistical forecasts do not automatically imply profitable trading strategies \citep{gu2020empirical}.

This distinction is especially important at short horizons. Hourly returns have a low signal-to-noise ratio, are affected by volatility clustering and regime shifts, and can make high-capacity models vulnerable to overfitting \citep{campbell1997econometrics,bailey2017pbo}. Cryptocurrency markets provide a useful setting for examining these issues because they trade continuously, generate large volumes of high-frequency data, and experience repeated transitions between bull markets, crashes, and consolidation regimes. Bitcoin is particularly suitable for this purpose because it combines long data availability, high liquidity relative to other digital assets, and substantial volatility.

The main empirical problem is that a predictable return is not the same as a profitable trade. Models are often selected using statistical criteria such as mean squared error, mean absolute error, directional accuracy, or rank correlation. These criteria may be informative about forecast quality, but they do not measure whether a position generated from the forecast earns money after transaction costs. This gap is material. A model that changes position frequently can lose money even when its gross signal is positive, because small expected gains are repeatedly consumed by spread crossing, fees, slippage, and other execution frictions. For this reason, the relevant question is not only whether hourly BTC/USDT returns are predictable, but whether the resulting forecasts can be converted into economically meaningful net-of-cost trading performance.

This paper studies that question using approximately 70,000 hourly BTC/USDT observations and a 27-fold walk-forward protocol covering the 2018--2025 out-of-sample evaluation period. Three forecasting architectures are compared: XGBoost, Long Short-Term Memory networks, and iTransformer. These models represent three different approaches to the forecasting problem: tree-based tabular learning, recurrent sequence modelling, and attention-based multivariate sequence modelling. All models are evaluated under the same data, feature sets, walk-forward design, transaction-cost assumption, and trading rules.

The main methodological feature of the paper is a cost-aware execution filter. The baseline trading rule maps the sign of the forecast into either a long-only or long-short position. The cost-aware version allows a position change only when the forecast magnitude exceeds a threshold proportional to the transaction cost generated by the trade. The threshold is governed by a parameter $\lambda$, which controls how selective the execution rule is. This mechanism directly targets the main weakness of short-horizon trading systems: excessive turnover caused by weak forecasts fluctuating around zero.

The empirical analysis is organised around six tests. First, it examines whether naive sign-based machine-learning strategies remain profitable once proportional transaction costs are imposed. Second, it tests whether the cost-aware execution filter improves net returns and reduces turnover. Third, it evaluates whether enriching the predictor set with technical indicators and EGARCH-derived volatility features improves traded performance. Fourth, it compares the three model architectures under the same cost-aware trading protocol. Fifth, it examines whether the choice between MSE and MAE training losses materially affects results. Sixth, it tests whether the validation-set model-selection criterion changes realised out-of-sample performance.

The main finding is that the forecasts contain some useful information, but this information becomes valuable only when the trading rule accounts for the cost of acting on it. Naive sign-based strategies generate positive gross performance in selected configurations, yet they collapse after transaction costs because they trade too frequently. The cost-aware filter reduces turnover by more than an order of magnitude and restores positive net performance in selected XGBoost configurations. In the strongest long-only specification, annualised return exceeds 65\% and the Sharpe ratio is above unity. However, this result is not a simple claim that machine learning beats buy-and-hold: formal Sharpe-ratio comparisons against the passive benchmark do not reject after bootstrap adjustment, and fold-level diagnostics show that performance is uneven across market regimes.

The remaining results are more qualified. XGBoost is descriptively the strongest architecture, outperforming LSTM and iTransformer in the main cost-aware long-only comparison, but paired circular block-bootstrap tests do not support a claim of statistically robust dominance after multiple-testing correction. Technical indicators improve performance in selected configurations, while EGARCH-derived features do not deliver a uniformly robust incremental gain. MSE performs better than MAE descriptively in the main XGBoost specification, but the difference is statistically fragile. Similarly, the model-selection criterion changes point estimates and strategy rankings, but no selector pair shows robust dominance. The results point to execution design as the main driver of economic performance. Feature enrichment, architecture, loss function, and selector choice matter, but their effects are smaller and less stable.

The paper makes three contributions. First, it evaluates machine-learning forecasts directly through realised trading performance rather than treating the trading simulation as a secondary exercise after statistical forecast evaluation. Second, it introduces and tests a simple cost-aware execution rule that links trading decisions to forecast magnitude and transaction costs. Third, it provides a controlled comparison of XGBoost, LSTM, and iTransformer on hourly cryptocurrency data within a common walk-forward framework. The results suggest that the economic value of a forecast depends as much on execution discipline as on the forecasting model itself.

The remainder of the paper is organised as follows. Section~\ref{sec:literature} reviews the related literature. Section~\ref{sec:data} describes the data and empirical design. Section~\ref{sec:methodology} presents the forecasting models, trading rules, performance metrics, and inference procedure. Section~\ref{sec:results} reports the empirical results. Section~\ref{sec:robustness} presents robustness checks and diagnostics. Section~\ref{sec:conclusion} concludes.

% ====================================================================
% 2. RELATED LITERATURE
% ====================================================================
\section{Related literature}
\label{sec:literature}

\subsection{Return predictability and financial machine learning}
\label{sec:lit_predictability_ml}

The starting point for return forecasting is the tension between market efficiency and the empirical evidence on predictable structure in asset returns. The Efficient Market Hypothesis implies that publicly available information should already be embedded in prices, making systematic abnormal profits difficult to obtain \citep{fama1970efficient}. Yet the empirical literature has documented patterns such as momentum and value premia across asset classes, suggesting that predictability may exist even if it is weak, regime-dependent, and costly to exploit \citep{jegadeesh1993returns,asness2013value}. At short horizons this tension becomes sharper: the predictable component of returns is small relative to noise, and any apparent signal can be unstable across market regimes \citep{campbell1997econometrics,rapach2013forecasting}.

Machine learning methods have become attractive in this setting because they relax the linearity and low-dimensionality assumptions of classical forecasting models. Tree ensembles, recurrent networks, and attention-based architectures can represent nonlinear interactions among predictors and may therefore detect weak structure that simpler models miss \citep{gu2020empirical,liaras2024machine}. XGBoost is widely used in tabular financial prediction because it combines nonlinear tree-based learning with regularisation and variable selection through boosting \citep{chen2016xgboost}. Recurrent neural networks, especially Long Short-Term Memory models, are designed to capture sequential dependence through gated memory cells \citep{hochreiter1997long}. Transformer-based architectures extend this logic by using attention mechanisms to model complex dependencies, and iTransformer is particularly relevant for multivariate forecasting because it treats variables, rather than timestamps, as tokens \citep{liu2024itransformer}.

The empirical evidence does not identify one architecture as universally dominant. Tree-based models often remain strong on medium-sized tabular datasets, while neural networks can be sensitive to sample size, frequency, loss function, and the stability of the data-generating process \citep{grinsztajn2022tree,grudniewicz2023ml}. This motivates the architecture comparison in the present paper. XGBoost, LSTM, and iTransformer are evaluated under the same hourly BTC/USDT data, feature sets, walk-forward protocol, transaction-cost model, and trading rules, so that differences in realised performance can be attributed primarily to model design rather than to differences in experimental setup.

\subsection{Transaction costs and the prediction-to-trading gap}
\label{sec:lit_prediction_trading_gap}

A central lesson from quantitative finance is that statistical predictability does not automatically translate into trading profitability. A model may reduce forecast error while still generating a losing strategy if its signals are too small, too unstable, or too expensive to trade. This issue is familiar from portfolio allocation: \citet{demiguel2009optimal} show that sophisticated optimisation rules can underperform simple benchmarks out of sample once estimation error is taken into account. The same principle applies to short-horizon trading, where repeated position changes can convert a weak gross edge into a negative net return.

This gap is especially important in cryptocurrency markets. Intraday crypto strategies often generate high turnover, and even modest proportional costs can erase apparent alpha when every signal is converted mechanically into a position change \citep{cohen2023intraday}. The relevant object is therefore not only forecast accuracy, but the realised return sequence after forecasts have passed through position rules, transaction costs, and execution constraints. A slightly less accurate model may be economically superior if its forecasts are more stable around the trading threshold or if it triggers fewer unnecessary trades.

Recent financial machine learning work has begun to address this mismatch more directly. Direction-aware loss functions, such as Mean Absolute Directional Loss, modify the training objective to reward correct directional forecasts \citep{michankow2023madl}. Extensions such as Generalised Mean Absolute Directional Loss incorporate additional penalties related to trading behaviour and turnover \citep{michankow2024gmadl}. Other work improves the feature representation used by forecasting models, for example through supervised autoencoder-based feature extraction \citep{bieganowski2025autoencoder}. These studies share a common implication: forecast-error minimisation alone is not enough when the final objective is economic performance.

This paper follows the same trading-oriented principle, but addresses the prediction-to-trading gap through the execution rule rather than through a specialised loss function. The cost-aware filter allows position changes only when the forecast magnitude is large enough to justify the expected transaction cost. This makes the trading rule itself cost-sensitive and directly targets the turnover channel through which short-horizon strategies often fail.

\subsection{Walk-forward evaluation and statistical inference}
\label{sec:lit_wfo_inference}

Reliable evaluation is particularly difficult in financial machine learning because high-frequency time series combine serial dependence, non-stationarity, and a high risk of overfitting. Random cross-validation is inappropriate in this setting because it breaks temporal order and can leak future information into the model development process \citep{lopezdeprado2018advances,bailey2017pbo}. Walk-forward evaluation is therefore better aligned with the trading problem: models are repeatedly trained on past data, selected on a subsequent validation window, and evaluated only on later test data.

The walk-forward framework also matters economically. A single train/test split can make a strategy appear strong because it happens to match one favourable market regime. A rolling protocol exposes the model to multiple bull, bear, and consolidation periods, making it harder for results to depend entirely on one historical episode. This is particularly relevant for Bitcoin, whose sample path from 2018 to 2025 includes sharp drawdowns, rapid rallies, and prolonged sideways regimes.

Statistical inference must also respect the dependence structure of trading returns. Standard tests designed for independent observations can understate uncertainty when returns are serially correlated or clustered. For this reason, the present paper uses paired circular block-bootstrap tests applied to realised return differentials, following the dependent-data resampling logic of \citet{kunsch1989jackknife} and \citet{politis1992circular}. This choice aligns the statistical test with the economic object of interest: not raw forecast errors, but the difference between two traded return series after transaction costs.

\subsection{Positioning of this paper}
\label{sec:lit_positioning}

Table~\ref{tab:lit_comparison} positions the present paper relative to selected studies in financial machine learning and cryptocurrency trading. The comparison focuses on five dimensions: asset class, data frequency, model architectures, evaluation protocol, and treatment of transaction costs.

\begin{table}[H]
\centering
\captionsetup{justification=centering,singlelinecheck=false}
\caption{Methodological comparison with selected prior studies}
\label{tab:lit_comparison}
\footnotesize
\renewcommand{\arraystretch}{0.92}
\setlength{\tabcolsep}{3.6pt}

\makebox[\textwidth][c]{%
\begin{tabularx}{1.0\textwidth}{@{}l l c Y c c@{}}
\toprule
Study & Asset & Freq. & Models & Eval. & TC \\
\midrule
Gu et al.~(\citeyear{gu2020empirical}) & US equities & Monthly & Lasso, GBRT, RF, NN & Expanding & Post-hoc \\
Alessandretti et al.~(\citeyear{alessandretti2018crypto}) & Crypto basket & Daily & GB, RF, regression & Single split & Prop. \\
Sebastião et al.~(\citeyear{sebastiao2021crypto}) & BTC, ETH, LTC & Daily & RF, SVM, linear & Single split & Prop. \\
Michańków et al.~(\citeyear{michankow2022lstm}) & BTC, S\&P~500 & Daily--15m & LSTM & WFO & Prop. \\
Kryńska et al.~(\citeyear{krynska2023lstm}) & BTC, S\&P~500 & Daily--15m & LSTM variants & WFO & Prop. \\
Michańków et al.~(\citeyear{michankow2023madl}) & BTC, Crude Oil & Daily--1h & LSTM with MADL & WFO & Prop. \\
Michańków et al.~(\citeyear{michankow2024gmadl}) & BTC, equities & 1h--15m & LSTM, Transformer & WFO & Prop. \\
Bieganowski et al.~(\citeyear{bieganowski2025autoencoder}) & S\&P~500, EUR/USD, BTC/USD & 5--30m & SAE--MLP & WFO & Prop. \\
Cohen~(\citeyear{cohen2023intraday}) & BTC, ETH, BNB, ADA, XRP & 5--180m & RSI, MACD, Keltner & Single split & Prop. \\
Grudniewicz et al.~(\citeyear{grudniewicz2023ml}) & Global equities & Daily & SVM, RF, NN, GLM & Rolling/OOS & None \\
Kashif et al.~(\citeyear{KASHIF2025113563}) & S\&P~500, FTSE~100, CAC~40 & Daily & LSTM--ARIMA & WFO & Prop. \\
Stefaniuk et al.~(\citeyear{STEFANIUK2026131599}) & BTC & 5--30m & Informer with GMADL & WFO & Prop. \\
\bottomrule
\end{tabularx}%
}

\vspace{0.35em}
\begin{minipage}{0.96\textwidth}
\footnotesize \textit{Note:} Freq. = frequency. Eval. = evaluation protocol. TC = transaction costs. RF = random forest. NN = neural network. GB = gradient boosting. SAE = supervised autoencoder. GLM = generalised linear model. Prop. = proportional transaction-cost model. WFO = walk-forward optimisation. ``Post-hoc'' means that transaction costs are considered after signal generation rather than embedded directly into the trading rule.
\end{minipage}

\renewcommand{\arraystretch}{1.0}
\end{table}

\FloatBarrier

The comparison highlights three gaps. First, studies that focus on cryptocurrency markets often use either neural-network variants or traditional machine learning models, but fewer provide a controlled comparison of tree-based, recurrent, and attention-based architectures under the same trading protocol. Second, although walk-forward evaluation is increasingly used, single train/test splits remain common. Third, transaction costs are often included proportionally, but they are less often linked directly to the execution rule that determines whether a forecast should be traded.

This paper addresses these gaps by combining hourly BTC/USDT data, a 27-fold rolling walk-forward design, three architecture families, proportional transaction costs, and a cost-aware execution filter within a single evaluation framework. The contribution is therefore not only a comparison of forecasting models, but an assessment of whether their predictions remain economically useful once the cost of acting on them is built into the trading decision itself.

% ====================================================================
% 3. DATA AND EMPIRICAL DESIGN
% ====================================================================
\section{Data and empirical design}
\label{sec:data}

\subsection{Data source and sample}
\label{sec:data_source}

The empirical analysis uses hourly OHLCV data for the BTC/USDT USD-margined futures contract collected from Binance through its public REST API. Binance is used because it is one of the most liquid cryptocurrency venues and provides a sufficiently long and continuous price history for short-horizon walk-forward evaluation. The raw sample runs from 1~December~2017 to 1~January~2026. The first month is used as a burn-in period for rolling feature construction, so the effective modelling window begins on 1~January~2018.

All timestamps are treated in Coordinated Universal Time (UTC), and each observation corresponds to a completed one-hour futures bar. The modelling pipeline uses only historical open, high, low, close, and volume fields available at the close of each bar. No proprietary vendor adjustments, sentiment variables, funding-rate variables, order-book variables, or future market information are added at the data-source stage.

The sample is well suited to the research question for three reasons. First, Bitcoin is the most liquid and longest-lived cryptocurrency asset, reducing the risk that the results are driven by idiosyncratic microstructure noise in a thin market. Second, the hourly frequency provides more than 70,000 observations, which is sufficient for repeated walk-forward estimation while still avoiding the strongest bid--ask bounce and quote-flickering effects found at tick or minute frequency. Third, the 2018--2025 evaluation period contains multiple market regimes, including bear markets, rapid rallies, sharp crashes, and extended consolidation phases.

\subsection{Preprocessing and target variable}
\label{sec:preprocessing_target}

A complete hourly index is reconstructed over the full sample. Missing hourly bars are filled by carrying forward the previous close and setting volume to zero, producing a flat synthetic bar at the missing timestamp. This preserves the regular hourly grid required by rolling technical indicators and sequence-based neural networks while avoiding interpolation methods that would create artificial price paths. The missing-data audit is reported in Appendix~\ref{app:missing_timestamps}, Table~\ref{tab:gap_distribution}. Since the number of missing hourly observations is small relative to the full sample, the resulting distortion is expected to be limited, although observations near gap timestamps may contain mechanically smoothed returns.

The forecasting target is the one-step-ahead log return,
\begin{equation}
r_{t+1} = \ln\left(\frac{P_{t+1}}{P_t}\right),
\label{eq:target_return}
\end{equation}
where $P_t$ denotes the BTC/USDT closing price at hour~$t$. The model receives information available up to and including time~$t$ and produces a forecast of $r_{t+1}$. This construction enforces strict temporal causality: no feature used to predict the next-hour return depends on information observed after the forecast time.

After preprocessing, the dataset contains 70,872 hourly observations. The effective walk-forward evaluation sample, net of the December~2017 burn-in period, contains 70,128 observations.

\subsection{Descriptive statistics}
\label{sec:descriptive_stats}

Table~\ref{tab:descriptive} reports descriptive statistics for BTC/USDT prices and hourly log returns over the cleaned sample.

\begin{table}[H]
\centering
\captionsetup{justification=centering,singlelinecheck=false}
\caption{Descriptive statistics of Bitcoin price and hourly log returns}
\label{tab:descriptive}
\small
\renewcommand{\arraystretch}{1.15}
\setlength{\tabcolsep}{12pt}

\begin{tabular}{lcc}
\toprule
Statistic & Price (USD) & Hourly log return (\%) \\
\midrule
Mean & 37,797.02 & 0.0025 \\
Median & 27,646.18 & 0.0060 \\
Standard deviation & 32,500.90 & 0.7278 \\
Minimum & 3,172.05 & -20.10 \\
Maximum & 126,011.18 & 16.03 \\
Skewness & 0.9474 & -0.4981 \\
Kurtosis & 2.7955 & 42.3227 \\
Observations & 70,872 & 70,872 \\
\bottomrule
\end{tabular}

\vspace{0.4em}
\begin{minipage}{0.85\textwidth}
\footnotesize \textit{Note:} Returns are expressed in percentage terms. The sample runs from 1~December~2017 to 1~January~2026. December~2017 is used as a burn-in period for rolling feature construction; the effective walk-forward modelling window begins on 1~January~2018.
\end{minipage}

\renewcommand{\arraystretch}{1.0}
\end{table}

The price series reflects the large boom--bust cycles that characterise Bitcoin over the sample period, ranging from approximately USD~3,172 to more than USD~126,000. Price-level statistics are reported for descriptive purposes only; the models are trained on returns rather than raw prices.

Hourly log returns are centred close to zero but display substantial volatility and strong non-normality. The standard deviation of hourly returns is 0.7278\%, corresponding to high annualised volatility. The return distribution is negatively skewed and extremely fat-tailed, with kurtosis above 42. The most extreme hourly observations, $-20.10\%$ and $+16.03\%$, illustrate the importance of tail risk in cryptocurrency markets.

\begin{figure}[H]
\centering
\captionsetup{justification=centering,singlelinecheck=false}
\caption{BTC/USDT price and hourly log-return series}
\includegraphics[width=\textwidth]{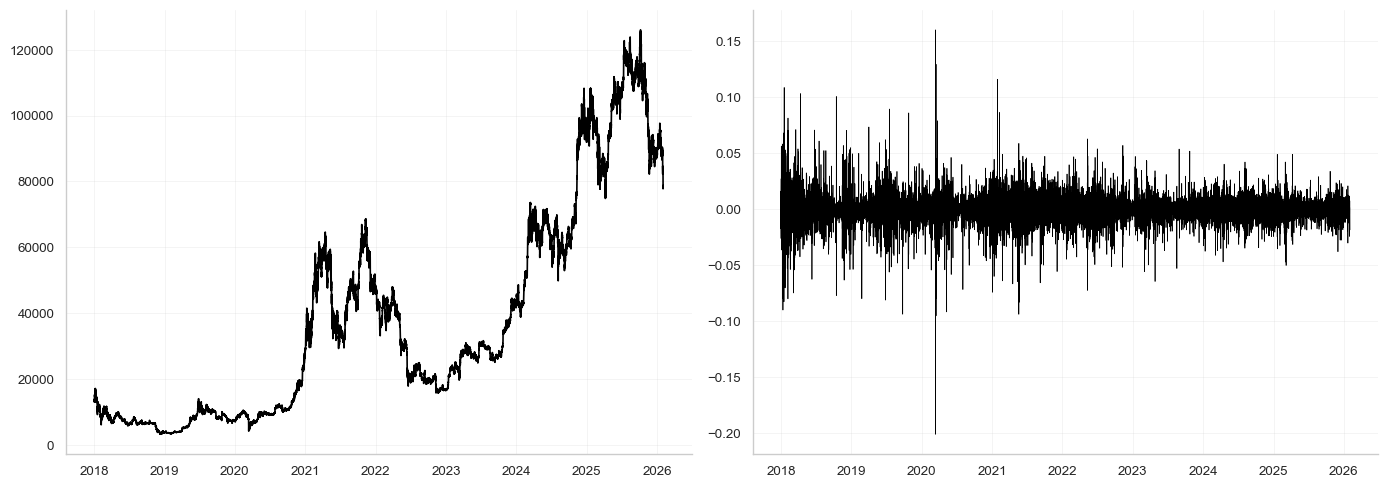}

\vspace{0.4em}
\begin{minipage}{0.95\textwidth}
\footnotesize \textit{Note:} The left panel shows the Bitcoin closing price, while the right panel shows hourly log returns over the effective walk-forward evaluation period beginning on 1~January~2018.
\end{minipage}

\label{fig:btc_series}
\end{figure}

Figure~\ref{fig:btc_series} compares the BTC/USDT price series with the corresponding hourly log-return series over the effective walk-forward evaluation window.

\FloatBarrier

The figure highlights the contrast between the non-stationary price level and the stationary return series. Returns fluctuate around zero but exhibit clear volatility clustering, with turbulent periods corresponding to major market events. Formal ADF and KPSS tests are reported in Appendix~\ref{app:stationarity}, Table~\ref{tab:stationarity}, and the Jarque--Bera test is reported in Appendix~\ref{app:normality}, Table~\ref{tab:normality}. These tests confirm that price levels are non-stationary, log returns are stationary, and the return distribution strongly departs from normality. These properties motivate the use of log returns as the prediction target, the inclusion of volatility-regime features, and the use of robust out-of-sample evaluation rather than reliance on in-sample fit.

\subsection{Walk-forward empirical design}
\label{sec:data_wfo_design}

The empirical evaluation uses a non-anchored rolling walk-forward design. Each fold contains a 12-month training window, a 3-month validation window, and a 3-month test window. After each fold, the entire window advances by three months. This produces 27 sequential test folds over the out-of-sample evaluation period. The validation segment is used for hyperparameter tuning, early stopping, and model selection, while the test segment is reserved exclusively for final out-of-sample evaluation.

\begin{figure}[H]
\centering
\captionsetup{justification=centering,singlelinecheck=false}
\caption{Walk-forward optimisation scheme}
\label{fig:wfo}

\includegraphics[width=\textwidth]{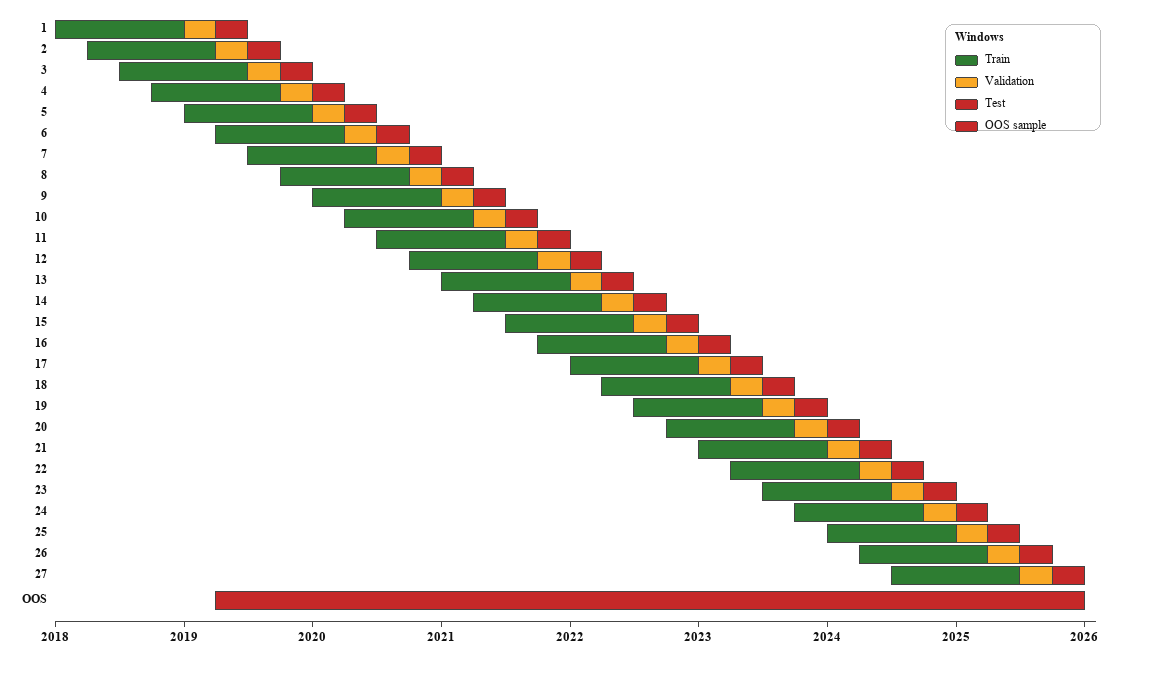}

\vspace{0.4em}
\begin{minipage}{0.95\textwidth}
\footnotesize \textit{Note:} In each fold, the model is trained on 12~months of data, validated on the following 3~months, and tested on the next 3~months. The window advances by 3~months after each fold, producing 27 sequential test folds. The consolidated out-of-sample evaluation is the union of all held-out test windows.
\end{minipage}
\end{figure}

This design serves two purposes. First, it preserves temporal ordering and prevents look-ahead bias, since every forecast is generated by a model trained only on past information. Second, it evaluates model performance across multiple market regimes rather than relying on a single train/test split. The detailed walk-forward protocol and the corresponding scheme are described in Section~\ref{sec:wfo} and illustrated in Figure~\ref{fig:wfo}.

% ====================================================================
% 4. FORECASTING MODELS, TRADING RULES, AND INFERENCE
% ====================================================================
\section{Forecasting models, trading rules, and inference}
\label{sec:methodology}

\subsection{Trading strategy and cost-aware execution}
\label{sec:trading_strategy}

At each hour~$t$, a forecasting model produces a point prediction $\hat{r}_{t+1}$ for the next-hour log return. The prediction is converted into a trading position through a sign-based rule. In the long-only specification, the strategy holds Bitcoin when the forecast is positive and remains in cash otherwise:
\begin{equation}
\mathrm{pos}_t =
\begin{cases}
1, & \hat{r}_{t+1} > 0, \\
0, & \hat{r}_{t+1} \leq 0.
\end{cases}
\label{eq:lo_position}
\end{equation}
In the long-short specification, negative forecasts generate short positions:
\begin{equation}
\mathrm{pos}_t =
\begin{cases}
1, & \hat{r}_{t+1} > 0, \\
-1, & \hat{r}_{t+1} \leq 0.
\end{cases}
\label{eq:ls_position}
\end{equation}

The position selected at hour~$t$ is applied to the return realised over the next hour. Net strategy returns are computed as:
\begin{equation}
r_{t+1}^{\mathrm{net}} =
\mathrm{pos}_t \cdot \left(e^{r_{t+1}}-1\right)
- c \cdot |\mathrm{pos}_t - \mathrm{pos}_{t-1}|,
\label{eq:net_return}
\end{equation}
where $c=0.001$ denotes the proportional transaction cost per unit of turnover. This corresponds to 10~basis points per position change and is interpreted as a conservative effective cost that includes explicit exchange fees, spread crossing, slippage, and other execution frictions. The assumption is deliberately above the lowest posted Binance USD-M futures fee tiers and is therefore intended to proxy an all-in effective trading cost rather than the exchange fee alone \citep{binancefees}. The backtest does not simulate order-book depth, partial fills, maker--taker order routing, or time-varying bid--ask spreads; transaction-cost sensitivity is therefore examined in Section~\ref{sec:tc_sensitivity}.

The baseline sign rule can generate excessive turnover because hourly forecasts often fluctuate around zero. A cost-aware execution filter is therefore introduced. Let $\mathrm{pos}_t^{*}$ denote the position implied by the sign of the forecast and let $\mathrm{pos}_{t-1}$ denote the current position. The strategy updates the position only if the forecast magnitude is large enough to compensate for the transaction cost implied by the trade:
\begin{equation}
|\hat{r}_{t+1}| >
\lambda \cdot c \cdot |\mathrm{pos}_t^{*} - \mathrm{pos}_{t-1}|,
\label{eq:cost_aware_condition}
\end{equation}
where $\lambda>0$ controls the strictness of the filter. If condition~\eqref{eq:cost_aware_condition} is satisfied, the strategy moves to the forecast-implied position:
\begin{equation}
\mathrm{pos}_t = \mathrm{pos}_t^{*}.
\end{equation}
Otherwise, it keeps the previous position:
\begin{equation}
\mathrm{pos}_t = \mathrm{pos}_{t-1}.
\end{equation}

The rule has a direct economic interpretation. The right-hand side of equation~\eqref{eq:cost_aware_condition} is the minimum forecast magnitude required to justify the cost of changing the position. In the main specification, $c=0.001$ and $\lambda=2.0$. Thus, a long-only entry or exit requires a forecast magnitude above 0.20\%, while a long-short reversal requires a forecast magnitude above 0.40\% because the turnover is two units. The filter is designed to suppress weak signals and concentrate trading activity on higher-confidence forecasts.

\subsection{Walk-forward optimisation}
\label{sec:wfo}

The empirical design uses a rolling walk-forward optimisation protocol. Standard random cross-validation is inappropriate for financial time series because it violates temporal ordering and can leak future information into model selection. In each fold, models are trained on a 12-month window, selected using the following 3-month validation window, and evaluated on a subsequent 3-month test window. The full window then advances by three months. The procedure produces 27 sequential out-of-sample test folds.

Within each fold, temporal integrity is maintained at every stage. Model parameters are estimated only on the training segment. The validation segment is used for hyperparameter tuning, early stopping, and model selection. The test segment is used only once, after the final model has been selected. Feature construction, scaling, target standardisation, and model estimation are repeated independently inside each fold using only information available up to the relevant time boundary.

\FloatBarrier

After the best configuration has been selected on the validation segment, the model is retrained on the combined training-plus-validation sample and then applied to the held-out test window. This final retraining step uses all information available before the test period without compromising the out-of-sample nature of the evaluation.

\subsection{Feature engineering}
\label{sec:features}

Three nested feature sets are evaluated. The first contains only OHLCV variables. The second augments OHLCV data with selected technical-analysis features. The third further adds EGARCH-derived volatility-regime features:
\begin{itemize}
    \item \textbf{OHLCV}: raw open, high, low, close, and volume-related variables.
    \item \textbf{OHLCV+TA}: OHLCV variables plus selected technical indicators.
    \item \textbf{OHLCV+TA+EGARCH}: OHLCV+TA plus conditional-volatility features.
\end{itemize}

Technical indicators are computed causally within each walk-forward fold. For each fold, the feature-engineering routine uses a historical slice with a warm-up buffer sufficient to initialise rolling calculations, computes indicators using information available up to each timestamp, discards the warm-up rows, and then splits the remaining data into training, validation, and test segments. This prevents future observations from entering rolling features.

The technical-indicator candidate pool consists of rolling indicators computed over windows of \(w \in \{3,6,12,24,48,72,168,336\}\) hours, together with six lagged-return variables. Table~\ref{tab:ta_summary_main} summarises the rolling technical-indicator families used in the implementation. The full generated TA-related candidate pool contains 94 variables before selection. To reduce redundancy, candidates are mapped into 10 configured selection groups. Within each walk-forward fold, selection is performed on the training sample only. The training sample is divided into four sequential blocks; within each block, candidates are ranked by absolute Spearman correlation with the aligned next-hour return target. For each candidate, ranks are averaged across blocks, and the candidate with the lowest average rank is retained within its group. The deployed OHLCV+TA dataset therefore adds 10 selected TA-related variables per fold. This procedure favours features with stable training-sample predictive content rather than features that perform well only in one sub-period.

\begin{table}[H]
\centering
\captionsetup{justification=centering,singlelinecheck=false}
\caption{Rolling technical-indicator candidate families}
\label{tab:ta_summary_main}
\small
\renewcommand{\arraystretch}{1.12}
\setlength{\tabcolsep}{9pt}

\begin{tabular}{lll}
\toprule
Group & Indicator & Prefix \\
\midrule
Momentum & Relative Strength Index & \texttt{RSI} \\
Momentum & Rate of Change & \texttt{ROC} \\
Trend & Distance to simple moving average & \texttt{DIST\_SMA} \\
Trend & Normalised MACD & \texttt{MACD} \\
Trend & MACD histogram & \texttt{MACD\_HIST} \\
Volatility & ATR ratio & \texttt{ATR\_RATIO} \\
Volatility & Rolling standard deviation & \texttt{ROLL\_STD} \\
Volatility & Bollinger Band position & \texttt{BB\_POS} \\
Price--volume & VWAP deviation & \texttt{VWAP\_DEV} \\
Volume & OBV slope & \texttt{OBV\_SLOPE} \\
Volume & Money Flow Index & \texttt{MFI} \\
\bottomrule
\end{tabular}

\vspace{0.4em}
\begin{minipage}{0.82\textwidth}
\footnotesize \textit{Note:} Indicators are computed over multiple hourly windows. Six lagged-return variables are also generated but are not listed because they are return-history variables rather than rolling technical-indicator families.
\end{minipage}

\renewcommand{\arraystretch}{1.0}
\end{table}

\FloatBarrier

The EGARCH component is included to provide a compact summary of the current volatility regime. Within each fold, an EGARCH model is fitted only on the training segment. The conditional mean is restricted to a constant, and Student-\(t\) innovations are used to accommodate the heavy-tailed return distribution documented in Section~\ref{sec:descriptive_stats} and Appendix~\ref{app:normality}, Table~\ref{tab:normality}. Four order combinations are considered,
\[
(p,o,q) \in \{(1,1,1),(2,1,1),(1,1,2),(2,1,2)\},
\]
and the preferred order is selected by AIC on the training sample.

The EGARCH log-variance equation is:
\begin{equation}
\log(\sigma_t^2)
=
\omega
+
\sum_{i=1}^{q}\beta_i\log(\sigma_{t-i}^2)
+
\sum_{j=1}^{p}\alpha_j
\left(
\frac{|z_{t-j}|}{\mathbb{E}|z_{t-j}|}
\right)
+
\sum_{k=1}^{o}\gamma_k z_{t-k},
\label{eq:egarch_variance}
\end{equation}
where \(z_t=\varepsilon_t/\sigma_t\) denotes the standardised residual.

The fitted parameters are then held fixed and propagated recursively through the validation and test segments using only information available up to time \(t-1\). No EGARCH refitting is performed inside the validation or test periods. The EGARCH-augmented dataset adds three variables: conditional volatility, log conditional volatility, and the standardised residual.

\subsection{Forecasting models}
\label{sec:models}

This paper compares three model classes that differ fundamentally in how they represent nonlinear structure and temporal dependence: a gradient-boosted tree ensemble (XGBoost), a gated recurrent neural network (LSTM), and an attention-based inverted transformer (iTransformer). All three models are configured as one-step-ahead regression models. At each forecast origin \(t\), the model receives information available up to time \(t\) and outputs a scalar forecast \(\hat{y}_{t+1}\) of the next-hour log return.

The comparison is designed to separate three modelling paradigms. XGBoost treats the problem as tabular prediction and relies on engineered predictors to encode temporal structure. LSTM processes rolling sequences and is therefore able to learn temporal dependence through recurrent hidden states. iTransformer also receives rolling sequences, but it applies attention across input variables rather than across timestamps. This makes the architecture comparison directly relevant to the empirical question: whether the economic value of hourly BTC/USDT forecasts depends on the model class used to extract information from the same market history.

\subsubsection{XGBoost}

XGBoost is used as the tree-based tabular benchmark \citep{chen2016xgboost}. It is a regularised gradient-boosting algorithm that constructs an additive ensemble of regression trees. For a feature vector \(\mathbf{X}_t\), the prediction is written as:
\begin{equation}
\hat{y}_t = \sum_{k=1}^{K} f_k(\mathbf{X}_t),
\qquad
f_k \in \mathcal{F},
\label{eq:xgb_prediction}
\end{equation}
where \(\mathcal{F}\) denotes the space of regression trees and \(K\) is the number of trees in the ensemble.

The model is estimated by minimising a regularised objective:
\begin{equation}
\mathcal{L}
=
\sum_{t=1}^{N}\ell(y_t,\hat{y}_t)
+
\sum_{k=1}^{K}\Omega(f_k),
\label{eq:xgb_objective}
\end{equation}
where \(\ell(y_t,\hat{y}_t)\) is either the MSE or MAE loss. The regularisation term penalises tree complexity:
\begin{equation}
\Omega(f_k)
=
\gamma T_k
+
\frac{1}{2}\lambda_{\mathrm{xgb}}\|w_k\|^2,
\label{eq:xgb_regularisation}
\end{equation}
where \(T_k\) is the number of terminal leaves and \(w_k\) denotes the vector of leaf weights.

XGBoost does not process ordered sequences internally. Any temporal information must therefore be encoded in the input vector through lagged returns, rolling technical indicators, OHLCV-derived variables, and EGARCH-based volatility features. This makes XGBoost a useful benchmark for testing whether engineered tabular predictors are sufficient in a noisy short-horizon cryptocurrency forecasting task. Early stopping is applied on the validation segment within each walk-forward fold.

\subsubsection{Long Short-Term Memory}

LSTM is used as the recurrent neural-network benchmark \citep{hochreiter1997long}. Unlike XGBoost, the LSTM receives a rolling window of past observations and updates a hidden state through time. This allows the model to represent sequential dependence directly rather than relying only on pre-engineered lagged predictors.

For input vector \(x_t\), previous hidden state \(h_{t-1}\), and previous cell state \(c_{t-1}\), the forget gate, input gate, and output gate are:
\begin{equation}
f_t = \sigma(W_f x_t + U_f h_{t-1} + b_f),
\label{eq:lstm_forget_gate}
\end{equation}
\begin{equation}
i_t = \sigma(W_i x_t + U_i h_{t-1} + b_i),
\label{eq:lstm_input_gate}
\end{equation}
\begin{equation}
o_t = \sigma(W_o x_t + U_o h_{t-1} + b_o).
\label{eq:lstm_output_gate}
\end{equation}
The candidate cell update is:
\begin{equation}
\tilde{c}_t =
\tanh(W_c x_t + U_c h_{t-1} + b_c).
\label{eq:lstm_candidate_cell}
\end{equation}
The cell state and hidden state are then updated as:
\begin{equation}
c_t =
f_t \odot c_{t-1}
+
i_t \odot \tilde{c}_t,
\label{eq:lstm_cell_state}
\end{equation}
\begin{equation}
h_t =
o_t \odot \tanh(c_t),
\label{eq:lstm_hidden_state}
\end{equation}
where \(\odot\) denotes element-wise multiplication.

In the implementation used here, one or two LSTM layers are stacked depending on the selected hyperparameters. The final hidden state \(h_t^{(\mathrm{last})}\) is passed through a compact regression head:
\begin{equation}
z_t =
\mathrm{GELU}(W_1 h_t^{(\mathrm{last})} + b_1),
\label{eq:lstm_regression_hidden}
\end{equation}
\begin{equation}
\hat{y}_{t+1}
=
W_2\,\mathrm{Dropout}(z_t) + b_2.
\label{eq:lstm_regression_output}
\end{equation}

The LSTM benchmark is included because recurrent networks are a standard architecture in financial time-series forecasting. In this paper, the model tests whether explicit sequence learning adds economic value beyond the tabular representation used by XGBoost.

\subsubsection{iTransformer}

iTransformer is used as the attention-based neural benchmark \citep{liu2024itransformer}. Its design differs from conventional time-series Transformers. Instead of treating timestamps or temporal patches as tokens, iTransformer treats input variables as tokens and applies self-attention across the feature dimension. This is useful in the present setting because the task is one-step-ahead multivariate forecasting: the relevant information may lie in the interaction between price, volume, technical indicators, and volatility-regime variables rather than in long-horizon trajectory extrapolation.

Let
\[
X_t \in \mathbb{R}^{L \times K}
\]
denote the input sequence at time \(t\), where \(L\) is the lookback length and \(K\) is the number of predictors. A conventional time-series Transformer usually treats each timestamp as a token. iTransformer instead transposes the input:
\begin{equation}
\tilde{X}_t = X_t^\top \in \mathbb{R}^{K \times L}.
\label{eq:itransformer_transpose}
\end{equation}
Thus, each predictor becomes one token represented by its full length-\(L\) history.

The feature-specific histories are projected into a latent representation:
\begin{equation}
Z_0 = \tilde{X}_t W_e + b_e,
\label{eq:itransformer_embedding}
\end{equation}
where \(W_e \in \mathbb{R}^{L \times d}\), \(b_e \in \mathbb{R}^{d}\), and \(Z_0 \in \mathbb{R}^{K \times d}\). In implementation terms, this is a linear projection from the lookback dimension into the model dimension, applied separately to each feature token.

Before the embedding step, per-variable instance normalisation is applied within each input window. For feature \(k\), the mean and standard deviation are computed only over the current lookback window:
\begin{equation}
\mu_{t,k}
=
\frac{1}{L}
\sum_{\ell=1}^{L}
X_{t,\ell,k},
\qquad
s_{t,k}
=
\sqrt{
\frac{1}{L}
\sum_{\ell=1}^{L}
(X_{t,\ell,k}-\mu_{t,k})^2
+
\epsilon
}.
\label{eq:itransformer_instance_stats}
\end{equation}
The normalised input is:
\begin{equation}
X^{\mathrm{norm}}_{t,\ell,k}
=
\frac{
X_{t,\ell,k}-\mu_{t,k}
}{
s_{t,k}
}.
\label{eq:itransformer_instance_norm}
\end{equation}
Because the normalisation is computed only inside the current lookback window, it does not use future information.

The embedded feature tokens are passed through Transformer encoder layers. Multi-head self-attention is applied across the \(K\) feature tokens:
\begin{equation}
\mathrm{Attention}(Q,K,V)
=
\mathrm{softmax}
\left(
\frac{QK^\top}{\sqrt{d_k}}
\right)V.
\label{eq:itransformer_attention}
\end{equation}
Since attention operates across predictors rather than across timestamps, the model learns relationships between variables within the same historical window.

After the encoder stack, the encoded feature representations are mean-pooled across the feature dimension:
\begin{equation}
h_t =
\frac{1}{K}
\sum_{k=1}^{K}
Z_{t,k}^{\mathrm{enc}}.
\label{eq:itransformer_pooling}
\end{equation}
The pooled representation is passed through a regression head:
\begin{equation}
\hat{y}_{t+1}
=
f_{\mathrm{MLP}}(h_t),
\label{eq:itransformer_output}
\end{equation}
which produces a scalar forecast of the next-hour log return.

This architecture is included to test whether cross-variable attention improves realised trading performance in the low-signal, high-noise environment of hourly BTC/USDT returns. The comparison with XGBoost and LSTM therefore evaluates not only model complexity, but also the way each architecture represents information: engineered tabular predictors, recurrent temporal states, or attention across variables.

\subsection{Hyperparameter optimisation}
\label{sec:hpo}

Hyperparameters are tuned independently within each walk-forward fold using Optuna with a Tree-structured Parzen Estimator sampler. Validation loss is used as the Optuna objective, while additional diagnostics such as prediction variance, rank correlation, and realised validation trading performance are recorded for later model-selection analysis.

The search budget is fixed by architecture. XGBoost uses 50 trials per fold, while LSTM and iTransformer use 40 trials per fold. Across the 27 walk-forward folds, this corresponds to 1,350 XGBoost trials and 1,080 trials for each neural architecture per model--feature--loss configuration. The experiment is repeated across three feature tiers and two loss functions.

Early stopping is applied inside each fold using the validation segment. XGBoost uses early stopping with patience of 50 boosting rounds, while the neural-network models use early stopping with patience of 10 epochs. After the preferred configuration is selected on the validation segment, the model is retrained on the combined training-plus-validation sample and then applied once to the held-out test segment.

The full hyperparameter search spaces are reported in Appendix~\ref{app:hpo_space}, Table~\ref{tab:hpo_space}.

\subsection{Performance metrics}
\label{sec:metrics}

Strategy performance is evaluated on the consolidated out-of-sample test returns after transaction costs. Let $r_t$ denote the realised net strategy return. The main reported metrics are annualised compounded return (ARC), annualised standard deviation (ASD), maximum drawdown (MD), maximum loss duration (MLD), Sharpe ratio, $IR^{*}$, $IR^{**}$, and the number of trades.

Annualised compounded return is computed as:
\begin{equation}
\mathrm{ARC} =
\left(\prod_{t=1}^{N}(1+r_t)\right)^{8760/N} - 1,
\label{eq:arc}
\end{equation}
where 8,760 is the number of hours in a calendar year. Annualised standard deviation is:
\begin{equation}
\mathrm{ASD} =
\sqrt{8760}\,
\sqrt{\frac{1}{N-1}\sum_{t=1}^{N}(r_t-\bar{r})^2}.
\label{eq:asd}
\end{equation}

Maximum drawdown is computed from the cumulative equity curve
\[
E_t=\prod_{i=1}^{t}(1+r_i).
\]
It is defined as:
\begin{equation}
\mathrm{MD}
=
\max_{t}
\left(
\frac{\max_{s\leq t}E_s-E_t}{\max_{s\leq t}E_s}
\right).
\label{eq:max_drawdown}
\end{equation}
The formula defines drawdown as a non-negative loss magnitude. In the empirical tables, MD is reported with a negative sign to follow the common reporting convention in which drawdowns are displayed as percentage losses.

Maximum loss duration measures the longest consecutive period during which the equity curve remains below its previous high-water mark. Let
\[
H_t=\max_{s\leq t}E_s
\]
denote the running maximum of the equity curve. The drawdown-state indicator is:
\begin{equation}
I_t=\mathbf{1}\{E_t<H_t\}.
\label{eq:drawdown_indicator}
\end{equation}
MLD is then the length of the longest consecutive sequence for which \(I_t=1\), expressed in years.

The Sharpe ratio is computed as:
\begin{equation}
\mathrm{Sharpe} =
\frac{\mathrm{ARC}-r_f}{\mathrm{ASD}},
\label{eq:sharpe}
\end{equation}
where the annual risk-free rate is set to $r_f=4.2\%$.

The paper also reports two information-ratio-style measures. The first is:
\begin{equation}
IR^{*} = \frac{\mathrm{ARC}}{\mathrm{ASD}}.
\label{eq:ir_star}
\end{equation}
The second is the modified information ratio:
\begin{equation}
IR^{**} =
\frac{\mathrm{ARC}^2 \cdot \mathrm{sign}(\mathrm{ARC})}
{\mathrm{ASD}\cdot|\mathrm{MD}|}.
\label{eq:ir_starstar}
\end{equation}
This metric rewards positive compounded returns while penalising both volatility and drawdown. Since $IR^{**}$ is non-standard and can become numerically unstable when maximum drawdown is small, it is interpreted only as a trading-oriented ranking heuristic. Formal comparisons against buy-and-hold are based on Sharpe-ratio differences, which are easier to interpret and more standard.

A minimum-trade rule is applied throughout the empirical interpretation. Strategies with fewer than 20 trades over the full out-of-sample evaluation period are marked with \(^{\dagger}\), reported for transparency, and excluded from formal inference and substantive conclusions. This rule prevents single-trade or near-degenerate strategies from being interpreted as reliable evidence.

\subsection{Model selection criteria}
\label{sec:selection}

Three validation-set selection rules are considered. The first is loss-based selection, which chooses the configuration with the lowest validation loss:
\begin{equation}
\theta^{*}_{\mathrm{loss}}
=
\arg\min_{\theta} L_{\mathrm{val}}(\theta),
\label{eq:loss_selection}
\end{equation}
where $L_{\mathrm{val}}$ is either MSE or MAE.

The second is information-coefficient selection. It ranks configurations by the Spearman correlation between validation forecasts and realised validation returns:
\begin{equation}
IC_{\mathrm{val}} =
\mathrm{corr}_{\mathrm{Spearman}}(\hat{y}_t,y_t).
\label{eq:ic_selection}
\end{equation}
The configuration with the highest validation information coefficient is selected, with validation loss used as a tie-breaker.

The third is trading-metric selection based on validation-period $IR^{**}$. Validation forecasts are converted into trading positions, transaction costs are deducted, and the resulting validation strategy is scored using equation~\eqref{eq:ir_starstar}. The procedure is applied separately to the long-only and long-short modes because the same forecast sequence can have different economic value under different position mappings.

In implementation, the \(IR^{**}\)-based selector applies a simple admissibility screen. A candidate configuration is treated as admissible if
\[
IR^{**}_{\mathrm{train}}>0,
\qquad
IR^{**}_{\mathrm{val}}>0,
\qquad
IR^{**}_{\mathrm{train}}\geq IR^{**}_{\mathrm{val}}.
\]
These restrictions are used only as a validation-stage stability screen; they are not interpreted as evidence of true profitability. Among admissible candidates, the configuration with the highest validation \(IR^{**}\) is selected. Ties are resolved by choosing the candidate with the smaller absolute train--validation gap,
\[
|IR^{**}_{\mathrm{train}}-IR^{**}_{\mathrm{val}}|.
\]

If no candidate satisfies the admissibility conditions, the selector relaxes the screen in a fixed order. It first selects the candidate with the highest positive validation \(IR^{**}\), regardless of the training score. If no positive validation score is available, it selects the candidate with the highest finite validation \(IR^{**}\). If no finite value is available, it falls back to the loss-based selector.

This selection framework should be interpreted as a transparent validation heuristic rather than a fully nested optimisation procedure. Optuna first searches candidate configurations using validation loss; the same validation segment is then used to compare loss-based, IC-based, and $IR^{**}$-based selectors. Substantive conclusions are therefore based only on held-out test-period realised returns, which remain untouched during hyperparameter tuning, early stopping, and model selection.

\subsection{Statistical inference}
\label{sec:inference}

Statistical inference is based on paired circular block-bootstrap tests applied to realised hourly return differentials. For two strategies $A$ and $B$, the return differential is:
\begin{equation}
d_t = r_t^{A} - r_t^{B}.
\label{eq:return_differential}
\end{equation}
The bootstrap resamples contiguous blocks of the observed differential series with replacement, wrapping around the sample boundary when needed. This preserves local dependence in hourly returns while generating pseudo-series of the same length as the original sample.

The procedure is repeated for block lengths of 24, 72, and 168 hours, corresponding to one-day, three-day, and seven-day dependence windows. Each test uses 10,000 bootstrap replications, so the minimum non-zero $p$-value resolution is approximately 0.0001. Where multiple related comparisons are conducted, Holm correction is applied to control the family-wise error rate.

The direction of the test depends on the empirical question. Transaction-cost and cost-aware execution comparisons use one-sided tests because the hypotheses specify a directional effect. Comparisons of loss functions and model-selection criteria use two-sided tests because the question is whether the modelling choice changes realised performance, not whether one specific choice must dominate. In addition to mean return differentials, the paper reports bootstrap confidence intervals for Sharpe-ratio differences as complementary evidence.

% ====================================================================
% 5. EMPIRICAL RESULTS
% ====================================================================
\section{Empirical results}
\label{sec:results}

\subsection{H1: Transaction costs and naive machine-learning trading}
\label{sec:h1}

The first empirical test examines whether naive sign-based machine-learning strategies remain profitable once proportional transaction costs are imposed. The comparison uses the OHLCV+TA+EGARCH feature set, MSE loss, and the loss-best selection rule. Forecasts are converted into positions using the baseline sign rule, without the cost-aware execution filter.

\begin{figure}[H]
\centering
\captionsetup{justification=centering,singlelinecheck=false}
\caption{Impact of transaction costs on naive strategy performance}
\includegraphics[width=\textwidth]{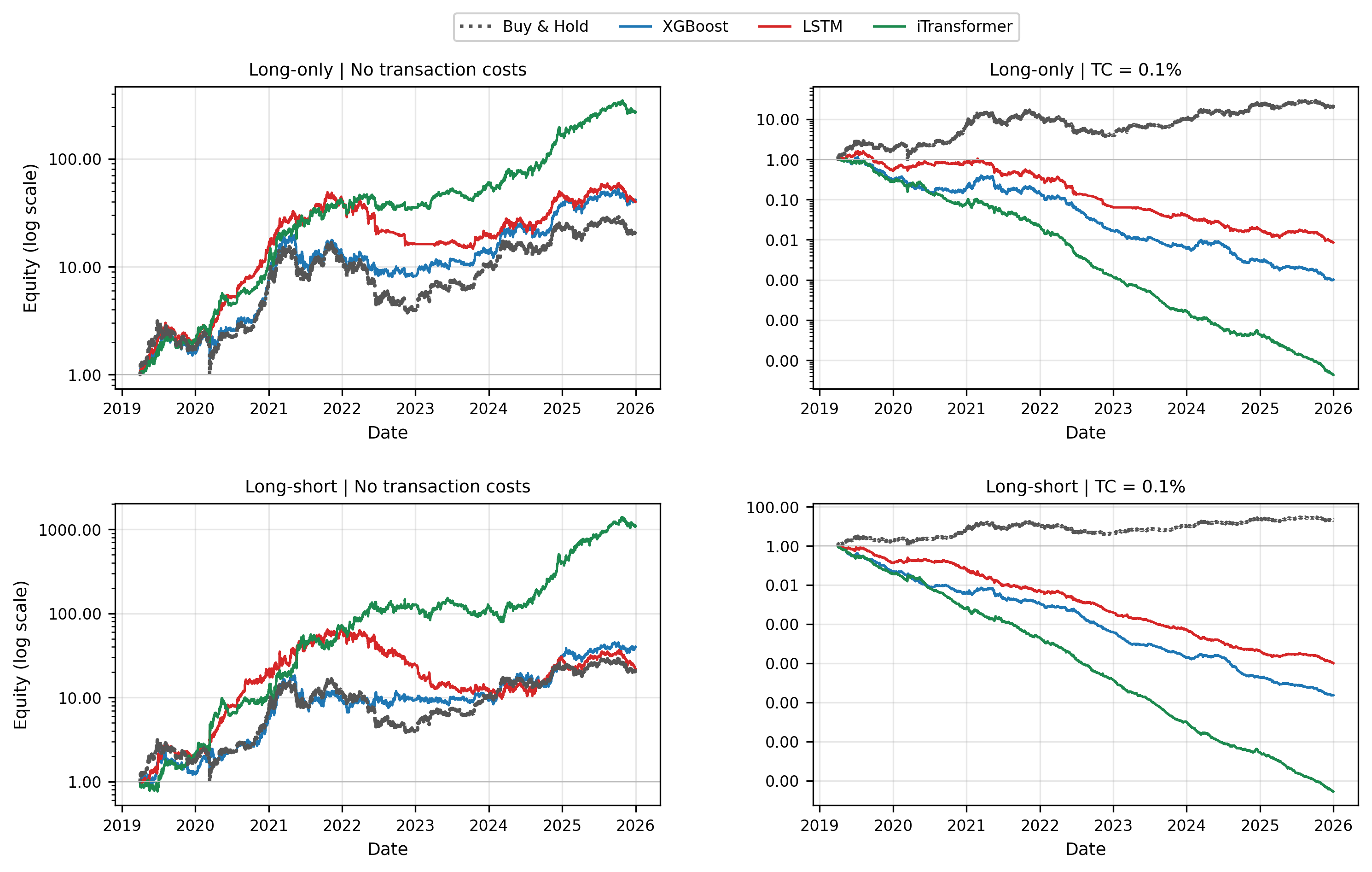}

\vspace{0.4em}
\begin{minipage}{0.95\textwidth}
\footnotesize \textit{Note:} Equity curves for long-only and long-short strategies. The left column shows results without transaction costs, while the right column reports results with transaction costs of 10~basis points per unit of turnover. Results use the OHLCV+TA+EGARCH feature set, MSE loss, and loss-best selection.
\end{minipage}

\label{fig:h1_equity}
\end{figure}

Figure~\ref{fig:h1_equity} shows the central result. In the frictionless setting, all three models generate positive gross equity curves in at least one trading mode. Once transaction costs of 10~basis points are imposed, the picture reverses: every naive ML strategy collapses into a persistent decline. The buy-and-hold benchmark, which generates almost no turnover, is largely unaffected.

\FloatBarrier

Table~\ref{tab:h1_metrics} quantifies the deterioration. Under zero transaction costs, XGBoost in the long-only mode delivers an annualised return of 73.50\%, while iTransformer in the long-short mode reaches 181.76\%. After transaction costs are applied, the same configurations fall to $-64.00\%$ and $-98.62\%$, respectively. The damage is largest for strategies with high turnover, especially iTransformer, whose frequent signal reversals produce the highest cost drag. The 24-hour momentum benchmark also performs poorly after costs, confirming that the turnover problem is not specific to machine-learning forecasts.
\begin{table}[H]
\centering
\captionsetup{justification=centering,singlelinecheck=false}
\caption{Naive strategy performance before and after transaction costs}
\label{tab:h1_metrics}
\footnotesize
\renewcommand{\arraystretch}{0.86}
\setlength{\tabcolsep}{2.4pt}

\begin{tabular}{llrrrrrrrr}
\toprule
Model & TC & ARC & ASD & MD & MLD & SR & IR$^{*}$ & IR$^{**}$ & Trades \\
\midrule

\multicolumn{10}{l}{\textit{LO, no transaction costs}} \\
Buy \& Hold & -- & 56.35 & 63.83 & -77.80 & 2.33 & 0.82 & 0.88 & 0.64 & 1 \\
24h Momentum & 0 & 31.16 & 42.75 & -60.68 & 0.15 & 0.63 & 0.73 & 0.37 & 5984 \\
XGBoost & 0 & 73.50 & 54.58 & -59.81 & 2.81 & 1.27 & 1.35 & 1.66 & 10619 \\
LSTM & 0 & 72.43 & 47.11 & -70.16 & 3.16 & 1.45 & 1.54 & 1.59 & 8448 \\
iTransformer & 0 & \textbf{129.07} & 48.24 & \textbf{-33.44} & 0.95 & \textbf{2.59} & \textbf{2.68} & \textbf{10.33} & 17942 \\

\addlinespace[1pt]
\multicolumn{10}{l}{\textit{LO, with transaction costs}} \\
Buy \& Hold & -- & \textbf{56.30} & 63.83 & \textbf{-77.80} & 2.33 & \textbf{0.82} & \textbf{0.88} & \textbf{0.64} & 1 \\
24h Momentum & 0.1\% & -45.93 & 42.84 & -99.26 & 6.52 & -1.17 & -1.07 & -0.50 & 5984 \\
XGBoost & 0.1\% & -64.00 & 54.71 & -99.92 & 6.48 & -1.25 & -1.17 & -0.75 & 10619 \\
LSTM & 0.1\% & -50.65 & 47.21 & -99.47 & 6.48 & -1.16 & -1.07 & -0.55 & 8448 \\
iTransformer & 0.1\% & -83.93 & 48.45 & -100.00 & 6.75 & -1.82 & -1.73 & -1.45 & 17942 \\

\addlinespace[1pt]
\multicolumn{10}{l}{\textit{LS, no transaction costs}} \\
Buy \& Hold & -- & 56.35 & 63.83 & -77.80 & 2.33 & 0.82 & 0.88 & 0.64 & 1 \\
24h Momentum & 0 & -12.04 & 63.82 & -91.14 & 0.21 & -0.25 & -0.19 & -0.02 & 5984 \\
XGBoost & 0 & 72.55 & 63.82 & -64.13 & 3.08 & 1.07 & 1.14 & 1.29 & 10620 \\
LSTM & 0 & 57.93 & 63.83 & -85.60 & 4.00 & 0.84 & 0.91 & 0.61 & 8449 \\
iTransformer & 0 & \textbf{181.76} & 63.82 & \textbf{-48.78} & 1.16 & \textbf{2.78} & \textbf{2.85} & \textbf{10.61} & 17943 \\

\addlinespace[1pt]
\multicolumn{10}{l}{\textit{LS, with transaction costs}} \\
Buy \& Hold & -- & \textbf{56.30} & \textbf{63.83} & \textbf{-77.80} & 2.33 & \textbf{0.82} & \textbf{0.88} & \textbf{0.64} & 1 \\
24h Momentum & 0.1\% & -85.06 & 64.05 & -100.00 & 6.47 & -1.39 & -1.33 & -1.13 & 5984 \\
XGBoost & 0.1\% & -92.58 & 64.27 & -100.00 & 6.75 & -1.51 & -1.44 & -1.33 & 10620 \\
LSTM & 0.1\% & -87.08 & 64.14 & -100.00 & 6.75 & -1.42 & -1.36 & -1.18 & 8449 \\
iTransformer & 0.1\% & -98.62 & 64.40 & -100.00 & 6.76 & -1.60 & -1.53 & -1.51 & 17943 \\

\bottomrule
\end{tabular}

\vspace{0.4em}
\begin{minipage}{0.95\textwidth}
\footnotesize \textit{Note:} LO = long-only; LS = long-short; TC = transaction costs. ARC, ASD, and MD are percentages; MLD is reported in years; SR, IR$^{*}$, and IR$^{**}$ are ratios. Results use the OHLCV+TA+EGARCH feature set, MSE loss, and loss-best selection. Bold values indicate the strongest result within each panel. The 24h Momentum benchmark is long when the cumulative return over the previous 24 hours is positive; in the long-only mode it is otherwise flat, while in the long-short mode it is short when non-positive.
\end{minipage}

\renewcommand{\arraystretch}{1.0}
\end{table}

\FloatBarrier

Inference follows the paired circular block-bootstrap procedure described in Section~\ref{sec:inference}. For H1, the tested return differential is defined as:
\begin{equation}
d_t = r_t^{\mathrm{NoTC}} - r_t^{\mathrm{TC}}.
\label{eq:h1_tc_diff}
\end{equation}
The one-sided alternative is that the no-cost strategy has a higher mean realised return than the transaction-cost-adjusted strategy. Rejection therefore indicates that transaction costs significantly reduce realised strategy returns.

Table~\ref{tab:h1_bootstrap} reports the conservative 168-hour block specification, which corresponds to a weekly-dependence assumption. Shorter 24-hour and 72-hour block lengths lead to the same conclusion and are reported in Appendix~\ref{app:h1_bootstrap_robustness}. All model--mode comparisons reject after Holm correction, confirming that the transaction-cost effect is both economically large and statistically robust.

\begin{table}[H]
\centering
\captionsetup{justification=centering,singlelinecheck=false}
\caption{Bootstrap evidence on transaction-cost impact, 168-hour blocks}
\label{tab:h1_bootstrap}
\footnotesize
\renewcommand{\arraystretch}{0.90}
\setlength{\tabcolsep}{6pt}

\begin{tabular}{llrrc}
\toprule
Model & Mode & Mean diff. (bps) & 95\% CI (bps) & Holm-Rej. \\
\midrule
XGBoost & LO & 1.79 & [1.67, 1.92] & Yes$^{***}$ \\
XGBoost & LS & 3.59 & [3.33, 3.84] & Yes$^{***}$ \\
LSTM & LO & 1.43 & [1.33, 1.52] & Yes$^{***}$ \\
LSTM & LS & 2.86 & [2.66, 3.05] & Yes$^{***}$ \\
iTransformer & LO & 3.03 & [2.97, 3.09] & Yes$^{***}$ \\
iTransformer & LS & 6.06 & [5.94, 6.18] & Yes$^{***}$ \\
\bottomrule
\end{tabular}

\vspace{0.4em}
\begin{minipage}{0.88\textwidth}
\footnotesize \textit{Note:} LO = long-only; LS = long-short. The table reports paired circular block-bootstrap tests using 168-hour blocks and 10,000 replications. Mean differences are expressed in basis points per hour and are computed as no-cost strategy returns minus transaction-cost-adjusted strategy returns. Holm-Rej. reports Holm-adjusted rejection decisions. $^{***}$ denotes rejection at the 1\% level.
\end{minipage}

\renewcommand{\arraystretch}{1.0}
\end{table}

\FloatBarrier

H1 is supported. Gross machine-learning signals exist, but naive sign-based execution converts them into high-turnover strategies that cannot survive realistic trading frictions. The hourly cost drag is positive for every model and trading mode, ranging from 1.43 to 6.06 basis points per hour, and is systematically larger in the long-short setting because position reversals generate greater turnover. This result establishes the central motivation for the cost-aware execution rule tested next.

\subsection{H2: Cost-aware execution}
\label{sec:h2}

Having established that transaction costs destroy naive sign-based strategies, the next test asks whether trading only on sufficiently strong forecasts can restore net-of-cost performance. The cost-aware filter introduced in Section~\ref{sec:trading_strategy} suppresses weak signals whose forecast magnitude is unlikely to compensate for the transaction cost generated by the position change.

\begin{figure}[H]
\centering
\captionsetup{justification=centering,singlelinecheck=false}
\caption{Impact of cost-aware filtering on strategy performance}
\includegraphics[width=\textwidth]{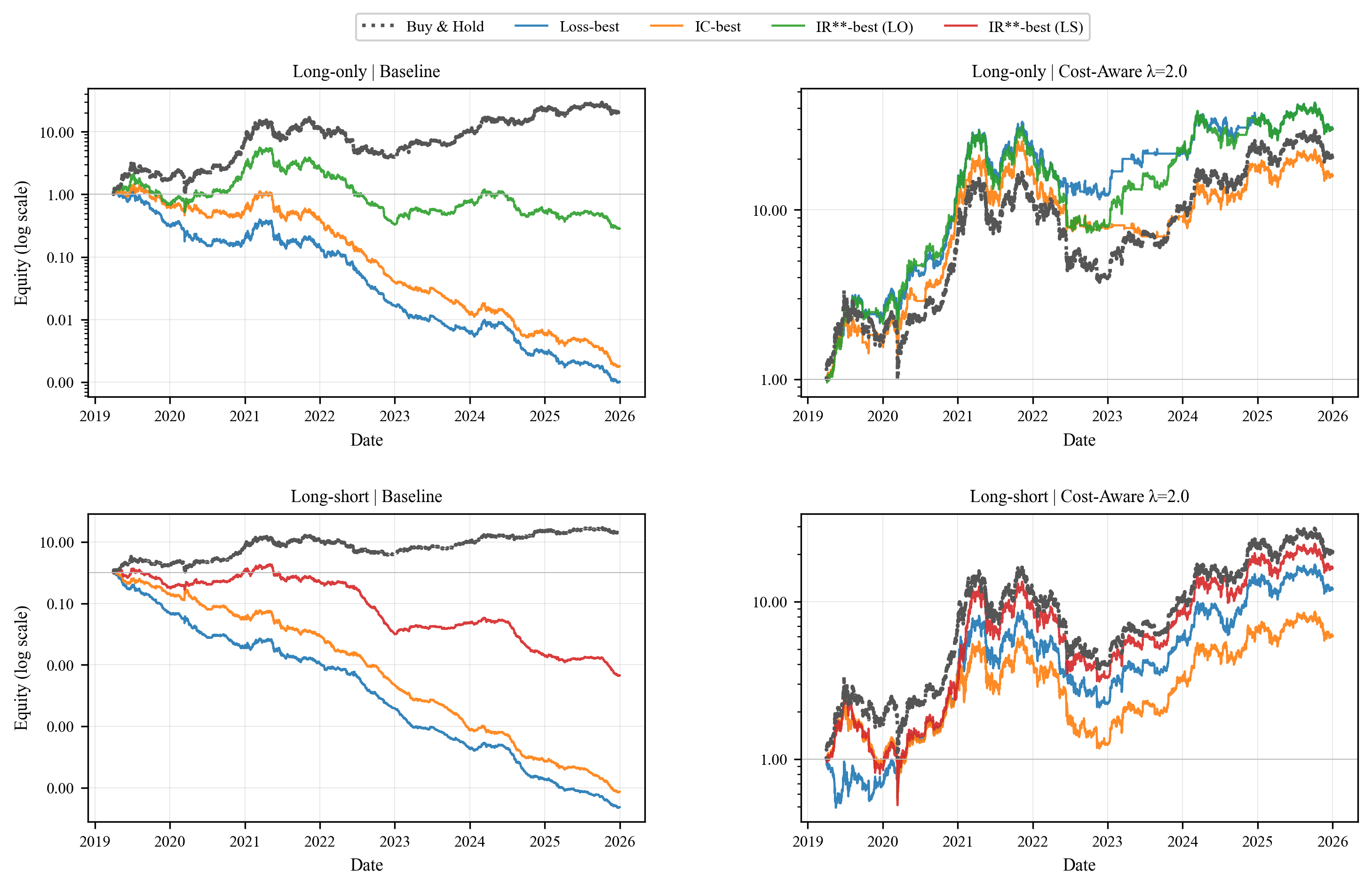}

\vspace{0.4em}
\begin{minipage}{0.95\textwidth}
\footnotesize \textit{Note:} Equity curves for XGBoost strategies under baseline execution and cost-aware filtering with $\lambda=2.0$. Results use the OHLCV+TA+EGARCH feature set, MSE loss, and loss-best, IC-best, and mode-specific IR$^{**}$-best selection rules. The filter suppresses weak forecast signals before transaction costs are applied.
\end{minipage}

\label{fig:h2_equity}
\end{figure}

Figure~\ref{fig:h2_equity} compares baseline and cost-aware XGBoost strategies under the OHLCV+TA+EGARCH feature set, MSE loss, and three validation-set selection rules. The visual pattern is clear: baseline strategies deteriorate after costs, while the cost-aware versions generate substantially more stable equity curves.

\FloatBarrier

Table~\ref{tab:h2_metrics} reports the corresponding performance metrics. In the long-only loss-best specification, ARC changes from $-64.00\%$ under baseline execution to $65.40\%$ under cost-aware execution, while the number of trades falls from 10,619 to 251. Similar baseline-to-filter improvements appear under the IC-best and IR$^{**}$-best selectors. In the long-short setting, the filter also restores positive performance relative to the baseline, although the filtered strategies remain weaker than buy-and-hold in risk-adjusted terms.

\begin{table}[H]
\centering
\captionsetup{justification=centering,singlelinecheck=false}
\caption{Performance comparison: baseline, momentum, and cost-aware strategies}
\label{tab:h2_metrics}
\footnotesize
\renewcommand{\arraystretch}{0.86}
\setlength{\tabcolsep}{2.2pt}

\begin{tabular}{lllrrrrrrrr}
\toprule
Selector & Mode & Type & ARC & ASD & MD & MLD & SR & IR$^{*}$ & IR$^{**}$ & Trades \\
\midrule

\multicolumn{11}{l}{\textit{LO}} \\
Buy \& Hold & LO & --
& 56.30 & 63.83 & -77.80 & 2.33 & 0.82 & 0.88 & 0.64 & 1 \\

24h Momentum & LO & --
& -45.93 & \textbf{42.84} & -99.26 & 6.52 & -1.17 & -1.07 & -0.50 & 5984 \\

Loss-best & LO & BL
& -64.00 & 54.71 & -99.92 & 6.48 & -1.25 & -1.17 & -0.75 & 10619 \\

Loss-best & LO & CA
& \textbf{65.40} & 56.22 & \textbf{-67.42} & 2.31 & \textbf{1.09} & \textbf{1.16} & \textbf{1.13} & 251 \\

IC-best & LO & BL
& -60.81 & 54.78 & -99.88 & 6.48 & -1.19 & -1.11 & -0.68 & 10347 \\

IC-best & LO & CA
& 50.48 & 55.57 & -74.01 & 4.20 & 0.83 & 0.91 & 0.62 & 261 \\

IR$^{**}$-best & LO & BL
& -17.08 & 58.02 & -95.01 & 4.80 & -0.37 & -0.29 & -0.05 & 5327 \\

IR$^{**}$-best & LO & CA
& 65.34 & 57.94 & -76.64 & \textbf{2.30} & 1.06 & 1.13 & 0.96 & 197 \\

\addlinespace[1pt]
\multicolumn{11}{l}{\textit{LS}} \\
Buy \& Hold & LS & --
& \textbf{56.30} & 63.83 & -77.80 & 2.33 & \textbf{0.82} & \textbf{0.88} & \textbf{0.64} & 1 \\

24h Momentum & LS & --
& -85.06 & 64.05 & -100.00 & 6.47 & -1.39 & -1.33 & -1.13 & 5984 \\

Loss-best & LS & BL
& -92.58 & 64.27 & -100.00 & 6.75 & -1.51 & -1.44 & -1.33 & 10620 \\

Loss-best & LS & CA
& 44.37 & 63.73 & \textbf{-75.73} & \textbf{2.31} & 0.63 & 0.70 & 0.41 & 60 \\

IC-best & LS & BL
& -91.20 & 64.24 & -100.00 & 6.75 & -1.48 & -1.42 & -1.29 & 10347 \\

IC-best & LS & CA
& 30.41 & 63.70 & -80.80 & 3.00 & 0.41 & 0.48 & 0.18 & 47 \\

IR$^{**}$-best & LS & BL
& -68.02 & 64.09 & -99.98 & 4.67 & -1.13 & -1.06 & -0.72 & 6141 \\

IR$^{**}$-best & LS & CA
& 51.05 & \textbf{63.67} & -82.60 & 2.31 & 0.74 & 0.80 & 0.50 & 47 \\

\bottomrule
\end{tabular}

\vspace{0.4em}
\begin{minipage}{0.95\textwidth}
\footnotesize \textit{Note:} LO = long-only; LS = long-short; BL = baseline execution; CA = cost-aware execution. Results are for XGBoost with the OHLCV+TA+EGARCH feature set, MSE loss, transaction costs of 10 basis points, and cost-aware threshold $\lambda=2.0$. ARC, ASD, and MD are percentages; MLD is reported in years; SR, IR$^{*}$, and IR$^{**}$ are ratios. Bold values indicate the strongest outcome within each trading mode. The 24h Momentum benchmark uses the cumulative return over the previous 24 hours and is evaluated under the same transaction-cost assumptions.
\end{minipage}

\renewcommand{\arraystretch}{1.0}
\end{table}

\FloatBarrier

The comparison shows that the predictive signal was not sufficient on its own. What mattered was whether the strategy avoided acting on weak forecasts. The 24-hour momentum benchmark reinforces this interpretation: although it is a natural trend-following rule for hourly BTC data, it loses heavily after transaction costs because it trades too often. Cost-aware execution therefore improves the machine-learning strategies not by adding a new predictive signal, but by changing which forecasts are allowed to become trades.

Inference follows the paired circular block-bootstrap procedure described in Section~\ref{sec:inference}. For each selector--mode pair, the tested return differential is:
\begin{equation}
d_t = r_t^{\mathrm{CA}} - r_t^{\mathrm{BL}}.
\label{eq:h2_return_diff}
\end{equation}
The one-sided alternative is that cost-aware execution produces a higher mean realised return than baseline execution. Sharpe-ratio differences are tested as complementary evidence.

Table~\ref{tab:h2_bootstrap} reports the 168-hour block specification. Appendix~\ref{app:h2_bootstrap_robustness} reports the 24-hour and 72-hour block specifications, which lead to the same qualitative conclusion as Table~\ref{tab:h2_bootstrap}. Cost-aware execution improves both mean realised returns and Sharpe ratios for every selector--mode pair, with all Holm-adjusted tests rejecting at the 1\% level.

\begin{table}[H]
\centering
\captionsetup{justification=centering,singlelinecheck=false}
\caption{Bootstrap evidence for cost-aware improvement, 168-hour blocks}
\label{tab:h2_bootstrap}
\footnotesize
\renewcommand{\arraystretch}{0.88}
\setlength{\tabcolsep}{3.0pt}

\begin{tabular}{llrrrrrr}
\toprule
Selector & Mode & Trades CA & Diff. (bps) & CI (bps) & $\Delta SR$ & SR CI & Holm-Rej. \\
\midrule

Loss-best & LO
& 251 & 1.75 & [1.42, 2.07] & 2.34 & [1.52, 3.55] & Yes$^{***}$ \\

IC-best & LO
& 261 & 1.54 & [1.24, 1.85] & 2.02 & [1.30, 3.13] & Yes$^{***}$ \\

IR$^{**}$-best & LO
& 197 & 0.79 & [0.50, 1.07] & 1.42 & [0.81, 2.34] & Yes$^{***}$ \\

Loss-best & LS
& 60 & 3.38 & [2.83, 3.93] & 2.14 & [1.30, 3.48] & Yes$^{***}$ \\

IC-best & LS
& 47 & 3.08 & [2.37, 3.74] & 1.90 & [1.08, 3.18] & Yes$^{***}$ \\

IR$^{**}$-best & LS
& 47 & 1.77 & [1.24, 2.32] & 1.86 & [1.05, 3.17] & Yes$^{***}$ \\

\bottomrule
\end{tabular}

\vspace{0.4em}
\begin{minipage}{0.95\textwidth}
\footnotesize \textit{Note:} LO = long-only; LS = long-short; CA = cost-aware execution. The table reports paired circular block-bootstrap results using 168-hour blocks and 10,000 replications. Diff. is the mean hourly return difference in basis points, computed as cost-aware minus baseline returns. CI (bps) is the 95\% confidence interval for the mean return difference. $\Delta SR$ is the Sharpe ratio of the cost-aware strategy minus the Sharpe ratio of the baseline strategy. SR CI is the 95\% confidence interval for $\Delta SR$. Holm-Rej. reports whether both the mean return-difference test and the Sharpe-difference test reject after Holm correction. $^{***}$ denotes rejection at the 1\% level.
\end{minipage}

\renewcommand{\arraystretch}{1.0}
\end{table}

\FloatBarrier

These results support H2. Cost-aware filtering significantly improves realised net returns and Sharpe ratios relative to naive baseline execution in both trading modes and under all three selection rules. The improvement is economically large: in the long-only loss-best case, the filter changes the strategy from a severe post-cost loss to positive annualised performance with a Sharpe ratio above one. In the long-short case, the improvement is also substantial, although the resulting strategies remain less attractive than the strongest long-only specifications.

A final benchmark check compares the reported cost-aware strategies against buy-and-hold. Since $IR^{**}$ is non-standard, this comparison is based on Sharpe-ratio differences:
\begin{equation}
\Delta SR = SR^{\mathrm{CA}} - SR^{\mathrm{BH}}.
\label{eq:h2_bh_sharpe_diff}
\end{equation}
Table~\ref{tab:h2_bh_bootstrap} reports the 168-hour block-bootstrap results. None of the reported cost-aware strategies significantly outperforms buy-and-hold in Sharpe-ratio terms after Holm correction.

\begin{table}[H]
\centering
\captionsetup{justification=centering,singlelinecheck=false}
\caption{Bootstrap comparison of cost-aware strategies against buy-and-hold, 168-hour blocks}
\label{tab:h2_bh_bootstrap}
\footnotesize
\renewcommand{\arraystretch}{0.90}
\setlength{\tabcolsep}{7pt}

\begin{tabular}{llrrrr}
\toprule
Selector & Mode & $\Delta SR$ & SR CI & $p_{\mathrm{Holm}}$ & Holm-Rej. \\
\midrule
IR$^{**}$-best & LO & 0.24 & [-0.36, 0.85] & 0.89 & No \\
Loss-best & LO & 0.27 & [-0.46, 0.95] & 0.89 & No \\
IR$^{**}$-best & LS & -0.08 & [-0.82, 0.57] & 1.00 & No \\
Loss-best & LS & -0.19 & [-1.22, 0.75] & 1.00 & No \\
\bottomrule
\end{tabular}

\vspace{0.4em}
\begin{minipage}{0.88\textwidth}
\footnotesize \textit{Note:} LO = long-only; LS = long-short. The table reports paired circular block-bootstrap results using 168-hour blocks and 10,000 replications. $\Delta SR$ is computed as the Sharpe ratio of the cost-aware strategy minus the Sharpe ratio of buy-and-hold. SR CI is the 95\% confidence interval for $\Delta SR$. Holm-adjusted $p$-values are reported. This benchmark comparison is interpreted as a supplementary check rather than as the main H2 test, which compares cost-aware execution against unfiltered baseline execution.
\end{minipage}

\renewcommand{\arraystretch}{1.0}
\end{table}

\FloatBarrier

This distinction matters. H2 is strongly supported in its direct form: the cost-aware filter materially improves the unfiltered machine-learning trading rule. However, the broader investment claim is more cautious. The filter restores the viability of ML-based trading relative to naive execution, but it does not establish statistically significant Sharpe-ratio outperformance over buy-and-hold. The main contribution of the filter is therefore execution discipline rather than a guaranteed benchmark-beating strategy.

\subsection{H3: Feature enrichment}
\label{sec:h3}

The third empirical test examines whether richer input features improve realised trading performance once transaction costs and cost-aware execution are already in place. Three nested feature sets are compared: OHLCV, OHLCV+TA, and OHLCV+TA+EGARCH. The comparison uses XGBoost with MSE loss and cost-aware execution with $\lambda=2.0$. Results are reported for loss-based selection and mode-specific IR$^{**}$-based selection.

\begin{figure}[H]
\centering
\captionsetup{justification=centering,singlelinecheck=false}
\caption{Impact of feature richness on cost-aware strategy performance}
\includegraphics[width=\textwidth]{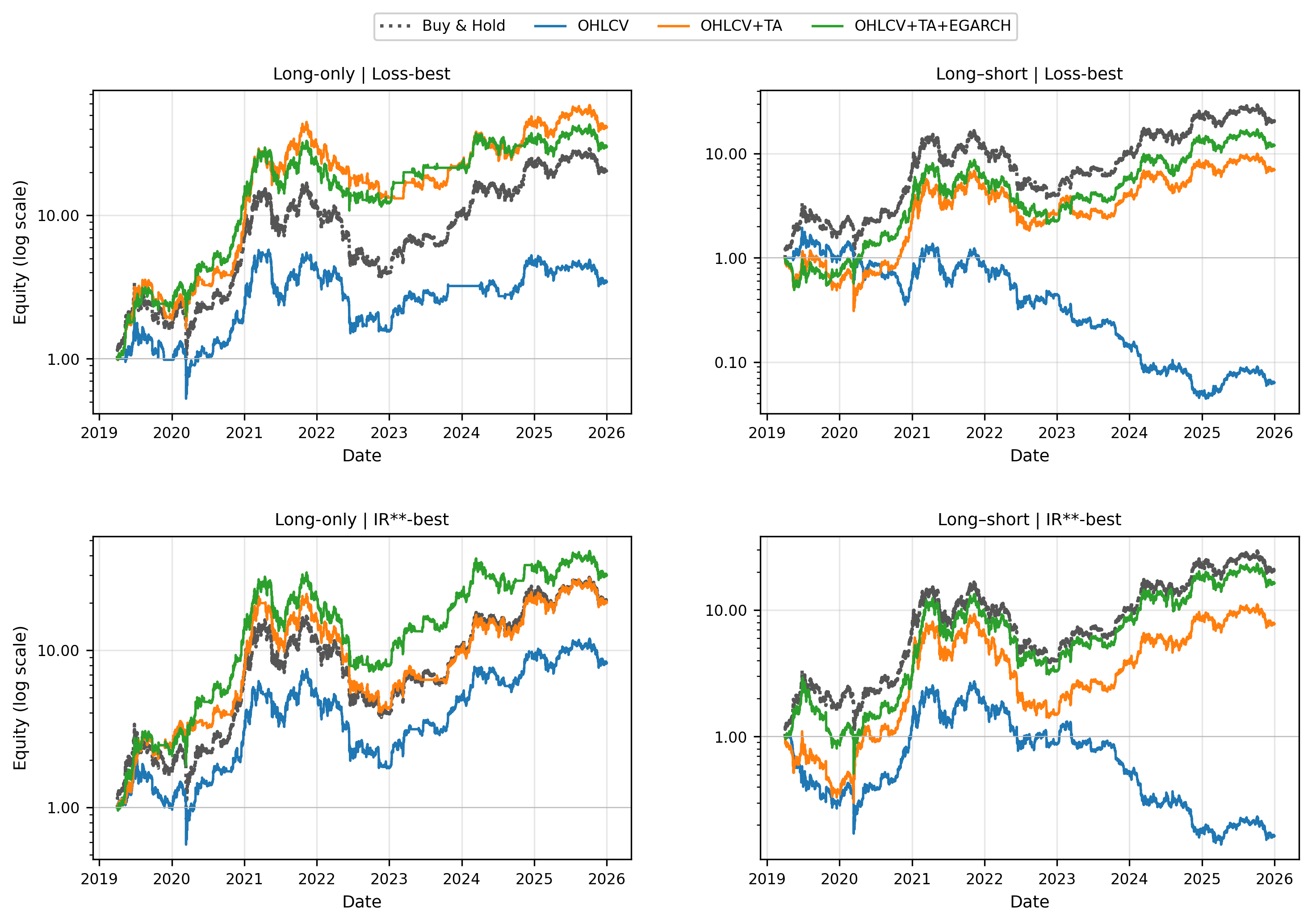}

\vspace{0.4em}
\begin{minipage}{0.95\textwidth}
\footnotesize \textit{Note:} Equity curves for OHLCV, OHLCV+TA, and OHLCV+TA+EGARCH specifications under cost-aware execution with $\lambda=2.0$. Results use XGBoost with MSE loss. The top row reports loss-based selection, while the bottom row reports mode-specific IR$^{**}$-based selection. The left column shows long-only strategies and the right column shows long-short strategies.
\end{minipage}

\label{fig:h3_equity}
\end{figure}

Figure~\ref{fig:h3_equity} shows that feature enrichment can improve performance, but the effect is conditional rather than uniform. In the long-only setting, adding technical indicators improves the equity curve relative to raw OHLCV. Adding EGARCH-derived volatility features produces mixed results: it improves the IR$^{**}$-selected long-only strategy, but does not uniformly dominate OHLCV+TA under loss-based selection. In the long-short setting, the raw OHLCV strategies are too sparse or too weak to support a reliable conclusion, while the richer feature sets generate positive performance in selected configurations.

\FloatBarrier

Table~\ref{tab:h3_metrics} reports the corresponding performance metrics. In the long-only loss-best comparison, OHLCV+TA delivers the strongest result, with ARC of 73.25\%, Sharpe ratio of 1.22, and IR$^{**}$ of 1.34. The EGARCH-augmented specification remains profitable but is slightly weaker in this selector. Under IR$^{**}$-based selection, however, OHLCV+TA+EGARCH becomes the strongest long-only feature tier, with ARC of 65.34\% and IR$^{**}$ of 0.96. In the long-short setting, raw OHLCV configurations fall below the minimum-trade threshold and are therefore treated as descriptive only. Once technical indicators are included, long-short performance becomes positive, with the strongest point estimates obtained by the EGARCH-augmented tier.

\begin{table}[H]
\centering
\captionsetup{justification=centering,singlelinecheck=false}
\caption{Performance comparison across feature specifications}
\label{tab:h3_metrics}
\footnotesize
\renewcommand{\arraystretch}{0.86}
\setlength{\tabcolsep}{2.2pt}

\begin{tabular}{llrrrrrrrr}
\toprule
Dataset & Selector & ARC & ASD & MD & MLD & SR & IR$^{*}$ & IR$^{**}$ & Trades \\
\midrule

\multicolumn{10}{l}{\textit{LO}} \\
Buy \& Hold & --
& 56.30 & 63.83 & -77.80 & 2.33 & 0.82 & 0.88 & 0.64 & 1 \\

OHLCV & Loss-best
& 19.96 & \textbf{56.75} & -74.27 & 4.67 & 0.27 & 0.34 & 0.09 & 117 \\

OHLCV+TA & Loss-best
& \textbf{73.25} & 56.75 & -70.72 & 3.03 & \textbf{1.22} & \textbf{1.29} & \textbf{1.34} & 277 \\

OHLCV+TA+EGARCH & Loss-best
& 65.40 & 56.22 & \textbf{-67.42} & \textbf{2.31} & 1.09 & 1.16 & 1.13 & 251 \\

\addlinespace[1pt]

OHLCV & IR$^{**}$-best
& 36.69 & \textbf{57.54} & \textbf{-76.64} & 2.31 & 0.54 & 0.61 & 0.29 & 143 \\

OHLCV+TA & IR$^{**}$-best
& 55.87 & 57.54 & -82.36 & 3.10 & 0.90 & 0.97 & 0.66 & 215 \\

OHLCV+TA+EGARCH & IR$^{**}$-best
& \textbf{65.34} & 57.94 & \textbf{-76.64} & \textbf{2.30} & \textbf{1.06} & \textbf{1.13} & \textbf{0.96} & 197 \\

\addlinespace[1pt]
\multicolumn{10}{l}{\textit{LS}} \\
Buy \& Hold & --
& 56.30 & 63.83 & -77.80 & 2.33 & 0.82 & 0.88 & 0.64 & 1 \\

OHLCV$^{\dagger}$ & Loss-best
& -33.58 & 63.38 & -97.72 & 6.52 & -0.60 & -0.53 & -0.18 & 13 \\

OHLCV+TA & Loss-best
& 33.29 & 63.75 & \textbf{-73.87} & 2.34 & 0.46 & 0.52 & 0.24 & 66 \\

OHLCV+TA+EGARCH & Loss-best
& \textbf{44.37} & \textbf{63.73} & -75.73 & \textbf{2.31} & \textbf{0.63} & \textbf{0.70} & \textbf{0.41} & 60 \\

\addlinespace[1pt]

OHLCV$^{\dagger}$ & IR$^{**}$-best
& -23.56 & 63.53 & -94.90 & 4.14 & -0.44 & -0.37 & -0.09 & 10 \\

OHLCV+TA & IR$^{**}$-best
& 35.37 & 63.74 & \textbf{-82.60} & 3.10 & 0.49 & 0.55 & 0.23 & 64 \\

OHLCV+TA+EGARCH & IR$^{**}$-best
& \textbf{51.05} & \textbf{63.67} & \textbf{-82.60} & \textbf{2.31} & \textbf{0.74} & \textbf{0.80} & \textbf{0.50} & 47 \\

\bottomrule
\end{tabular}

\vspace{0.4em}
\begin{minipage}{0.95\textwidth}
\footnotesize \textit{Note:} LO = long-only; LS = long-short. Results are for XGBoost with MSE loss under cost-aware execution with $\lambda=2.0$. ARC, ASD, and MD are percentages; MLD is reported in years; SR, IR$^{*}$, and IR$^{**}$ are ratios. $^{\dagger}$ denotes fewer than 20 trades; these rows are reported for transparency but excluded from formal inference. Bold values indicate the strongest reliable result within each selector--mode group; buy-and-hold is shown only as a passive benchmark.
\end{minipage}

\renewcommand{\arraystretch}{1.0}
\end{table}

\FloatBarrier

The evidence suggests that feature enrichment helps, but only conditionally. Technical indicators add economically useful information in the long-only loss-best setting. EGARCH-derived features are more ambiguous: they improve some selected configurations, especially under IR$^{**}$-based selection, but do not provide a uniformly robust gain beyond technical indicators. The long-short results should be interpreted more cautiously because some raw-OHLCV configurations generate fewer than 20 trades.

A further caveat concerns the interaction between feature richness and the cost-aware threshold. Under a fixed $\lambda=2.0$, a model that produces larger forecast magnitudes will pass the filter more often and therefore trade more frequently. Performance differences across feature tiers may therefore reflect both incremental predictive content and changes in the frequency with which forecasts clear the execution hurdle. These two channels cannot be fully separated under a fixed threshold and are treated as part of the limitation of the feature-enrichment comparison.

Inference follows the paired circular block-bootstrap procedure described in Section~\ref{sec:inference}. For each comparison, the richer feature set is compared with the simpler one:
\begin{equation}
d_t = r_t^{A} - r_t^{B},
\label{eq:h3_feature_diff}
\end{equation}
where $A$ denotes the richer feature specification and $B$ denotes the comparison specification. The one-sided alternative is that the richer feature set produces higher mean realised returns. Sharpe-ratio differences are again used as complementary evidence.

Table~\ref{tab:h3_bootstrap} reports the 72-hour block specification. Appendix~\ref{app:h3_bootstrap_robustness} reports the 24-hour and 168-hour robustness specifications in Tables~\ref{tab:h3_bootstrap_24h} and~\ref{tab:h3_bootstrap_168h}. Comparisons involving long-short OHLCV strategies are excluded from formal testing because the raw OHLCV strategies have fewer than 20 trades.

\begin{table}[H]
\centering
\captionsetup{justification=centering,singlelinecheck=false}
\caption{Bootstrap evidence for feature-enrichment effects, 72-hour blocks}
\label{tab:h3_bootstrap}
\footnotesize
\renewcommand{\arraystretch}{0.86}
\setlength{\tabcolsep}{2.1pt}

\begin{tabular}{lllrrrrrc}
\toprule
Selector & Mode & Comparison & Trades & Diff. (bps) & CI (bps) & $\Delta SR$ & SR CI & Holm \\
\midrule

\multirow{4}{*}{Loss-best}
& LO & OHLCV+TA / OHLCV
& 277 / 117 & 0.41 & [0.10, 0.74] & 0.95 & [0.23, 1.88] & Yes \\

& LO & EGARCH / OHLCV+TA
& 251 / 277 & -0.06 & [-0.30, 0.19] & -0.13 & [-0.84, 0.52] & No \\

& LO & EGARCH / OHLCV
& 251 / 117 & 0.35 & [0.04, 0.67] & 0.82 & [0.14, 1.67] & No \\

& LS & EGARCH / OHLCV+TA
& 60 / 66 & 0.09 & [-0.31, 0.50] & 0.17 & [-0.66, 1.00] & No \\

\addlinespace[1pt]

\multirow{4}{*}{IR$^{**}$-best}
& LO & OHLCV+TA / OHLCV
& 215 / 143 & 0.13 & [-0.11, 0.40] & 0.36 & [-0.21, 1.01] & No \\

& LO & EGARCH / OHLCV+TA
& 197 / 215 & 0.07 & [-0.16, 0.30] & 0.16 & [-0.42, 0.77] & No \\

& LO & EGARCH / OHLCV
& 197 / 143 & 0.20 & [-0.06, 0.49] & 0.51 & [-0.09, 1.25] & No \\

& LS & EGARCH / OHLCV+TA
& 47 / 64 & 0.13 & [-0.26, 0.50] & 0.25 & [-0.55, 1.09] & No \\

\bottomrule
\end{tabular}

\vspace{0.4em}
\begin{minipage}{0.95\textwidth}
\footnotesize \textit{Note:} LO = long-only; LS = long-short; EGARCH denotes the OHLCV+TA+EGARCH feature specification. The table reports paired circular block-bootstrap results using 72-hour blocks and 10,000 replications. Trades reports the number of trades in the richer / comparison specification. Diff. is the mean hourly return difference in basis points. CI (bps) is the 95\% confidence interval for the mean return difference. $\Delta SR$ is the Sharpe-ratio difference, and SR CI is its 95\% confidence interval. Holm indicates whether both the mean return-difference and Sharpe-difference tests reject after Holm correction. Long-short comparisons against raw OHLCV are not tested because the OHLCV long-short strategies have fewer than 20 trades.
\end{minipage}

\renewcommand{\arraystretch}{1.0}
\end{table}

\FloatBarrier

H3 should be interpreted cautiously. The strongest evidence is that adding technical indicators to raw OHLCV improves the long-only loss-best strategy. This comparison is positive in both mean-return and Sharpe-ratio terms and survives Holm correction under the 72-hour block specification. By contrast, adding EGARCH-derived features on top of technical indicators does not generate a statistically reliable improvement. Under IR$^{**}$-based selection, feature effects are mostly positive in direction but weak statistically. H3 is therefore only partially supported: richer features can improve realised performance in selected cases, but the effect is weaker and less stable than the cost-aware execution effect documented in H2.

\subsection{H4: Model architecture comparison}
\label{sec:h4}

The fourth empirical test examines whether model architecture materially affects net-of-cost trading performance. The comparison uses the OHLCV+TA+EGARCH feature set, MSE loss, loss-best selection, and cost-aware execution with $\lambda=2.0$. The three evaluated architectures are XGBoost, LSTM, and iTransformer, representing tree-based tabular learning, recurrent sequence modelling, and attention-based multivariate sequence modelling.

\begin{figure}[H]
\centering
\captionsetup{justification=centering,singlelinecheck=false}
\caption{Impact of model architecture on cost-aware strategy performance}
\includegraphics[width=\textwidth]{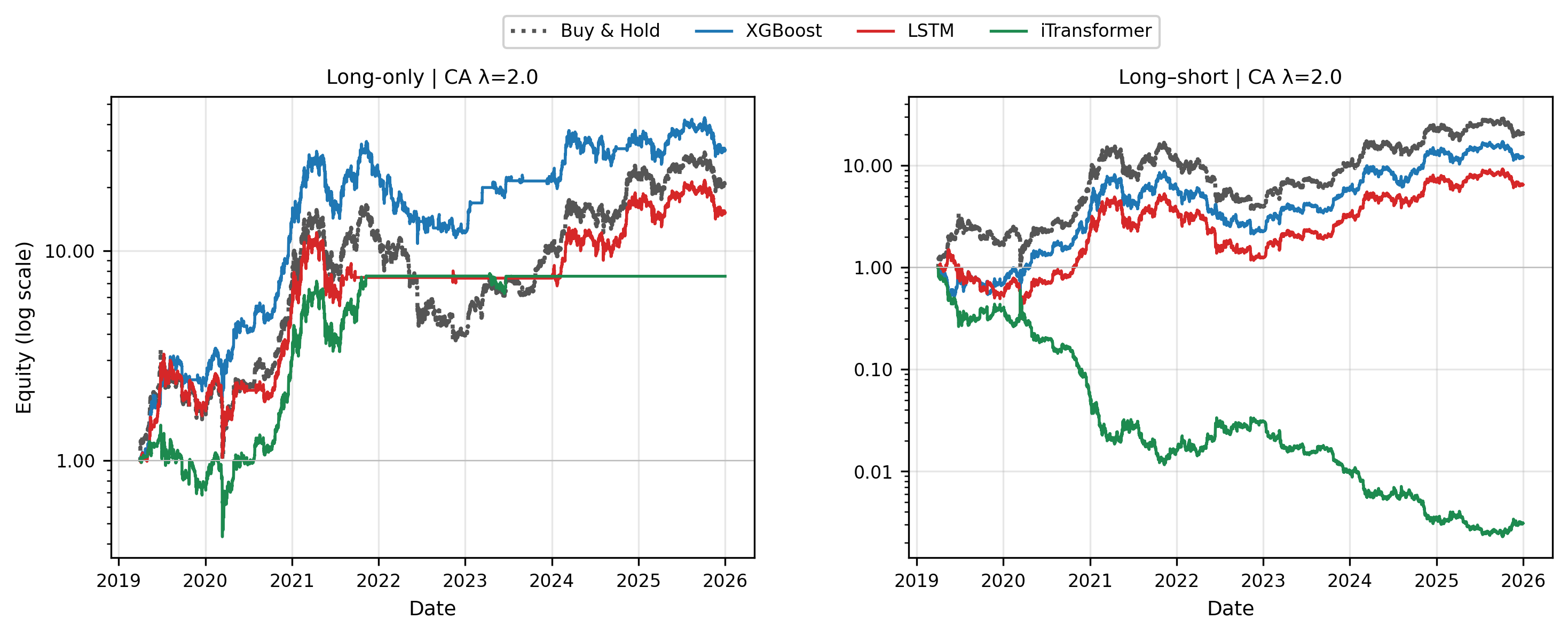}

\vspace{0.4em}
\begin{minipage}{0.95\textwidth}
\footnotesize \textit{Note:} Equity curves for XGBoost, LSTM, and iTransformer strategies with cost-aware execution and $\lambda=2.0$. The comparison uses the OHLCV+TA+EGARCH feature set, MSE loss, and loss-best selection. The left panel shows long-only strategies and the right panel shows long-short strategies.
\end{minipage}

\label{fig:h4_equity}
\end{figure}

Figure~\ref{fig:h4_equity} shows the cost-aware equity curves. In the long-only mode, all three architectures produce positive net performance after filtering, but XGBoost generates the strongest and most stable equity curve. In the long-short mode, XGBoost remains viable, while the neural-network strategies become difficult to interpret because cost-aware filtering suppresses almost all trades.

\FloatBarrier

Table~\ref{tab:h4_metrics} reports the performance metrics. Without cost-aware execution, all three architectures fail after transaction costs, confirming that the baseline sign rule is not viable regardless of model class. After cost-aware filtering, the long-only XGBoost strategy achieves the strongest result, with ARC of 65.40\%, Sharpe ratio of 1.09, and IR$^{**}$ of 1.13. LSTM remains profitable but weaker, with ARC of 49.41\% and Sharpe ratio of 0.84. iTransformer also improves relative to its baseline but produces the weakest long-only cost-aware result among the three architectures, with ARC of 34.87\% and Sharpe ratio of 0.64.

\begin{table}[H]
\centering
\captionsetup{justification=centering,singlelinecheck=false}
\caption{Performance comparison across model architectures}
\label{tab:h4_metrics}
\footnotesize
\renewcommand{\arraystretch}{0.86}
\setlength{\tabcolsep}{2.2pt}

\begin{tabular}{lllrrrrrrrr}
\toprule
Architecture & Mode & Type & ARC & ASD & MD & MLD & SR & IR$^{*}$ & IR$^{**}$ & Trades \\
\midrule

\multicolumn{11}{l}{\textit{LO}} \\
Buy \& Hold & LO & --
& 56.30 & 63.83 & -77.80 & 2.33 & 0.82 & 0.88 & 0.64 & 1 \\

24h Momentum & LO & --
& -45.93 & 42.84 & -99.26 & 6.52 & -1.17 & -1.07 & -0.50 & 5984 \\

XGBoost & LO & BL
& -64.00 & 54.71 & -99.92 & 6.48 & -1.25 & -1.17 & -0.75 & 10619 \\

XGBoost & LO & CA
& \textbf{65.40} & 56.22 & \textbf{-67.42} & \textbf{2.31} & \textbf{1.09} & \textbf{1.16} & \textbf{1.13} & 251 \\

LSTM & LO & BL
& -50.65 & 47.21 & -99.47 & 6.48 & -1.16 & -1.07 & -0.55 & 8448 \\

LSTM & LO & CA
& 49.41 & 53.66 & -67.88 & 2.91 & 0.84 & 0.92 & 0.67 & 69 \\

iTransformer & LO & BL
& -83.93 & 48.45 & -100.00 & 6.75 & -1.82 & -1.73 & -1.45 & 17942 \\

iTransformer & LO & CA
& 34.87 & \textbf{48.12} & -71.13 & 2.72 & 0.64 & 0.72 & 0.36 & 76 \\

\addlinespace[1pt]
\multicolumn{11}{l}{\textit{LS}} \\
Buy \& Hold & LS & --
& 56.30 & 63.83 & -77.80 & 2.33 & 0.82 & 0.88 & 0.64 & 1 \\

24h Momentum & LS & --
& -85.06 & 64.05 & -100.00 & 6.47 & -1.39 & -1.33 & -1.13 & 5984 \\

XGBoost & LS & BL
& -92.58 & 64.27 & -100.00 & 6.75 & -1.51 & -1.44 & -1.33 & 10620 \\

XGBoost & LS & CA
& \textbf{44.37} & \textbf{63.73} & \textbf{-75.73} & \textbf{2.31} & \textbf{0.63} & \textbf{0.70} & \textbf{0.41} & 60 \\

LSTM & LS & BL
& -87.08 & 64.14 & -100.00 & 6.75 & -1.42 & -1.36 & -1.18 & 8449 \\

LSTM$^{\dagger}$ & LS & CA
& 31.71 & 63.68 & -79.58 & 2.33 & 0.43 & 0.50 & 0.20 & 9 \\

iTransformer & LS & BL
& -98.62 & 64.40 & -100.00 & 6.76 & -1.60 & -1.53 & -1.51 & 17943 \\

iTransformer$^{\dagger}$ & LS & CA
& -57.43 & 63.83 & -99.77 & 6.75 & -0.97 & -0.90 & -0.52 & 1 \\

\bottomrule
\end{tabular}

\vspace{0.4em}
\begin{minipage}{0.95\textwidth}
\footnotesize \textit{Note:} LO = long-only; LS = long-short; BL = baseline execution; CA = cost-aware execution. CA uses $\lambda=2.0$. Results use OHLCV+TA+EGARCH features, MSE loss, and loss-best selection. ARC, ASD, and MD are percentages; MLD is reported in years; SR, IR$^{*}$, and IR$^{**}$ are ratios. Bold values indicate the strongest reliable cost-aware ML result within each trading mode. $^{\dagger}$ denotes fewer than 20 trades; these rows are descriptive and excluded from formal inference.
\end{minipage}

\renewcommand{\arraystretch}{1.0}
\end{table}

\FloatBarrier

The long-short comparison requires particular caution. XGBoost produces 60 trades and remains above the minimum-trade threshold, but the cost-aware LSTM and iTransformer long-short strategies produce only 9 and 1 trades, respectively. These neural-network rows are therefore treated as descriptive only. Under the minimum-trade rule, formal architecture inference is restricted to the long-only cost-aware strategies, where all three architectures produce at least 20 trades.

Inference follows the paired circular block-bootstrap procedure described in Section~\ref{sec:inference}. The tested return differential is:
\begin{equation}
d_t = r_t^{\mathrm{XGBoost}} - r_t^{M},
\qquad
M \in \{\mathrm{LSTM},\mathrm{iTransformer}\}.
\label{eq:h4_architecture_diff}
\end{equation}
The one-sided alternative is that XGBoost produces higher mean realised returns than the neural benchmark. Table~\ref{tab:h4_bootstrap} reports the 72-hour block specification. Appendix~\ref{app:h4_bootstrap_robustness} reports the 24-hour and 168-hour robustness specifications in Tables~\ref{tab:h4_bootstrap_24h} and~\ref{tab:h4_bootstrap_168h}.

\begin{table}[H]
\centering
\captionsetup{justification=centering,singlelinecheck=false}
\caption{Bootstrap evidence for architecture comparisons, 72-hour blocks}
\label{tab:h4_bootstrap}
\footnotesize
\renewcommand{\arraystretch}{0.90}
\setlength{\tabcolsep}{4.5pt}

\begin{tabular}{lrrrrrrr}
\toprule
Comparison & Trades & Diff. (bps) & CI (bps) & $p$ & $p_{\mathrm{Holm}}$ & $\Delta SR$ & SR CI \\
\midrule

XGBoost / LSTM
& 251 / 69
& 0.13
& [-0.19, 0.46]
& 0.215
& 0.215
& 0.25
& [-0.61, 1.12] \\

XGBoost / iTransformer
& 251 / 76
& 0.28
& [-0.09, 0.65]
& 0.067
& 0.134
& 0.45
& [-0.44, 1.47] \\

\bottomrule
\end{tabular}

\vspace{0.4em}
\begin{minipage}{0.92\textwidth}
\footnotesize \textit{Note:} The table reports paired circular block-bootstrap tests for long-only cost-aware strategies using OHLCV+TA+EGARCH features, MSE loss, loss-best selection, and $\lambda=2.0$. Comparisons are reported as first strategy / second strategy; Diff. and $\Delta SR$ are computed as first minus second. Diff. is expressed in basis points per hour. CI (bps) is the 95\% confidence interval for Diff.; SR CI is the 95\% confidence interval for $\Delta SR$. Positive values favour XGBoost. The reported $p$-values are one-sided centred-null bootstrap $p$-values with Holm correction.
\end{minipage}

\renewcommand{\arraystretch}{1.0}
\end{table}

\FloatBarrier

H4 receives partial support. XGBoost is the strongest architecture in descriptive economic terms: it delivers the highest long-only ARC, Sharpe ratio, IR$^{*}$, and IR$^{**}$, and it is the only architecture with a reliable long-short cost-aware result. However, the bootstrap evidence is not strong enough to establish statistically robust dominance over LSTM or iTransformer. The mean-return differences are positive, and the Sharpe-ratio differences also favour XGBoost, but all confidence intervals include zero and neither comparison survives Holm correction. The appropriate conclusion is therefore that architecture matters descriptively, with XGBoost performing best in this empirical setting, but the evidence does not justify a stronger claim of formal statistical superiority.

\subsection{H5: Loss function comparison}
\label{sec:h5}

The fifth empirical test examines whether the training loss function materially changes realised net-of-cost performance once the main modelling choices are fixed. The comparison is conducted for XGBoost with the OHLCV+TA+EGARCH feature set, loss-best selection, cost-aware execution with $\lambda=2.0$, and transaction costs of 10~basis points. The only element varied is the training objective: MSE versus MAE.

\begin{figure}[H]
\centering
\captionsetup{justification=centering,singlelinecheck=false}
\caption{Impact of loss function on cost-aware strategy performance}
\includegraphics[width=\textwidth]{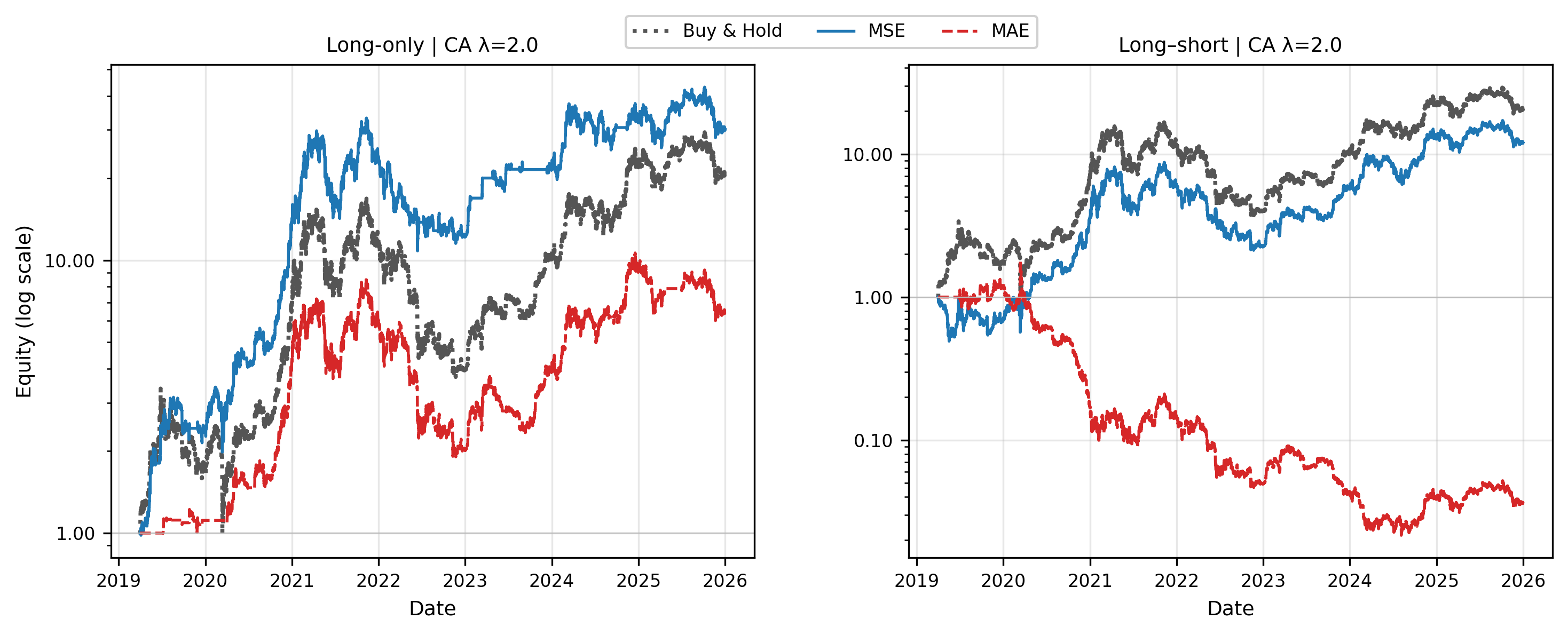}

\vspace{0.4em}
\begin{minipage}{0.95\textwidth}
\footnotesize \textit{Note:} Equity curves for XGBoost models trained under MSE and MAE objectives with cost-aware execution and $\lambda=2.0$. The comparison uses the OHLCV+TA+EGARCH feature set and loss-best selection. The left panel shows long-only strategies and the right panel shows long-short strategies.
\end{minipage}

\label{fig:h5_equity}
\end{figure}

Figure~\ref{fig:h5_equity} compares the equity curves. In the long-only mode, MSE produces a visibly stronger equity path than MAE under cost-aware execution. In the long-short mode, MSE remains viable while MAE deteriorates.

\FloatBarrier

Table~\ref{tab:h5_metrics} reports the corresponding performance metrics. In the long-only cost-aware strategy, MSE reaches an ARC of 65.40\%, a Sharpe ratio of 1.09, and an IR$^{**}$ of 1.13. The corresponding MAE values are lower, at 31.63\%, 0.53, and 0.25. The difference is economically visible. In the long-short mode, MSE remains profitable with 60 trades, while MAE produces only 8 trades and is therefore treated as descriptive only.

\begin{table}[H]
\centering
\captionsetup{justification=centering,singlelinecheck=false}
\caption{Performance comparison across loss functions}
\label{tab:h5_metrics}
\footnotesize
\renewcommand{\arraystretch}{0.86}
\setlength{\tabcolsep}{2.2pt}

\begin{tabular}{lllrrrrrrrr}
\toprule
Loss & Mode & Type & ARC & ASD & MD & MLD & SR & IR$^{*}$ & IR$^{**}$ & Trades \\
\midrule

\multicolumn{11}{l}{\textit{LO}} \\
Buy \& Hold & LO & --
& 56.30 & 63.83 & -77.80 & 2.33 & 0.82 & 0.88 & 0.64 & 1 \\

24h Momentum & LO & --
& -45.93 & \textbf{42.84} & -99.26 & 6.52 & -1.17 & -1.07 & -0.50 & 5984 \\

MSE & LO & BL
& -64.00 & 54.71 & -99.92 & 6.48 & -1.25 & -1.17 & -0.75 & 10619 \\

MSE & LO & CA
& \textbf{65.40} & 56.22 & \textbf{-67.42} & \textbf{2.31} & \textbf{1.09} & \textbf{1.16} & \textbf{1.13} & 251 \\

MAE & LO & BL
& -85.67 & 51.46 & -100.00 & 6.75 & -1.75 & -1.66 & -1.43 & 16598 \\

MAE & LO & CA
& 31.63 & 52.13 & -77.80 & 3.00 & 0.53 & 0.61 & 0.25 & 285 \\

\addlinespace[1pt]
\multicolumn{11}{l}{\textit{LS}} \\
Buy \& Hold & LS & --
& \textbf{56.30} & 63.83 & -77.80 & 2.33 & \textbf{0.82} & \textbf{0.88} & \textbf{0.64} & 1 \\

24h Momentum & LS & --
& -85.06 & 64.05 & -100.00 & 6.47 & -1.39 & -1.33 & -1.13 & 5984 \\

MSE & LS & BL
& -92.58 & 64.27 & -100.00 & 6.75 & -1.51 & -1.44 & -1.33 & 10620 \\

MSE & LS & CA
& 44.37 & 63.73 & \textbf{-75.73} & \textbf{2.31} & 0.63 & 0.70 & 0.41 & 60 \\

MAE & LS & BL
& -98.87 & 64.43 & -100.00 & 6.76 & -1.60 & -1.53 & -1.52 & 16599 \\

MAE$^{\dagger}$ & LS & CA
& -38.83 & \textbf{62.00} & -98.85 & 5.81 & -0.69 & -0.63 & -0.25 & 8 \\

\bottomrule
\end{tabular}

\vspace{0.4em}
\begin{minipage}{0.95\textwidth}
\footnotesize \textit{Note:} LO = long-only; LS = long-short; BL = baseline execution; CA = cost-aware execution. CA uses $\lambda=2.0$. Results are for XGBoost with OHLCV+TA+EGARCH features under loss-best selection. ARC, ASD, and MD are percentages; MLD is reported in years; SR, IR$^{*}$, and IR$^{**}$ are ratios. Bold values indicate the strongest outcome within each mode. $^{\dagger}$ denotes fewer than 20 trades; this row is descriptive and excluded from formal inference.
\end{minipage}

\renewcommand{\arraystretch}{1.0}
\end{table}

\FloatBarrier

The formal H5 comparison is restricted to the long-only cost-aware strategies, where both MSE and MAE exceed the minimum-trade threshold. The tested differential is:
\begin{equation}
d_t = r_t^{\mathrm{MSE}} - r_t^{\mathrm{MAE}}.
\label{eq:h5_loss_diff}
\end{equation}
Because H5 asks whether the training loss matters, rather than whether one loss must always dominate, the bootstrap test is two-sided. Table~\ref{tab:h5_bootstrap} reports 24-hour, 72-hour, and 168-hour block specifications together, since this comparison is compact and does not require a separate appendix table.

\begin{table}[H]
\centering
\captionsetup{justification=centering,singlelinecheck=false}
\caption{Bootstrap evidence for loss-function comparison}
\label{tab:h5_bootstrap}
\footnotesize
\renewcommand{\arraystretch}{0.90}
\setlength{\tabcolsep}{6pt}

\begin{tabular}{rrrrrrr}
\toprule
Block (h) & Trades & Diff. (bps) & CI (bps) & $p$ & $\Delta SR$ & SR CI \\
\midrule
24  & 251 / 285 & 0.286 & [-0.044, 0.632] & 0.094 & 0.562 & [-0.236, 1.530] \\
72  & 251 / 285 & 0.286 & [-0.045, 0.622] & 0.092 & 0.562 & [-0.233, 1.468] \\
168 & 251 / 285 & 0.286 & [-0.061, 0.641] & 0.113 & 0.562 & [-0.283, 1.510] \\
\bottomrule
\end{tabular}

\vspace{0.4em}
\begin{minipage}{0.92\textwidth}
\footnotesize \textit{Note:} The table reports paired circular block-bootstrap tests for long-only cost-aware XGBoost strategies using OHLCV+TA+EGARCH features, loss-best selection, and $\lambda=2.0$. Diff. and $\Delta SR$ are computed as MSE minus MAE; positive values favour MSE. Diff. is expressed in basis points per hour. CI (bps) is the 95\% confidence interval for Diff.; SR CI is the 95\% confidence interval for $\Delta SR$. Reported $p$-values are two-sided centred-null bootstrap $p$-values.
\end{minipage}

\renewcommand{\arraystretch}{1.0}
\end{table}

\FloatBarrier

The results point toward MSE, but not strongly enough to make the loss function a primary driver of performance. MSE exceeds MAE by 0.286 basis points per hour, and the Sharpe-ratio difference is positive at 0.562 across all block lengths. However, the confidence intervals include zero and the two-sided bootstrap $p$-values range from 0.092 to 0.113. H5 receives only weak support. The loss function changes the point estimates, with MSE outperforming MAE descriptively, but the effect is statistically fragile and secondary to execution design.

\subsection{H6: Model selection criterion}
\label{sec:h6}

The sixth empirical test examines whether the validation-set model-selection criterion affects realised out-of-sample trading performance. The comparison is conducted for XGBoost with the OHLCV+TA+EGARCH feature set, MSE loss, transaction costs of 10~basis points, and cost-aware execution with $\lambda=2.0$. The three compared selectors are loss-best, information-coefficient-best, and mode-specific IR$^{**}$-best selection.

\begin{figure}[H]
\centering
\captionsetup{justification=centering,singlelinecheck=false}
\caption{Impact of model-selection criterion on cost-aware strategy performance}
\includegraphics[width=\textwidth]{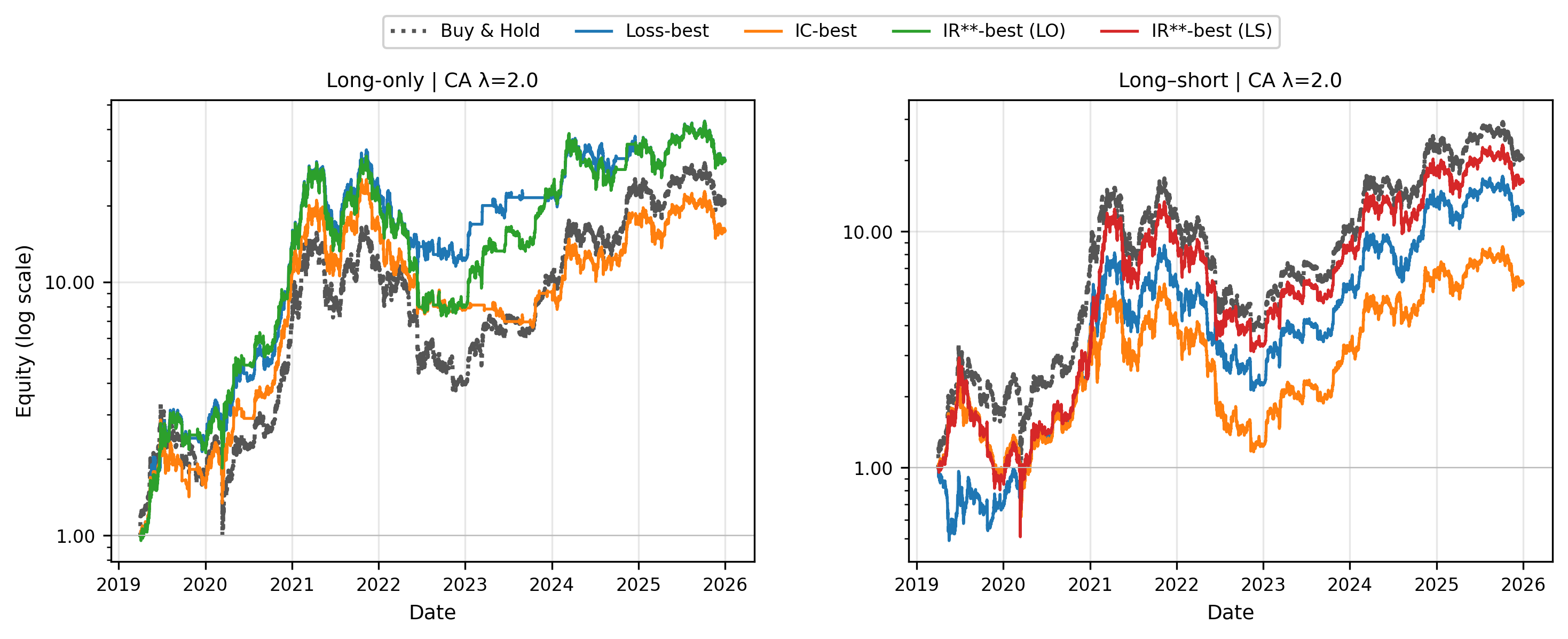}

\vspace{0.4em}
\begin{minipage}{0.95\textwidth}
\footnotesize \textit{Note:} Equity curves for XGBoost under loss-best, IC-best, and mode-specific IR$^{**}$-best selection. The comparison uses the OHLCV+TA+EGARCH feature set, MSE loss, and cost-aware execution with $\lambda=2.0$. The left panel shows long-only strategies and the right panel shows long-short strategies.
\end{minipage}

\label{fig:h6_equity}
\end{figure}

Figure~\ref{fig:h6_equity} compares the equity curves across selectors. In the long-only mode, loss-best and IR$^{**}$-best produce similar trajectories, while IC-best is visibly weaker. In the long-short mode, IR$^{**}$-best is descriptively strongest.

\FloatBarrier

The descriptive performance metrics for these selector--mode combinations are already reported in Table~\ref{tab:h2_metrics}, because H6 uses the same XGBoost OHLCV+TA+EGARCH MSE configurations as the cost-aware comparison in Section~\ref{sec:h2}. The selector comparison should therefore be read from the cost-aware rows of that table. In the long-only setting, loss-best and IR$^{**}$-best selection produce nearly identical annualised returns, 65.40\% and 65.34\%, while IC-best is lower at 50.48\%. In the long-short setting, IR$^{**}$-best selection is descriptively strongest, with ARC of 51.05\% and IR$^{**}$ of 0.50, followed by loss-best and IC-best. The selector therefore changes point estimates and ranking, but the magnitude of the effect is substantially smaller than the execution-filter effect reported in Section~\ref{sec:h2}.

Inference follows the paired circular block-bootstrap procedure described in Section~\ref{sec:inference}. For two selection rules $A$ and $B$, the tested differential is:
\begin{equation}
d_t = r_t^{A} - r_t^{B}.
\label{eq:h6_selector_diff}
\end{equation}
The tests are two-sided because the question is whether the selector changes realised performance, not whether one selector must dominate by construction. Table~\ref{tab:h6_bootstrap} reports the 72-hour block specification. Appendix~\ref{app:h6_bootstrap_robustness} reports the 24-hour and 168-hour robustness specifications. The inference is unchanged across block lengths.

\begin{table}[H]
\centering
\captionsetup{justification=centering,singlelinecheck=false}
\caption{Bootstrap evidence for model-selection criterion comparisons, 72-hour blocks}
\label{tab:h6_bootstrap}
\footnotesize
\renewcommand{\arraystretch}{0.86}
\setlength{\tabcolsep}{2.0pt}

\begin{tabular}{lllrrrrrrr}
\toprule
Mode & Comparison & Trades & Diff. (bps) & CI (bps) & $p$ & $p_{\mathrm{Holm}}$ & $\Delta SR$ & SR CI & Holm \\
\midrule

LO & Loss / IC
& 251 / 261 & 0.112 & [-0.132, 0.359] & 0.370 & 1.000 & 0.256 & [-0.333, 0.918] & No \\

LO & Loss / IR$^{**}$
& 251 / 197 & -0.011 & [-0.222, 0.194] & 0.920 & 1.000 & 0.033 & [-0.516, 0.560] & No \\

LO & IR$^{**}$ / IC
& 197 / 261 & 0.123 & [-0.117, 0.364] & 0.310 & 1.000 & 0.222 & [-0.346, 0.886] & No \\

LS & Loss / IC
& 60 / 47 & 0.116 & [-0.259, 0.509] & 0.556 & 1.000 & 0.219 & [-0.538, 1.022] & No \\

LS & IR$^{**}$ / Loss
& 47 / 60 & 0.052 & [-0.344, 0.465] & 0.805 & 1.000 & 0.106 & [-0.751, 0.988] & No \\

LS & IR$^{**}$ / IC
& 47 / 47 & 0.167 & [-0.196, 0.533] & 0.350 & 1.000 & 0.324 & [-0.372, 1.127] & No \\

\bottomrule
\end{tabular}

\vspace{0.4em}
\begin{minipage}{0.95\textwidth}
\footnotesize \textit{Note:} LO = long-only; LS = long-short. Loss = loss-best selector; IC = information-coefficient-best selector; IR$^{**}$ = mode-specific IR$^{**}$-best selector. The table reports paired circular block-bootstrap tests using 72-hour blocks and 10,000 replications. Results use XGBoost with OHLCV+TA+EGARCH features, MSE loss, and cost-aware execution with $\lambda=2.0$. Comparisons are reported as first selector / second selector; Diff. and $\Delta SR$ are computed as first minus second. Diff. is expressed in basis points per hour. CI (bps) is the 95\% confidence interval for Diff.; SR CI is the 95\% confidence interval for $\Delta SR$. Reported $p$-values are two-sided centred-null bootstrap $p$-values with Holm correction.
\end{minipage}

\renewcommand{\arraystretch}{1.0}
\end{table}

\FloatBarrier

The bootstrap results qualify the descriptive selector comparison. All mean-return confidence intervals include zero, all Sharpe-ratio confidence intervals include zero, and no pairwise comparison survives Holm correction. H6 receives partial support. The selection criterion changes point estimates and can alter the ranking of deployed strategies, but the effect is not statistically robust. In this specification, selector choice is best interpreted as a secondary design choice rather than a primary source of net-of-cost profitability.

% ====================================================================
% 6. ROBUSTNESS AND DIAGNOSTICS
% ====================================================================
\section{Robustness and diagnostics}
\label{sec:robustness}

\subsection{Cost-aware threshold sensitivity}
\label{sec:lambda_sensitivity}

The first robustness check examines whether the cost-aware result depends on the specific threshold value used in the main specification. The parameter $\lambda$ controls the strictness of the execution filter. Lower values allow more forecasts to become trades, while higher values require stronger predicted returns before a position change is executed. The baseline case $\lambda=0$ corresponds to the unfiltered strategy.

\begin{figure}[H]
\centering
\captionsetup{justification=centering,singlelinecheck=false}
\caption{Sensitivity of strategy performance to the cost-aware threshold $\lambda$}
\includegraphics[width=\textwidth]{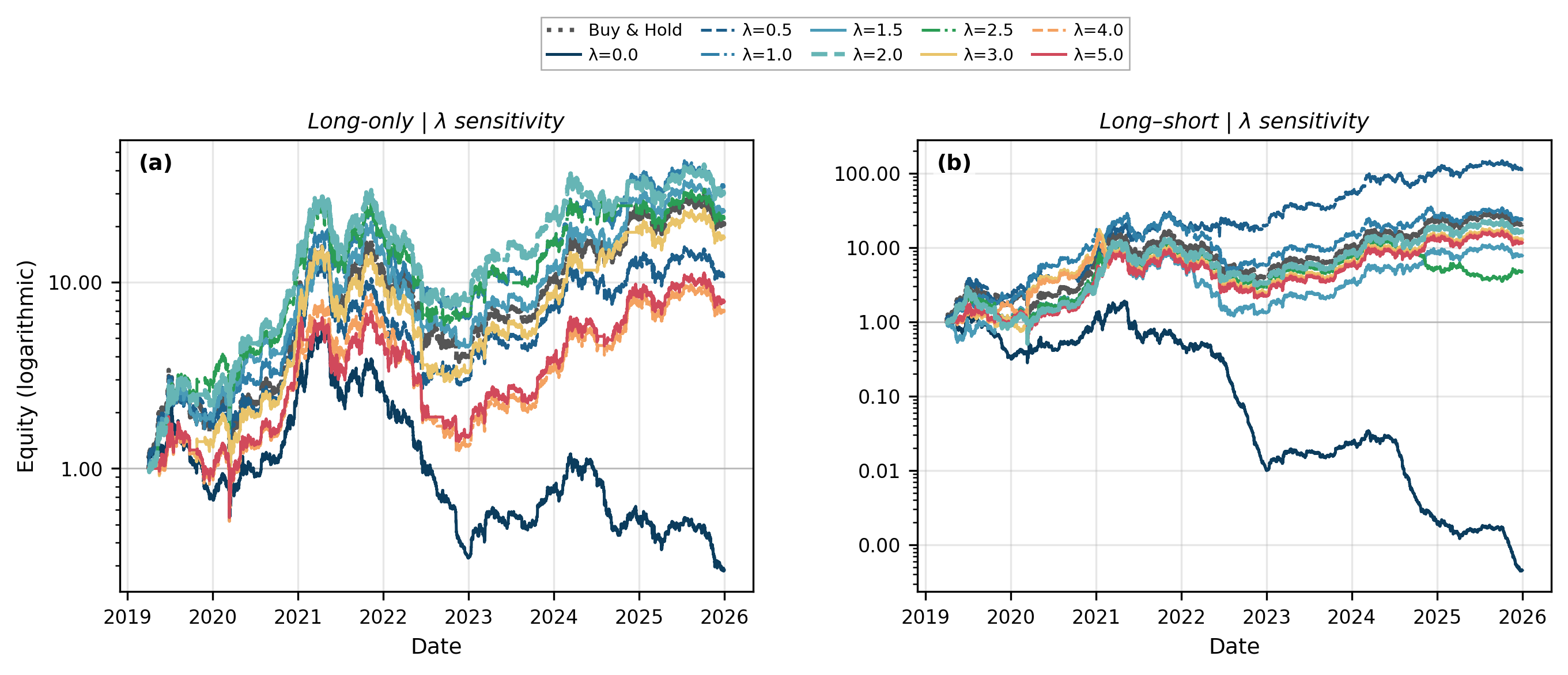}

\vspace{0.4em}
\begin{minipage}{0.95\textwidth}
\footnotesize \textit{Note:} The figure shows equity curves for different values of the cost-aware threshold $\lambda$. Lower values imply weaker filtering and more frequent trading, whereas higher values impose stricter execution thresholds and may suppress trading activity. The row $\lambda=0$ corresponds to the unfiltered baseline strategy.
\end{minipage}

\label{fig:sensitivity_lambda}
\end{figure}

Figure~\ref{fig:sensitivity_lambda} and Table~\ref{tab:sensitivity_lambda} report the performance of the XGBoost OHLCV+TA+EGARCH MSE loss-best specification across $\lambda \in \{0,0.5,1,1.5,2,2.5,3,4,5\}$. The main transaction-cost assumption remains fixed at 10~basis points.

\FloatBarrier

\begin{table}[H]
\centering
\captionsetup{justification=centering,singlelinecheck=false}
\caption{Sensitivity of performance to the cost-aware threshold $\lambda$}
\label{tab:sensitivity_lambda}
\footnotesize
\renewcommand{\arraystretch}{0.86}
\setlength{\tabcolsep}{2.2pt}

\begin{tabular}{llrrrrrrrr}
\toprule
Mode & $\lambda$ & ARC & ASD & MD & MLD & SR & IR$^{*}$ & IR$^{**}$ & Trades \\
\midrule

\multicolumn{10}{l}{\textit{LO}} \\
LO & Buy \& Hold
& 56.30 & 63.83 & -77.80 & 2.33 & 0.82 & 0.88 & 0.64 & 1 \\

LO & 24h Momentum
& -45.93 & \textbf{42.84} & -99.26 & 6.52 & -1.17 & -1.07 & -0.50 & 5984 \\

LO & 0.0
& -64.00 & 54.71 & -99.92 & 6.48 & -1.25 & -1.17 & -0.75 & 10619 \\

LO & 0.5
& 35.75 & 53.98 & -70.14 & 4.67 & 0.58 & 0.66 & 0.34 & 1679 \\

LO & 1.0
& 36.17 & 54.28 & -70.57 & 3.09 & 0.59 & 0.67 & 0.34 & 757 \\

LO & 1.5
& 34.31 & 54.05 & \textbf{-62.63} & 2.33 & 0.56 & 0.63 & 0.35 & 411 \\

LO & 2.0
& \textbf{65.40} & 56.22 & -67.42 & 2.31 & \textbf{1.09} & \textbf{1.16} & \textbf{1.13} & 251 \\

LO & 2.5
& 43.18 & 56.71 & -76.06 & 4.72 & 0.69 & 0.76 & 0.43 & 179 \\

LO & 3.0
& 62.77 & 57.14 & -76.10 & 2.91 & 1.03 & 1.10 & 0.91 & 103 \\

LO & 4.0
& 54.56 & 57.55 & -75.01 & \textbf{2.30} & 0.88 & 0.95 & 0.69 & 59 \\

LO & 5.0
& 42.83 & 57.59 & -75.94 & 2.31 & 0.67 & 0.74 & 0.42 & 41 \\

\addlinespace[1pt]
\midrule
\multicolumn{10}{l}{\textit{LS}} \\
LS & Buy \& Hold
& 56.30 & 63.83 & -77.80 & 2.33 & 0.82 & 0.88 & 0.64 & 1 \\

LS & 24h Momentum
& -85.06 & 64.05 & -100.00 & 6.47 & -1.39 & -1.33 & -1.13 & 5984 \\

LS & 0.0
& -92.58 & 64.27 & -100.00 & 6.75 & -1.51 & -1.44 & -1.33 & 10620 \\

LS & 0.5
& 8.26 & 63.77 & -80.28 & 4.20 & 0.06 & 0.13 & 0.01 & 759 \\

LS & 1.0
& \textbf{62.60} & 63.73 & \textbf{-65.84} & 4.13 & \textbf{0.92} & \textbf{0.98} & \textbf{0.93} & 252 \\

LS & 1.5
& 59.35 & 63.73 & -82.18 & 4.72 & 0.87 & 0.93 & 0.67 & 104 \\

LS & 2.0
& 44.37 & 63.73 & -75.73 & \textbf{2.31} & 0.63 & 0.70 & 0.41 & 60 \\

LS & 2.5
& 41.11 & 63.70 & -86.25 & 4.58 & 0.58 & 0.65 & 0.31 & 41 \\

LS & 3.0
& 39.48 & 63.70 & -86.79 & 4.49 & 0.55 & 0.62 & 0.28 & 25 \\

LS & 4.0$^{\dagger}$
& 41.14 & 63.24 & -77.80 & 2.33 & 0.58 & 0.65 & 0.34 & 7 \\

LS & 5.0$^{\dagger}$
& 42.78 & 63.15 & -77.80 & 2.33 & 0.61 & 0.68 & 0.37 & 3 \\

\bottomrule
\end{tabular}

\vspace{0.4em}
\begin{minipage}{0.95\textwidth}
\footnotesize \textit{Note:} LO = long-only; LS = long-short. Results are for XGBoost with OHLCV+TA+EGARCH features, MSE loss, loss-best selection, and transaction costs of 10 basis points. The row $\lambda=0$ corresponds to the unfiltered baseline strategy. ARC, ASD, and MD are percentages; MLD is reported in years; SR, IR$^{*}$, and IR$^{**}$ are ratios. Bold values indicate the strongest reliable result within each mode. $^{\dagger}$ denotes fewer than 20 trades; these rows are descriptive and excluded from formal conclusions.
\end{minipage}

\renewcommand{\arraystretch}{1.0}
\end{table}

\FloatBarrier

The sensitivity analysis shows that the cost-aware filter changes the economics of the strategy, rather than merely relabelling the baseline rule. When $\lambda=0$, the filter is inactive and both trading modes suffer large losses after transaction costs: ARC equals $-64.00\%$ in the long-only case and $-92.58\%$ in the long-short case. Introducing a positive threshold sharply reduces turnover and improves net performance.

In the long-only setting, the strongest result is obtained at the main threshold $\lambda=2.0$, with ARC of 65.40\%, Sharpe ratio of 1.09, IR$^{**}$ of 1.13, and 251 trades. Performance remains positive across all positive threshold values, suggesting that the result is not driven by a single fragile threshold. In the long-short setting, the best reliable result occurs at $\lambda=1.0$, with ARC of 62.60\%, Sharpe ratio of 0.92, and 252 trades. At very high thresholds, long-short strategies become too sparse, and the $\lambda=4.0$ and $\lambda=5.0$ cases fall below the minimum-trade threshold.

The threshold analysis supports the main H2 conclusion. The unfiltered strategy fails because it trades too often, while a positive cost-aware threshold restores net performance by suppressing weak signals. The exact optimal threshold differs across trading modes, but the qualitative result is stable: execution discipline is essential for converting forecasts into viable post-cost trading returns.

\subsection{Transaction-cost sensitivity}
\label{sec:tc_sensitivity}

The second robustness check examines whether the cost-aware results survive under alternative effective transaction-cost assumptions. This diagnostic is conducted for the mode-specific IR$^{**}$-best XGBoost specification with OHLCV+TA+EGARCH features, MSE loss, and $\lambda=2.0$. This selector is used because the transaction-cost grid is intended to stress-test the trading-oriented specification selected by realised validation performance. The main specification uses 10~basis points per position change. The sensitivity grid extends this assumption to 0, 5, 10, 15, 20, and 25~basis points. For each cost level, positions are regenerated using the same cost-aware rule and the corresponding transaction cost. The analysis therefore captures both the direct cost drag and the endogenous change in trading frequency caused by a higher execution hurdle.

\begin{figure}[H]
\centering
\captionsetup{justification=centering,singlelinecheck=false}
\caption{Sensitivity of strategy performance to transaction costs}
\includegraphics[width=\textwidth]{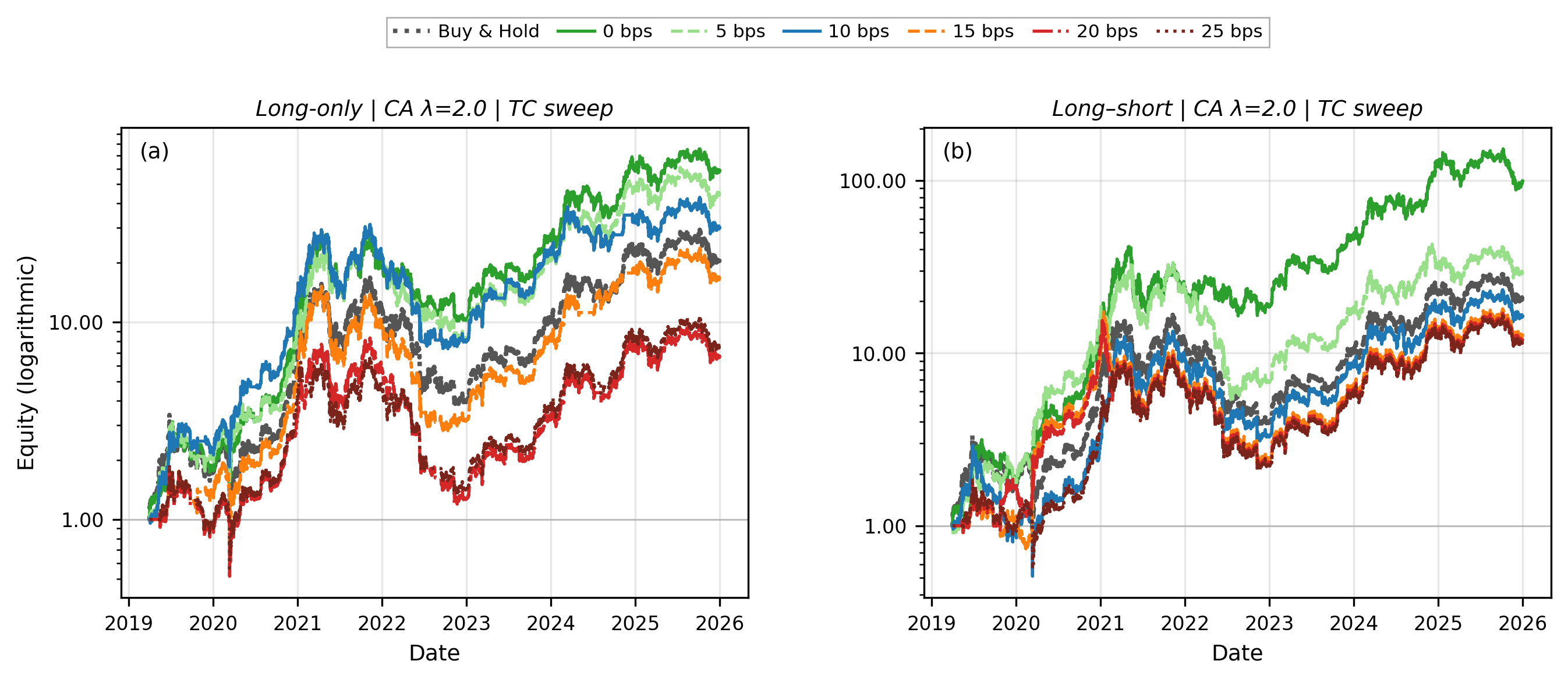}

\vspace{0.4em}
\begin{minipage}{0.95\textwidth}
\footnotesize \textit{Note:} The figure shows equity curves under alternative transaction-cost assumptions. For each cost level, positions are regenerated using the corresponding transaction cost and the same cost-aware threshold $\lambda=2.0$. The figure therefore captures both direct cost drag and the endogenous reduction in trading frequency caused by a higher execution hurdle.
\end{minipage}

\label{fig:sensitivity_tc}
\end{figure}

Figure~\ref{fig:sensitivity_tc} and Table~\ref{tab:sensitivity_tc} report the results.

\FloatBarrier

\begin{table}[H]
\centering
\captionsetup{justification=centering,singlelinecheck=false}
\caption{Performance across different transaction-cost levels}
\label{tab:sensitivity_tc}
\footnotesize
\renewcommand{\arraystretch}{0.86}
\setlength{\tabcolsep}{2.2pt}

\begin{tabular}{llrrrrrrrr}
\toprule
Mode & TC & ARC & ASD & MD & MLD & SR & IR$^{*}$ & IR$^{**}$ & Trades \\
\midrule

\multicolumn{10}{l}{\textit{LO}} \\
LO & 0 bps
& \textbf{82.51} & 57.95 & \textbf{-64.09} & 2.67 & \textbf{1.35} & \textbf{1.42} & \textbf{1.83} & 5327 \\

LO & 5 bps
& 75.23 & \textbf{57.86} & -74.21 & \textbf{2.30} & 1.23 & 1.30 & 1.32 & 611 \\

LO & 10 bps
& 65.34 & 57.94 & -76.64 & \textbf{2.30} & 1.06 & 1.13 & 0.96 & 197 \\

LO & 15 bps
& 51.44 & 59.15 & -81.51 & 3.44 & 0.80 & 0.87 & 0.55 & 93 \\

LO & 20 bps
& 32.34 & 60.01 & -85.62 & 3.53 & 0.47 & 0.54 & 0.20 & 49 \\

LO & 25 bps
& 34.54 & 59.49 & -80.02 & 3.00 & 0.51 & 0.58 & 0.25 & 35 \\

\addlinespace[1pt]
\midrule
\multicolumn{10}{l}{\textit{LS}} \\
LS & 0 bps
& \textbf{97.29} & 63.82 & \textbf{-60.84} & 2.52 & \textbf{1.46} & \textbf{1.52} & \textbf{2.44} & 6141 \\

LS & 5 bps
& 64.65 & 63.73 & -83.99 & 2.97 & 0.95 & 1.01 & 0.78 & 200 \\

LS & 10 bps
& 51.05 & 63.67 & -82.60 & \textbf{2.31} & 0.74 & 0.80 & 0.50 & 47 \\

LS & 15 bps$^{\dagger}$
& 45.45 & 63.28 & -86.84 & 4.49 & 0.65 & 0.72 & 0.38 & 17 \\

LS & 20 bps$^{\dagger}$
& 44.20 & 63.30 & -86.02 & 4.50 & 0.63 & 0.70 & 0.36 & 7 \\

LS & 25 bps$^{\dagger}$
& 43.48 & 63.16 & -77.80 & 2.33 & 0.62 & 0.69 & 0.38 & 1 \\

\bottomrule
\end{tabular}

\vspace{0.4em}
\begin{minipage}{0.95\textwidth}
\footnotesize \textit{Note:} LO = long-only; LS = long-short; TC = transaction costs. Results are for the mode-specific IR$^{**}$-best cost-aware XGBoost strategy with OHLCV+TA+EGARCH features, MSE loss, and $\lambda=2.0$. Positions are regenerated separately for each transaction-cost level. ARC, ASD, and MD are percentages; MLD is reported in years; SR, IR$^{*}$, and IR$^{**}$ are ratios. Bold values indicate the strongest reliable outcome within each mode. $^{\dagger}$ denotes fewer than 20 trades; these rows are descriptive and excluded from formal conclusions.
\end{minipage}

\renewcommand{\arraystretch}{1.0}
\end{table}

\FloatBarrier

The results show that transaction costs materially affect performance, but the adaptive cost-aware filter prevents the strategy from collapsing immediately as costs rise. In the long-only case, ARC falls from 82.51\% at 0~basis points to 65.34\% at the main 10~basis point assumption, and remains positive throughout the full cost grid. At 25~basis points, the strategy still records ARC of 34.54\% with 35 trades.

The small improvement from 20~basis points to 25~basis points should not be interpreted as transaction costs improving the strategy. It reflects the endogenous regeneration of the position path: a higher cost assumption raises the execution hurdle, suppresses additional trades, and can therefore avoid some losing position changes. The broader pattern remains clear: higher effective costs reduce trading activity and weaken risk-adjusted performance relative to the low-cost cases.

The long-short results are more fragile. Performance is strongest at 0~basis points and remains positive at 5 and 10~basis points, but the strategy becomes sparse at higher costs. At 15~basis points and above, the long-short rows fall below the 20-trade threshold and are therefore descriptive only. This confirms that long-short execution is more sensitive to trading frictions because position reversals generate greater turnover than long-only entries and exits.

The transaction-cost sensitivity points in the same direction as the main results. Cost-aware execution substantially improves robustness to trading frictions, but it does not eliminate cost sensitivity. The long-only strategy remains economically viable across the tested cost grid, while long-short performance becomes unreliable once higher effective costs suppress the number of trades.

\subsection{Fold-level stability}
\label{sec:fold_stability}

The final diagnostic decomposes the flagship cost-aware strategy across the 27 walk-forward test folds. The purpose is to assess whether the aggregate result is broadly distributed across the evaluation period or concentrated in a small number of favourable market regimes. The diagnostic is conducted for the main long-only XGBoost specification: OHLCV+TA+EGARCH features, MSE loss, loss-best selection, cost-aware execution with $\lambda=2.0$, and transaction costs of 10~basis points.

Table~\ref{tab:fold_stability_summary} reports the fold-level summary statistics. The full fold-by-fold decomposition is reported in Appendix~\ref{app:folds}, Tables~\ref{tab:fold_decomposition_returns} and~\ref{tab:fold_decomposition_execution}.

\begin{table}[H]
\centering
\captionsetup{justification=centering,singlelinecheck=false}
\caption{Fold-level stability summary for the flagship cost-aware XGBoost strategy}
\label{tab:fold_stability_summary}
\footnotesize
\renewcommand{\arraystretch}{0.92}
\setlength{\tabcolsep}{5pt}

\begin{tabular}{p{4.2cm} c p{6.8cm}}
\toprule
Diagnostic & Value & Interpretation \\
\midrule
Number of test folds & 27 & Quarterly walk-forward test windows \\
Profitable folds & 16 & 59.26\% of folds have positive ARC \\
Positive-Sharpe folds & 14 & 51.85\% of folds have positive Sharpe ratios \\
Mean fold ARC & 331.43\% & Inflated by short-window annualisation \\
Median fold ARC & 7.98\% & More informative than the mean \\
Std. dev. of fold ARC & 866.66\% & High cross-fold dispersion \\
Minimum fold ARC & -81.29\% & Worst quarterly annualised fold result \\
Maximum fold ARC & 3316.76\% & Best quarterly annualised fold result \\
Mean trades per fold & 9.8 & Strategy is sparse at fold level \\
Median trades per fold & 4.0 & Most folds contain few trades \\
Low-trade folds & 20 & 74.07\% of folds have fewer than 20 trades \\
Full-sample trades & 251 & Above the formal full-sample threshold \\
Full-sample ARC & 65.40\% & Main aggregate result \\
Full-sample Sharpe ratio & 1.09 & Main aggregate risk-adjusted result \\
Full-sample IR$^{**}$ & 1.13 & Main trading-oriented metric \\
\bottomrule
\end{tabular}

\vspace{0.4em}
\begin{minipage}{0.95\textwidth}
\footnotesize \textit{Note:} Diagnostics are computed for the flagship long-only cost-aware XGBoost strategy using OHLCV+TA+EGARCH features, MSE loss, loss-best selection, $\lambda=2.0$, and 10 basis point transaction costs. Fold-level ARC is annualised from three-month test windows and can therefore become mechanically large. The low-trade flag is applied at the fold level only; formal inference is based on the full out-of-sample return series.
\end{minipage}

\renewcommand{\arraystretch}{1.0}
\end{table}

\FloatBarrier

The diagnostic shows that the aggregate result is not uniformly distributed across all quarters. Sixteen of the 27 test folds are profitable, and 14 have positive Sharpe ratios. This means that the strategy is not driven entirely by a single isolated episode, but it also does not generate stable profitability in every market regime. The hit rate is positive but moderate.

The dispersion across folds is substantial. The median fold ARC is only 7.98\%, while the mean fold ARC is much larger because several three-month windows compound strongly and therefore receive very large annualised ARC values. This is a mechanical consequence of annualising short-window returns and should not be interpreted as evidence that such performance would persist at the same rate over a full year.

Trade counts also qualify the interpretation. Although the full out-of-sample strategy produces 251 trades and therefore exceeds the minimum-trade threshold used for formal inference, many individual quarterly folds contain fewer than 20 trades. The strategy is therefore sparse at the fold level. This is consistent with the design of the cost-aware filter: by suppressing weak forecasts, it improves aggregate net-of-cost performance but concentrates trading activity in periods where predicted returns are sufficiently large.

The fold-level diagnostic calls for a cautious interpretation of the main result. Cost-aware execution restores full-sample viability and produces positive aggregate performance, but the realised gains are uneven across market regimes. The evidence therefore supports the paper's central claim that execution discipline is crucial, while also ruling out an overly strong claim of stable quarter-by-quarter profitability.

% ====================================================================
% 7. CONCLUSION
% ====================================================================
\section{Conclusion}
\label{sec:conclusion}

This paper asked whether machine-learning forecasts of hourly BTC/USDT returns can be converted into economically meaningful trading performance after transaction costs. The evidence is conditional. The models contain some useful predictive information, but this information has little economic value when forecasts are translated mechanically into positions through a naive sign rule. Once proportional transaction costs are imposed, the baseline machine-learning strategies collapse because they trade too frequently.

The main result is that prediction quality matters only after the execution rule decides which forecasts are worth trading. A simple cost-aware filter, which allows position changes only when the forecast magnitude exceeds a transaction-cost-based threshold, substantially reduces turnover and prevents many weak forecasts from becoming costly trades in selected configurations. In the main long-only XGBoost specification, the filter changes the strategy from a large post-cost loss into a positive annualised return above 65\%, with a Sharpe ratio above unity. This result shows that the practical value of short-horizon machine-learning forecasts depends less on the existence of weak statistical predictability alone and more on whether the trading rule avoids acting on weak signals.

The remaining empirical results qualify this conclusion. XGBoost is the strongest architecture in descriptive economic terms, outperforming LSTM and iTransformer in the main cost-aware long-only comparison. However, paired circular block-bootstrap tests do not support a claim of statistically robust dominance after multiple-testing correction. Feature enrichment also has a conditional effect: technical indicators improve performance in selected long-only configurations, while EGARCH-derived volatility features do not deliver a uniformly robust incremental gain beyond technical indicators. The choice of MSE versus MAE loss function and the choice of validation-set model-selection criterion both affect point estimates, but neither emerges as a primary or statistically robust driver of performance.

The benchmark and diagnostic results also impose important limits on the interpretation. Although cost-aware execution strongly improves the machine-learning strategy relative to the unfiltered baseline, the best cost-aware specifications do not significantly outperform buy-and-hold in Sharpe-ratio terms after bootstrap adjustment. Fold-level diagnostics further show that the aggregate result is uneven across the 27 walk-forward test folds. The flagship strategy is profitable in 16 folds and has a positive Sharpe ratio in 14 folds, but many quarterly folds contain few trades and performance is concentrated in favourable regimes. The result should therefore not be read as evidence of stable quarter-by-quarter profitability.

The paper shows that the prediction-to-trading gap is economically large, not a minor implementation detail. In hourly cryptocurrency trading, transaction costs can completely reverse the conclusion drawn from gross equity curves. A model that appears profitable before costs may be economically useless once turnover is priced correctly. Conversely, a relatively simple execution filter can recover much of the useful signal by suppressing low-conviction trades. The evidence therefore supports a more cautious view of machine learning in short-horizon crypto markets: model complexity is secondary to the discipline with which forecasts are converted into positions.

Several limitations remain. The analysis uses one asset, BTC/USDT, and one exchange-based data source. The transaction-cost model is proportional and does not simulate order-book depth, partial fills, maker--taker routing, or time-varying bid--ask spreads. The study also excludes funding rates, order-book variables, liquidations, sentiment measures, and cross-asset predictors, all of which may contain information relevant to intraday cryptocurrency trading. Finally, the cost-aware threshold is simple and parametric; a more adaptive execution rule could adjust the threshold to market volatility, liquidity, or recent model calibration.

Future research should therefore extend the framework in four directions. First, the analysis should be repeated across additional crypto assets and market regimes to test whether the cost-aware result generalises beyond Bitcoin. Second, richer execution models should incorporate time-varying spreads, order-book liquidity, and funding-rate effects. Third, trading-oriented loss functions could be compared directly with the execution-filter approach to determine whether costs are better handled during model training or during trade execution. Fourth, adaptive threshold rules could be developed to vary the cost-aware filter with volatility, liquidity, or forecast uncertainty.

The evidence suggests that machine-learning forecasts for hourly BTC/USDT returns can have economic value, but only when the execution rule is disciplined. The main lesson is not that a particular architecture reliably beats the market. Rather, it is that short-horizon predictive signals are fragile, transaction costs are powerful, and the path from forecast to trade matters as much as the forecasting model itself.

% ====================================================================
% APPENDICES
% ====================================================================
\appendix

% ====================================================================
% APPENDIX A. DATA PREPROCESSING DETAILS
% ====================================================================
\section{Data preprocessing details}
\label{app:data}

This appendix reports additional data-quality and distributional diagnostics for the BTC/USDT hourly sample. These diagnostics support the main data-design choices described in Section~\ref{sec:data}: preserving a regular hourly grid, modelling log returns rather than price levels, and treating the return distribution as heavy-tailed.

\subsection{Missing timestamp audit}
\label{app:missing_timestamps}

A complete hourly time index was reconstructed over the modelling period beginning on 1~January~2018. The audit identifies 122 missing hourly timestamps within the modelling period. These gaps are filled using the procedure described in Section~\ref{sec:preprocessing_target}: the most recent close is carried forward and volume is set to zero, creating a flat synthetic bar at the missing timestamp.

Table~\ref{tab:gap_distribution} reports the temporal distribution of missing timestamps. Most gaps occur in the early part of the sample, particularly in 2018 and 2019. Since the total number of missing observations is small relative to the full sample of 70,872 hourly observations, the effect on aggregate results is expected to be limited. Nevertheless, observations immediately around gap timestamps may contain mechanically smoothed returns.

\begin{table}[H]
\centering
\captionsetup{justification=centering,singlelinecheck=false}
\caption{Temporal distribution of missing hourly timestamps}
\label{tab:gap_distribution}
\small
\renewcommand{\arraystretch}{1.05}
\setlength{\tabcolsep}{7pt}

\begin{tabular}{lrrrrrrrr}
\toprule
Year & 2018 & 2019 & 2020 & 2021 & 2022 & 2023 & 2024 & 2025 \\
\midrule
Missing timestamps & 62 & 28 & 18 & 13 & 0 & 1 & 0 & 0 \\
\bottomrule
\end{tabular}

\vspace{0.4em}
\begin{minipage}{0.88\textwidth}
\footnotesize \textit{Note:} The table reports missing timestamps from the walk-forward modelling period beginning on 1~January~2018. Six additional gaps occur in 2017 before the effective evaluation window and are excluded from this count.
\end{minipage}

\renewcommand{\arraystretch}{1.0}
\end{table}

\FloatBarrier

\subsection{Stationarity diagnostics}
\label{app:stationarity}

Stationarity is assessed using the Augmented Dickey--Fuller (ADF) test and the Kwiatkowski--Phillips--Schmidt--Shin (KPSS) test. The ADF test evaluates the null hypothesis of a unit root, while the KPSS test evaluates the null hypothesis of stationarity. Table~\ref{tab:stationarity} reports the results for BTC/USDT price levels and hourly log returns.

\begin{table}[H]
\centering
\captionsetup{justification=centering,singlelinecheck=false}
\caption{Stationarity tests for price and log returns}
\label{tab:stationarity}
\small
\renewcommand{\arraystretch}{1.05}
\setlength{\tabcolsep}{9pt}

\begin{threeparttable}
\begin{tabular}{lccc}
\toprule
Test & Statistic & Critical value (5\%) & $p$-value \\
\midrule
ADF (price) & -0.7929 & -2.8616 & 0.8212 \\
KPSS (price) & 32.9990 & 0.4630 & 0.0100 \\
ADF (log returns) & -37.5188 & -2.8616 & 0.0000 \\
KPSS (log returns) & 0.1314 & 0.4630 & 0.1000 \\
\bottomrule
\end{tabular}

\vspace{0.3em}
\begin{tablenotes}[flushleft]
\footnotesize
\item \textit{Note:} ADF tests the null hypothesis of a unit root. KPSS tests the null hypothesis of stationarity.
\end{tablenotes}
\end{threeparttable}

\renewcommand{\arraystretch}{1.0}
\end{table}

\FloatBarrier

The two tests support the modelling choice used in the main paper. For price levels, the ADF test fails to reject the unit-root null, while the KPSS test rejects stationarity. For log returns, the ADF test strongly rejects the unit-root null, while the KPSS test does not reject stationarity. This confirms that the forecasting target should be based on returns rather than raw prices.

\subsection{Normality diagnostic}
\label{app:normality}

Table~\ref{tab:normality} reports the Jarque--Bera test for hourly log returns. The test strongly rejects the null hypothesis of normality.

\begin{table}[H]
\centering
\captionsetup{justification=centering,singlelinecheck=false}
\caption{Jarque--Bera test for hourly log returns}
\label{tab:normality}
\small
\renewcommand{\arraystretch}{1.05}
\setlength{\tabcolsep}{12pt}

\begin{threeparttable}
\begin{tabular}{lc}
\toprule
Statistic & Value \\
\midrule
Jarque--Bera statistic & 4,569,076.78 \\
$p$-value & 0.0000 \\
\bottomrule
\end{tabular}

\vspace{0.3em}
\begin{tablenotes}[flushleft]
\footnotesize
\item \textit{Note:} The Jarque--Bera test evaluates the null hypothesis of normality.
\end{tablenotes}
\end{threeparttable}

\renewcommand{\arraystretch}{1.0}
\end{table}

\FloatBarrier

The rejection is driven by the extreme skewness and kurtosis documented in Table~\ref{tab:descriptive}. This supports two design choices in the empirical framework. First, performance evaluation should not rely only on Gaussian assumptions or in-sample fit. Second, volatility-regime information may be relevant for the forecasting problem, motivating the EGARCH-augmented feature tier used in the main analysis.

% ====================================================================
% APPENDIX B. HYPERPARAMETER SEARCH SPACES AND MODEL DETAILS
% ====================================================================
\section{Hyperparameter search spaces and model details}
\label{app:hpo}

This appendix reports implementation details that are relevant for reproducibility but too detailed for the main text: full hyperparameter search spaces, feature-selection diagnostics, and additional model implementation details.

\subsection{Hyperparameter search spaces}
\label{app:hpo_space}

Table~\ref{tab:hpo_space} reports the hyperparameter search spaces used for XGBoost, LSTM, and iTransformer. The two loss functions, MSE and MAE, are evaluated across all architectures.

\begin{table}[H]
\centering
\captionsetup{justification=centering,singlelinecheck=false}
\caption{Hyperparameter search space by architecture}
\label{tab:hpo_space}

\begin{threeparttable}
\footnotesize
\renewcommand{\arraystretch}{0.90}
\setlength{\tabcolsep}{5pt}

\begin{tabular}{p{2.4cm} p{4.2cm} p{6.2cm}}
\toprule
Model & Hyperparameter & Search space \\
\midrule

\multicolumn{3}{l}{\textit{XGBoost}} \\
& Loss function & \{\texttt{MSE}, \texttt{MAE}\} \\
& Maximum tree depth & integer $[2,4]$ \\
& Learning rate & $[0.005,\,0.03]$ log-uniform \\
& Number of estimators & integer $[1000,\,2500]$ \\
& Minimum child weight & $[10.0,\,40.0]$ \\
& Subsample & $[0.60,\,0.90]$ \\
& Column sample by tree & $[0.60,\,0.90]$ \\
& L1 regularisation & $[10^{-4},\,0.05]$ log-uniform \\
& L2 regularisation & $[1.0,\,40.0]$ log-uniform \\

\addlinespace[1pt]
\multicolumn{3}{l}{\textit{LSTM}} \\
& Loss function & \{\texttt{MSE}, \texttt{MAE}\} \\
& Sequence length & $\{24,48,72,96,120,168\}$ \\
& Hidden size & $\{32,64,96,128\}$ \\
& Number of layers & integer $[1,2]$ \\
& Dropout & $\{0.1,0.2,0.3\}$ \\
& Optimiser & \{\texttt{Adam}, \texttt{NAdam}\} \\
& Learning rate & $[5\times10^{-5},\,10^{-3}]$ log-uniform \\
& Weight decay & $\{10^{-5},10^{-4},10^{-3}\}$ \\
& Gradient clipping & $\{0.5,1.0\}$ \\
& Batch size & $\{32,64\}$ \\

\addlinespace[1pt]
\multicolumn{3}{l}{\textit{iTransformer}} \\
& Loss function & \{\texttt{MSE}, \texttt{MAE}\} \\
& Sequence length & $\{48,72,96,120,168\}$ \\
& Embedding dimension & $\{64,96,128\}$ \\
& Number of attention heads & $\{2,4\}$ \\
& Number of encoder layers & integer $[1,3]$ \\
& Feed-forward width & $\{128,256,512\}$ \\
& Dropout & $\{0.1,0.2,0.3\}$ \\
& Learning rate & $[5\times10^{-5},\,5\times10^{-4}]$ log-uniform \\
& Weight decay & $[10^{-5},\,3\times10^{-4}]$ log-uniform \\
& Gradient clipping & $\{0.5,1.0\}$ \\
& Batch size & $\{32,64\}$ \\
\bottomrule
\end{tabular}

\vspace{0.4em}
\begin{tablenotes}[flushleft]
\footnotesize
\item \textit{Note:} Hyperparameters are tuned independently within each walk-forward fold. For iTransformer, infeasible configurations are pruned when the embedding dimension is not divisible by the number of attention heads. Neural-network training uses early stopping with patience of 10 epochs; XGBoost uses early stopping with patience of 50 boosting rounds.
\end{tablenotes}

\end{threeparttable}

\renewcommand{\arraystretch}{1.0}
\end{table}

\FloatBarrier

\subsection{Feature-selection diagnostics}
\label{app:feature_selection_diagnostics}

Within each fold, candidate TA-related features are selected using a rank-based procedure applied to the training sample only. Candidate features are grouped by indicator family. The training sample is divided into four sequential blocks, and features are ranked within each block by absolute Spearman correlation with the aligned return target. For each feature~$j$, the average rank is:
\begin{equation}
\bar{R}_j = \frac{1}{4}\sum_{b=1}^{4} R_{j,b}.
\end{equation}
The feature with the lowest average rank is retained within each configured selection group.

Selection stability is measured with the Jaccard similarity coefficient between adjacent fold-level selected feature sets:
\begin{equation}
J(a,b) = \frac{|S_a \cap S_b|}{|S_a \cup S_b|},
\end{equation}
where $S_a$ and $S_b$ denote the selected TA-related feature sets in two consecutive folds.

Table~\ref{tab:feature_selection_diagnostics} summarises the resulting feature counts and the stability of the selected TA-related feature sets across adjacent walk-forward folds.

\begin{table}[H]
\centering
\captionsetup{justification=centering,singlelinecheck=false}
\caption{Feature-selection diagnostics across walk-forward folds}
\label{tab:feature_selection_diagnostics}
\small
\renewcommand{\arraystretch}{1.05}
\setlength{\tabcolsep}{7pt}

\begin{tabular}{lcccccc}
\toprule
Dataset tier & OHLCV & TA & EGARCH & Total & Mean Jaccard & Range \\
\midrule
OHLCV & 15 & 0 & 0 & 15 & -- & -- \\
OHLCV+TA & 15 & 10 & 0 & 25 & 0.673 & 0.333--1.000 \\
OHLCV+TA+EGARCH & 15 & 10 & 3 & 28 & 0.673 & 0.333--1.000 \\
\bottomrule
\end{tabular}

\vspace{0.4em}
\begin{minipage}{0.90\textwidth}
\footnotesize \textit{Note:} Feature counts follow from the dataset construction. The OHLCV+TA tier adds one selected feature from each of the 10 configured TA-related selection groups, and the OHLCV+TA+EGARCH tier further adds three EGARCH-derived volatility-regime predictors. Jaccard similarity is computed only for the selected TA-related feature sets.
\end{minipage}

\renewcommand{\arraystretch}{1.0}
\end{table}

\FloatBarrier

The mean Jaccard similarity of 0.673 indicates moderate stability. The selector does not produce unrelated feature sets across neighbouring folds, but it still adapts to changing market conditions.

\subsection{Model implementation summary}
\label{app:model_architecture_details}

The forecasting models follow standard implementations of XGBoost, LSTM, and iTransformer \citep{chen2016xgboost,hochreiter1997long,liu2024itransformer}. The main text reports the architecture equations because H4 directly compares model classes. Since the contribution of the paper is the trading evaluation framework rather than a new model architecture, this appendix reports only implementation choices that affect empirical comparability: input representation, sequence length, loss function, early stopping, feature-selection diagnostics, and hyperparameter search ranges. These details are summarised in Table~\ref{tab:hpo_space}.

% ====================================================================
% APPENDIX C. BOOTSTRAP ROBUSTNESS CHECKS
% ====================================================================
\section{Bootstrap robustness checks}
\label{app:bootstrap}

This appendix reports additional circular block-bootstrap specifications for the main inferential tests. The main text reports the primary block-length specification for each hypothesis. The tables and summaries below show whether the conclusions change under alternative assumptions about serial dependence in hourly strategy returns.

\subsection{H1 bootstrap robustness checks}
\label{app:h1_bootstrap_robustness}

Table~\ref{tab:h1_bootstrap_all_blocks} reports the 24-hour, 72-hour, and 168-hour circular block-bootstrap specifications for the H1 transaction-cost comparison. The conclusion is unchanged across all block lengths: transaction costs significantly reduce realised returns for every model--mode pair.

\begin{table}[H]
\centering
\captionsetup{justification=centering,singlelinecheck=false}
\caption{H1 bootstrap robustness check across block lengths}
\label{tab:h1_bootstrap_all_blocks}
\footnotesize
\renewcommand{\arraystretch}{0.86}
\setlength{\tabcolsep}{3pt}

\begin{tabular}{llcrrc}
\toprule
Model & Mode & Block (h) & Mean diff. (bps) & 95\% CI (bps) & Holm-Rej. \\
\midrule

XGBoost & LO & 24  & 1.79 & [1.73, 1.86] & Yes$^{***}$ \\
XGBoost & LO & 72  & 1.79 & [1.71, 1.88] & Yes$^{***}$ \\
XGBoost & LO & 168 & 1.79 & [1.67, 1.92] & Yes$^{***}$ \\

XGBoost & LS & 24  & 3.59 & [3.47, 3.71] & Yes$^{***}$ \\
XGBoost & LS & 72  & 3.59 & [3.41, 3.77] & Yes$^{***}$ \\
XGBoost & LS & 168 & 3.59 & [3.33, 3.84] & Yes$^{***}$ \\

LSTM & LO & 24  & 1.43 & [1.38, 1.48] & Yes$^{***}$ \\
LSTM & LO & 72  & 1.43 & [1.36, 1.50] & Yes$^{***}$ \\
LSTM & LO & 168 & 1.43 & [1.33, 1.52] & Yes$^{***}$ \\

LSTM & LS & 24  & 2.86 & [2.76, 2.96] & Yes$^{***}$ \\
LSTM & LS & 72  & 2.86 & [2.71, 2.99] & Yes$^{***}$ \\
LSTM & LS & 168 & 2.86 & [2.66, 3.05] & Yes$^{***}$ \\

iTransformer & LO & 24  & 3.03 & [2.99, 3.08] & Yes$^{***}$ \\
iTransformer & LO & 72  & 3.03 & [2.98, 3.08] & Yes$^{***}$ \\
iTransformer & LO & 168 & 3.03 & [2.97, 3.09] & Yes$^{***}$ \\

iTransformer & LS & 24  & 6.06 & [5.97, 6.15] & Yes$^{***}$ \\
iTransformer & LS & 72  & 6.06 & [5.96, 6.17] & Yes$^{***}$ \\
iTransformer & LS & 168 & 6.06 & [5.94, 6.18] & Yes$^{***}$ \\

\bottomrule
\end{tabular}

\vspace{0.4em}
\begin{minipage}{0.95\textwidth}
\footnotesize \textit{Note:} LO = long-only; LS = long-short. Mean differences are expressed in basis points per hour and are computed as no-cost strategy returns minus transaction-cost-adjusted strategy returns. Confidence intervals and Holm-adjusted decisions are obtained from paired circular block bootstraps with 10,000 replications. Blocks of 24, 72, and 168 hours correspond to one-day, three-day, and seven-day dependence windows. $^{***}$ denotes rejection at the 1\% level.
\end{minipage}

\renewcommand{\arraystretch}{1.0}
\end{table}

\FloatBarrier

\subsection{H2 bootstrap robustness checks}
\label{app:h2_bootstrap_robustness}

The main text reports the 168-hour block specification for H2 in Table~\ref{tab:h2_bootstrap}. Table~\ref{tab:h2_bootstrap_short_blocks} reports the 24-hour and 72-hour block specifications. The inference is unchanged: cost-aware execution improves the unfiltered baseline across all selector--mode pairs.

\begin{table}[H]
\centering
\captionsetup{justification=centering,singlelinecheck=false}
\caption{H2 bootstrap robustness check, 24-hour and 72-hour blocks}
\label{tab:h2_bootstrap_short_blocks}
\footnotesize
\renewcommand{\arraystretch}{0.86}
\setlength{\tabcolsep}{2.5pt}

\begin{tabular}{llcrrrrr}
\toprule
Selector & Mode & Block (h) & Trades CA & Diff. (bps) & CI (bps) & $\Delta SR$ & SR CI \\
\midrule

Loss-best & LO & 24
& 251 & 1.75 & [1.47, 2.05] & 2.34 & [1.53, 3.58] \\

Loss-best & LO & 72
& 251 & 1.75 & [1.44, 2.06] & 2.34 & [1.54, 3.55] \\

IC-best & LO & 24
& 261 & 1.54 & [1.28, 1.82] & 2.02 & [1.31, 3.12] \\

IC-best & LO & 72
& 261 & 1.54 & [1.26, 1.84] & 2.02 & [1.31, 3.13] \\

IR$^{**}$-best & LO & 24
& 197 & 0.79 & [0.53, 1.06] & 1.42 & [0.83, 2.34] \\

IR$^{**}$-best & LO & 72
& 197 & 0.79 & [0.51, 1.07] & 1.42 & [0.82, 2.33] \\

Loss-best & LS & 24
& 60 & 3.38 & [2.88, 3.92] & 2.14 & [1.30, 3.50] \\

Loss-best & LS & 72
& 60 & 3.38 & [2.85, 3.93] & 2.14 & [1.30, 3.49] \\

IC-best & LS & 24
& 47 & 3.08 & [2.43, 3.71] & 1.90 & [1.09, 3.19] \\

IC-best & LS & 72
& 47 & 3.08 & [2.39, 3.72] & 1.90 & [1.08, 3.18] \\

IR$^{**}$-best & LS & 24
& 47 & 1.77 & [1.27, 2.29] & 1.86 & [1.07, 3.18] \\

IR$^{**}$-best & LS & 72
& 47 & 1.77 & [1.25, 2.31] & 1.86 & [1.06, 3.17] \\

\bottomrule
\end{tabular}

\vspace{0.4em}
\begin{minipage}{0.95\textwidth}
\footnotesize \textit{Note:} LO = long-only; LS = long-short; CA = cost-aware execution. The table reports paired circular block-bootstrap results using 24-hour and 72-hour blocks and 10,000 replications. Diff. is the mean hourly return difference in basis points, computed as cost-aware minus baseline returns. CI (bps) is the 95\% confidence interval for the mean return difference. $\Delta SR$ is the Sharpe ratio of the cost-aware strategy minus the Sharpe ratio of the baseline strategy. SR CI is the 95\% confidence interval for $\Delta SR$. All corresponding Holm-adjusted tests reject at the 1\% level, consistent with the 168-hour results in Table~\ref{tab:h2_bootstrap}.
\end{minipage}

\renewcommand{\arraystretch}{1.0}
\end{table}

\FloatBarrier

\subsection{H3 bootstrap robustness checks}
\label{app:h3_bootstrap_robustness}

The main text reports the 72-hour block specification for H3 in Table~\ref{tab:h3_bootstrap}. Tables~\ref{tab:h3_bootstrap_24h} and~\ref{tab:h3_bootstrap_168h} report the corresponding 24-hour and 168-hour robustness specifications. The 24-hour specification gives the same qualitative result as the 72-hour case: the strongest evidence is for adding technical indicators to raw OHLCV under loss-based long-only selection. Under the stricter 168-hour block length, this effect remains positive but no longer survives Holm correction.

\begin{table}[H]
\centering
\captionsetup{justification=centering,singlelinecheck=false}
\caption{H3 bootstrap robustness check, 24-hour blocks}
\label{tab:h3_bootstrap_24h}
\footnotesize
\renewcommand{\arraystretch}{0.86}
\setlength{\tabcolsep}{2.1pt}

\begin{tabular}{lllrrrrrc}
\toprule
Selector & Mode & Comparison & Trades & Diff. (bps) & CI (bps) & $\Delta SR$ & SR CI & Holm \\
\midrule

\multirow{4}{*}{Loss-best}
& LO & OHLCV+TA / OHLCV
& 277 / 117 & 0.41 & [0.11, 0.72] & 0.95 & [0.25, 1.89] & Yes \\

& LO & EGARCH / OHLCV+TA
& 251 / 277 & -0.06 & [-0.29, 0.18] & -0.13 & [-0.78, 0.49] & No \\

& LO & EGARCH / OHLCV
& 251 / 117 & 0.35 & [0.04, 0.65] & 0.82 & [0.14, 1.68] & No \\

& LS & EGARCH / OHLCV+TA
& 60 / 66 & 0.09 & [-0.32, 0.51] & 0.17 & [-0.65, 1.04] & No \\

\addlinespace[1pt]

\multirow{4}{*}{IR$^{**}$-best}
& LO & OHLCV+TA / OHLCV
& 215 / 143 & 0.13 & [-0.11, 0.39] & 0.36 & [-0.20, 1.02] & No \\

& LO & EGARCH / OHLCV+TA
& 197 / 215 & 0.07 & [-0.15, 0.29] & 0.16 & [-0.39, 0.76] & No \\

& LO & EGARCH / OHLCV
& 197 / 143 & 0.20 & [-0.05, 0.47] & 0.51 & [-0.07, 1.29] & No \\

& LS & EGARCH / OHLCV+TA
& 47 / 64 & 0.13 & [-0.24, 0.49] & 0.25 & [-0.53, 1.05] & No \\

\bottomrule
\end{tabular}

\vspace{0.4em}
\begin{minipage}{0.95\textwidth}
\footnotesize \textit{Note:} LO = long-only; LS = long-short; EGARCH denotes the OHLCV+TA+EGARCH feature specification. The table reports paired circular block-bootstrap results using 24-hour blocks and 10,000 replications. Trades reports the number of trades in the richer / comparison specification. Diff. is the mean hourly return difference in basis points. CI (bps) is the 95\% confidence interval for the mean return difference. $\Delta SR$ is the Sharpe-ratio difference, and SR CI is its 95\% confidence interval. Holm indicates whether both the mean return-difference and Sharpe-difference tests reject after Holm correction. Comparisons involving long-short OHLCV are excluded because the OHLCV long-short strategies have fewer than 20 trades.
\end{minipage}

\renewcommand{\arraystretch}{1.0}
\end{table}

\FloatBarrier

\begin{table}[H]
\centering
\captionsetup{justification=centering,singlelinecheck=false}
\caption{H3 bootstrap robustness check, 168-hour blocks}
\label{tab:h3_bootstrap_168h}
\footnotesize
\renewcommand{\arraystretch}{0.86}
\setlength{\tabcolsep}{2.1pt}

\begin{tabular}{lllrrrrrc}
\toprule
Selector & Mode & Comparison & Trades & Diff. (bps) & CI (bps) & $\Delta SR$ & SR CI & Holm \\
\midrule

\multirow{4}{*}{Loss-best}
& LO & OHLCV+TA / OHLCV
& 277 / 117 & 0.41 & [0.07, 0.76] & 0.95 & [0.19, 1.91] & No \\

& LO & EGARCH / OHLCV+TA
& 251 / 277 & -0.06 & [-0.33, 0.21] & -0.13 & [-0.90, 0.55] & No \\

& LO & EGARCH / OHLCV
& 251 / 117 & 0.35 & [0.01, 0.70] & 0.82 & [0.09, 1.71] & No \\

& LS & EGARCH / OHLCV+TA
& 60 / 66 & 0.09 & [-0.29, 0.48] & 0.17 & [-0.61, 0.98] & No \\

\addlinespace[1pt]

\multirow{4}{*}{IR$^{**}$-best}
& LO & OHLCV+TA / OHLCV
& 215 / 143 & 0.13 & [-0.12, 0.40] & 0.36 & [-0.22, 1.05] & No \\

& LO & EGARCH / OHLCV+TA
& 197 / 215 & 0.07 & [-0.16, 0.30] & 0.16 & [-0.42, 0.76] & No \\

& LO & EGARCH / OHLCV
& 197 / 143 & 0.20 & [-0.07, 0.49] & 0.51 & [-0.10, 1.25] & No \\

& LS & EGARCH / OHLCV+TA
& 47 / 64 & 0.13 & [-0.28, 0.54] & 0.25 & [-0.60, 1.17] & No \\

\bottomrule
\end{tabular}

\vspace{0.4em}
\begin{minipage}{0.95\textwidth}
\footnotesize \textit{Note:} LO = long-only; LS = long-short; EGARCH denotes the OHLCV+TA+EGARCH feature specification. The table reports paired circular block-bootstrap results using 168-hour blocks and 10,000 replications. Trades reports the number of trades in the richer / comparison specification. Diff. is the mean hourly return difference in basis points. CI (bps) is the 95\% confidence interval for the mean return difference. $\Delta SR$ is the Sharpe-ratio difference, and SR CI is its 95\% confidence interval. Holm indicates whether both the mean return-difference and Sharpe-difference tests reject after Holm correction. Comparisons involving long-short OHLCV are excluded because the OHLCV long-short strategies have fewer than 20 trades.
\end{minipage}

\renewcommand{\arraystretch}{1.0}
\end{table}

\FloatBarrier

The 168-hour results confirm the cautious H3 interpretation. Technical indicators improve the long-only loss-best strategy in point-estimate terms, but the result no longer survives Holm correction under the weekly dependence assumption. EGARCH-derived features do not provide a robust incremental improvement beyond technical indicators.

\subsection{H4 bootstrap robustness checks}
\label{app:h4_bootstrap_robustness}

The main text reports the 72-hour block specification for H4 in Table~\ref{tab:h4_bootstrap}. Tables~\ref{tab:h4_bootstrap_24h} and~\ref{tab:h4_bootstrap_168h} report the 24-hour and 168-hour alternatives. Both robustness checks lead to the same conclusion: XGBoost is descriptively stronger, but the bootstrap evidence is not sufficient to claim statistically robust dominance over LSTM or iTransformer.

\begin{table}[H]
\centering
\captionsetup{justification=centering,singlelinecheck=false}
\caption{H4 bootstrap robustness check, architecture comparisons, 24-hour blocks}
\label{tab:h4_bootstrap_24h}
\footnotesize
\renewcommand{\arraystretch}{0.90}
\setlength{\tabcolsep}{4.5pt}

\begin{tabular}{lrrrrrrr}
\toprule
Comparison & Trades & Diff. (bps) & CI (bps) & $p$ & $p_{\mathrm{Holm}}$ & $\Delta SR$ & SR CI \\
\midrule

XGBoost / LSTM
& 251 / 69
& 0.13
& [-0.19, 0.46]
& 0.211
& 0.211
& 0.25
& [-0.59, 1.15] \\

XGBoost / iTransformer
& 251 / 76
& 0.28
& [-0.08, 0.65]
& 0.065
& 0.130
& 0.45
& [-0.47, 1.53] \\

\bottomrule
\end{tabular}

\vspace{0.4em}
\begin{minipage}{0.92\textwidth}
\footnotesize \textit{Note:} The table reports paired circular block-bootstrap tests using 24-hour blocks and 10,000 replications. Comparisons are based on long-only cost-aware strategies under the OHLCV+TA+EGARCH feature specification, MSE loss, loss-best selection, and $\lambda=2.0$. Diff. and $\Delta SR$ are computed as XGBoost minus the comparison architecture. Diff. is expressed in basis points per hour. CI (bps) is the 95\% confidence interval for Diff.; SR CI is the 95\% confidence interval for $\Delta SR$. Positive values favour XGBoost.
\end{minipage}

\renewcommand{\arraystretch}{1.0}
\end{table}

\begin{table}[H]
\centering
\captionsetup{justification=centering,singlelinecheck=false}
\caption{H4 bootstrap robustness check, architecture comparisons, 168-hour blocks}
\label{tab:h4_bootstrap_168h}
\footnotesize
\renewcommand{\arraystretch}{0.90}
\setlength{\tabcolsep}{4.5pt}

\begin{tabular}{lrrrrrrr}
\toprule
Comparison & Trades & Diff. (bps) & CI (bps) & $p$ & $p_{\mathrm{Holm}}$ & $\Delta SR$ & SR CI \\
\midrule

XGBoost / LSTM
& 251 / 69
& 0.13
& [-0.19, 0.46]
& 0.208
& 0.208
& 0.25
& [-0.63, 1.14] \\

XGBoost / iTransformer
& 251 / 76
& 0.28
& [-0.08, 0.65]
& 0.068
& 0.136
& 0.45
& [-0.49, 1.50] \\

\bottomrule
\end{tabular}

\vspace{0.4em}
\begin{minipage}{0.92\textwidth}
\footnotesize \textit{Note:} The table reports paired circular block-bootstrap tests using 168-hour blocks and 10,000 replications. Comparisons are based on long-only cost-aware strategies under the OHLCV+TA+EGARCH feature specification, MSE loss, loss-best selection, and $\lambda=2.0$. Diff. and $\Delta SR$ are computed as XGBoost minus the comparison architecture. Diff. is expressed in basis points per hour. CI (bps) is the 95\% confidence interval for Diff.; SR CI is the 95\% confidence interval for $\Delta SR$. Positive values favour XGBoost.
\end{minipage}

\renewcommand{\arraystretch}{1.0}
\end{table}

\FloatBarrier

\subsection{H6 bootstrap robustness checks}
\label{app:h6_bootstrap_robustness}

The main text reports the 72-hour block specification for H6 in Table~\ref{tab:h6_bootstrap}. Table~\ref{tab:h6_bootstrap_robustness_blocks} reports the 24-hour and 168-hour robustness specifications. The inference is unchanged across block lengths: selector choice changes point estimates, but no pairwise selector comparison survives Holm correction.

\begin{table}[H]
\centering
\captionsetup{justification=centering,singlelinecheck=false}
\caption{H6 bootstrap robustness check, 24-hour and 168-hour blocks}
\label{tab:h6_bootstrap_robustness_blocks}
\footnotesize
\renewcommand{\arraystretch}{0.84}
\setlength{\tabcolsep}{1.8pt}

\begin{tabular}{llcrrrrrr}
\toprule
Mode & Comparison & Block (h) & Trades & Diff. (bps) & CI (bps) & $p_{\mathrm{Holm}}$ & $\Delta SR$ & SR CI \\
\midrule

LO & Loss / IC & 24
& 251 / 261 & 0.112 & [-0.129, 0.356] & 1.000 & 0.256 & [-0.335, 0.920] \\

LO & Loss / IC & 168
& 251 / 261 & 0.112 & [-0.137, 0.366] & 1.000 & 0.256 & [-0.344, 0.927] \\

LO & Loss / IR$^{**}$ & 24
& 251 / 197 & -0.011 & [-0.218, 0.193] & 1.000 & 0.033 & [-0.512, 0.558] \\

LO & Loss / IR$^{**}$ & 168
& 251 / 197 & -0.011 & [-0.226, 0.201] & 1.000 & 0.033 & [-0.526, 0.566] \\

LO & IR$^{**}$ / IC & 24
& 197 / 261 & 0.123 & [-0.114, 0.361] & 1.000 & 0.222 & [-0.343, 0.884] \\

LO & IR$^{**}$ / IC & 168
& 197 / 261 & 0.123 & [-0.123, 0.371] & 1.000 & 0.222 & [-0.356, 0.892] \\

LS & Loss / IC & 24
& 60 / 47 & 0.116 & [-0.252, 0.503] & 1.000 & 0.219 & [-0.532, 1.021] \\

LS & Loss / IC & 168
& 60 / 47 & 0.116 & [-0.268, 0.516] & 1.000 & 0.219 & [-0.548, 1.030] \\

LS & IR$^{**}$ / Loss & 24
& 47 / 60 & 0.052 & [-0.337, 0.458] & 1.000 & 0.106 & [-0.744, 0.984] \\

LS & IR$^{**}$ / Loss & 168
& 47 / 60 & 0.052 & [-0.352, 0.476] & 1.000 & 0.106 & [-0.762, 0.996] \\

LS & IR$^{**}$ / IC & 24
& 47 / 47 & 0.167 & [-0.189, 0.527] & 1.000 & 0.324 & [-0.365, 1.124] \\

LS & IR$^{**}$ / IC & 168
& 47 / 47 & 0.167 & [-0.205, 0.544] & 1.000 & 0.324 & [-0.384, 1.136] \\

\bottomrule
\end{tabular}

\vspace{0.4em}
\begin{minipage}{0.95\textwidth}
\footnotesize \textit{Note:} LO = long-only; LS = long-short. Loss = loss-best selector; IC = information-coefficient-best selector; IR$^{**}$ = mode-specific IR$^{**}$-best selector. The table reports paired circular block-bootstrap robustness checks using 24-hour and 168-hour blocks and 10,000 replications. Results use XGBoost with OHLCV+TA+EGARCH features, MSE loss, and cost-aware execution with $\lambda=2.0$. Comparisons are reported as first selector / second selector; Diff. and $\Delta SR$ are computed as first minus second. Diff. is expressed in basis points per hour. CI (bps) is the 95\% confidence interval for Diff.; SR CI is the 95\% confidence interval for $\Delta SR$. No comparison survives Holm correction.
\end{minipage}

\renewcommand{\arraystretch}{1.0}
\end{table}

\FloatBarrier

Across the alternative block lengths, the main conclusions remain unchanged. The H3 evidence is conditional and weakens under weekly blocks; the H4 architecture ranking favours XGBoost descriptively but not statistically; and the H6 selector comparisons show no robust dominance after Holm correction.

% ====================================================================
% APPENDIX D. FOLD-LEVEL DECOMPOSITION
% ====================================================================
\section{Fold-level decomposition}
\label{app:folds}

This appendix reports the fold-level decomposition for the flagship cost-aware strategy used in the main empirical discussion: XGBoost with OHLCV+TA+EGARCH features, MSE loss, loss-best selection, long-only cost-aware execution with $\lambda=2.0$, and transaction costs of 10~basis points. The purpose is diagnostic rather than inferential. Each fold corresponds to a three-month out-of-sample test window, so annualised fold-level metrics should be interpreted with caution.

The decomposition is split into two tables for readability. Table~\ref{tab:fold_decomposition_returns} reports fold-level return and risk metrics, while Table~\ref{tab:fold_decomposition_execution} reports execution diagnostics. Because each fold contains only one quarter of data, annualised ARC, Sharpe ratio, and IR$^{**}$ values can become mechanically large when the corresponding non-annualised quarterly return is high and drawdowns remain limited. The cumulative fold return and the number of trades are therefore the more informative diagnostics for assessing temporal stability.

\begin{table}[H]
\centering
\captionsetup{justification=centering,singlelinecheck=false}
\caption{Fold-level decomposition for the flagship cost-aware XGBoost strategy: returns and risk}
\label{tab:fold_decomposition_returns}
\footnotesize
\renewcommand{\arraystretch}{0.82}
\setlength{\tabcolsep}{2.5pt}

\begin{tabular}{rllrrrrrr}
\toprule
Fold & Start & End & ARC & ASD & MD & MLD & SR & IR$^{**}$ \\
     &       &     & (\%) & (\%) & (\%) & (yrs) &    &          \\
\midrule
1  & 2019-04-01 & 2019-06-30 & 3255.26 & 63.59 & -16.18 & 0.07 & 51.12 & 10295.34 \\
2  & 2019-07-01 & 2019-09-30 & 9.70    & 69.43 & -26.97 & 0.12 & 0.08  & 0.05 \\
3  & 2019-10-01 & 2019-12-31 & -22.53  & 29.25 & -16.22 & 0.09 & -0.91 & -1.07 \\
4  & 2020-01-01 & 2020-03-31 & 104.81  & 93.96 & -42.52 & 0.13 & 1.07  & 2.75 \\
5  & 2020-04-01 & 2020-06-30 & 235.97  & 60.51 & -17.46 & 0.08 & 3.83  & 52.69 \\
6  & 2020-07-01 & 2020-09-30 & 84.61   & 44.29 & -19.93 & 0.12 & 1.82  & 8.11 \\
7  & 2020-10-01 & 2020-12-31 & 3316.76 & 53.87 & -12.93 & 0.03 & 61.49 & 15797.99 \\
8  & 2021-01-01 & 2021-03-31 & 1057.89 & 88.60 & -21.02 & 0.05 & 11.89 & 600.84 \\
9  & 2021-04-01 & 2021-06-30 & -81.29  & 88.11 & -51.05 & 0.14 & -0.97 & -1.47 \\
10 & 2021-07-01 & 2021-09-30 & 69.79   & 58.68 & -23.13 & 0.07 & 1.12  & 3.59 \\
11 & 2021-10-01 & 2021-12-31 & -10.68  & 58.57 & -31.96 & 0.14 & -0.25 & -0.06 \\
12 & 2022-01-01 & 2022-03-31 & -53.51  & 61.31 & -35.53 & 0.14 & -0.94 & -1.31 \\
13 & 2022-04-01 & 2022-06-30 & -22.83  & 56.02 & -30.05 & 0.06 & -0.48 & -0.31 \\
14 & 2022-07-01 & 2022-09-30 & -21.99  & 50.56 & -17.74 & 0.09 & -0.52 & -0.54 \\
15 & 2022-10-01 & 2022-12-31 & -24.53  & 29.38 & -13.58 & 0.09 & -0.98 & -1.51 \\
16 & 2023-01-01 & 2023-03-31 & 433.71  & 28.94 & -3.59  & 0.13 & 14.84 & 1811.43 \\
17 & 2023-04-01 & 2023-06-30 & 42.22   & 33.90 & -16.29 & 0.15 & 1.12  & 3.23 \\
18 & 2023-07-01 & 2023-09-30 & -1.08   & 16.97 & -6.40  & 0.09 & -0.31 & -0.01 \\
19 & 2023-10-01 & 2023-12-31 & 2.76    & 19.16 & -5.83  & 0.03 & -0.08 & 0.07 \\
20 & 2024-01-01 & 2024-03-31 & 430.66  & 52.92 & -20.18 & 0.09 & 8.06  & 173.63 \\
21 & 2024-04-01 & 2024-06-30 & -35.02  & 43.20 & -17.28 & 0.10 & -0.91 & -1.64 \\
22 & 2024-07-01 & 2024-09-30 & 1.22    & 51.23 & -29.01 & 0.17 & -0.06 & 0.00 \\
23 & 2024-10-01 & 2024-12-31 & 25.74   & 39.89 & -15.13 & 0.04 & 0.54  & 1.10 \\
24 & 2025-01-01 & 2025-03-31 & -32.49  & 53.91 & -28.31 & 0.19 & -0.68 & -0.69 \\
25 & 2025-04-01 & 2025-06-30 & 236.51  & 39.33 & -11.54 & 0.11 & 5.91  & 123.27 \\
26 & 2025-07-01 & 2025-09-30 & 7.98    & 29.42 & -13.27 & 0.13 & 0.13  & 0.16 \\
27 & 2025-10-01 & 2025-12-31 & -61.13  & 44.13 & -29.16 & 0.18 & -1.48 & -2.90 \\
\bottomrule
\end{tabular}

\vspace{0.4em}
\begin{minipage}{0.95\textwidth}
\footnotesize \textit{Note:} Results are reported for XGBoost with OHLCV+TA+EGARCH features, MSE loss, loss-best selection, long-only cost-aware execution, $\lambda=2.0$, and 10 basis point transaction costs. ARC, ASD, and MD are percentages; MLD is reported in years. Fold-level ARC, SR, and IR$^{**}$ are annualised from three-month test windows and can therefore become mechanically large when a quarterly cumulative return is high. These fold-level values are diagnostic and should be interpreted together with the execution diagnostics in Table~\ref{tab:fold_decomposition_execution}.
\end{minipage}

\renewcommand{\arraystretch}{1.0}
\end{table}

\FloatBarrier

\begin{table}[H]
\centering
\captionsetup{justification=centering,singlelinecheck=false}
\caption{Fold-level decomposition for the flagship cost-aware XGBoost strategy: execution diagnostics}
\label{tab:fold_decomposition_execution}
\footnotesize
\renewcommand{\arraystretch}{0.82}
\setlength{\tabcolsep}{4pt}

\begin{tabular}{rrrrcc}
\toprule
Fold & Trades & Equity final & Cum. return & Profitable & Low-trade \\
     &        &              & (\%)        &            &           \\
\midrule
1  & 20 & 2.4000 & 140.00 & Yes & No \\
2  & 31 & 1.0236 & 2.36   & Yes & No \\
3  & 3  & 0.9377 & -6.23  & No  & Yes \\
4  & 35 & 1.1956 & 19.56  & Yes & No \\
5  & 5  & 1.3526 & 35.26  & Yes & Yes \\
6  & 1  & 1.1670 & 16.70  & Yes & Yes \\
7  & 9  & 2.4343 & 143.43 & Yes & Yes \\
8  & 39 & 1.8287 & 82.87  & Yes & No \\
9  & 23 & 0.6586 & -34.14 & No  & No \\
10 & 1  & 1.1427 & 14.27  & Yes & Yes \\
11 & 5  & 0.9719 & -2.81  & No  & Yes \\
12 & 5  & 0.8280 & -17.20 & No  & Yes \\
13 & 28 & 0.9375 & -6.25  & No  & No \\
14 & 11 & 0.9393 & -6.07  & No  & Yes \\
15 & 3  & 0.9316 & -6.84  & No  & Yes \\
16 & 4  & 1.5110 & 51.10  & Yes & Yes \\
17 & 1  & 1.0917 & 9.17   & Yes & Yes \\
18 & 2  & 0.9973 & -0.27  & No  & Yes \\
19 & 1  & 1.0069 & 0.69   & Yes & Yes \\
20 & 1  & 1.5157 & 51.57  & Yes & Yes \\
21 & 5  & 0.8981 & -10.19 & No  & Yes \\
22 & 26 & 1.0030 & 0.30   & Yes & No \\
23 & 1  & 1.0594 & 5.94   & Yes & Yes \\
24 & 1  & 0.9077 & -9.23  & No  & Yes \\
25 & 1  & 1.3531 & 35.31  & Yes & Yes \\
26 & 1  & 1.0195 & 1.95   & Yes & Yes \\
27 & 1  & 0.7882 & -21.18 & No  & Yes \\
\bottomrule
\end{tabular}

\vspace{0.4em}
\begin{minipage}{0.90\textwidth}
\footnotesize \textit{Note:} Equity final is the terminal equity value within each three-month test fold, starting from one. Cumulative return is computed as terminal equity minus one and expressed in percentage terms. Low-trade indicates fewer than 20 trades within the individual fold. The low-trade flag is diagnostic at the fold level; formal inference in the main paper is based on the full consolidated out-of-sample return series.
\end{minipage}

\renewcommand{\arraystretch}{1.0}
\end{table}

\FloatBarrier

The fold-level decomposition confirms the interpretation in Section~\ref{sec:fold_stability}. The flagship strategy is profitable in 16 of 27 folds, but the distribution of returns is uneven. Several profitable folds generate very large annualised ARC values because they compound strongly over a short three-month window, while many other folds contain very few trades. The result is therefore best interpreted as evidence of positive aggregate cost-aware performance with substantial regime dependence, not as evidence of stable profitability in every quarter.

% ====================================================================
% APPENDIX E. VALIDATION LOSS DIAGNOSTIC
% ====================================================================
\section{Validation loss diagnostic}
\label{app:validation_diagnostic}

This appendix reports an additional diagnostic on the relationship between validation loss and realised out-of-sample trading performance. The diagnostic is conducted for XGBoost with OHLCV+TA+EGARCH features, MSE loss, cost-aware execution with $\lambda=2.0$, and transaction costs of 10~basis points. The purpose is to assess whether lower validation loss is a reliable proxy for higher economic value after forecasts are converted into positions and transaction costs are applied.

For each trading mode, the selected validation loss is paired with realised out-of-sample trading metrics across fold--selector observations. Spearman rank correlations are used because both validation losses and realised trading outcomes are noisy, non-Gaussian, and affected by extreme market episodes.

\begin{equation}
\rho_s =
\mathrm{corr}_{\mathrm{Spearman}}
\left(L_{\mathrm{val}}, M_{\mathrm{test}}\right),
\end{equation}
where \(L_{\mathrm{val}}\) denotes the validation loss of the selected configuration and \(M_{\mathrm{test}}\) denotes the realised out-of-sample trading metric.

\begin{figure}[H]
\centering
\captionsetup{justification=centering,singlelinecheck=false}
\caption{Validation loss versus realised out-of-sample trading performance}
\includegraphics[width=\textwidth]{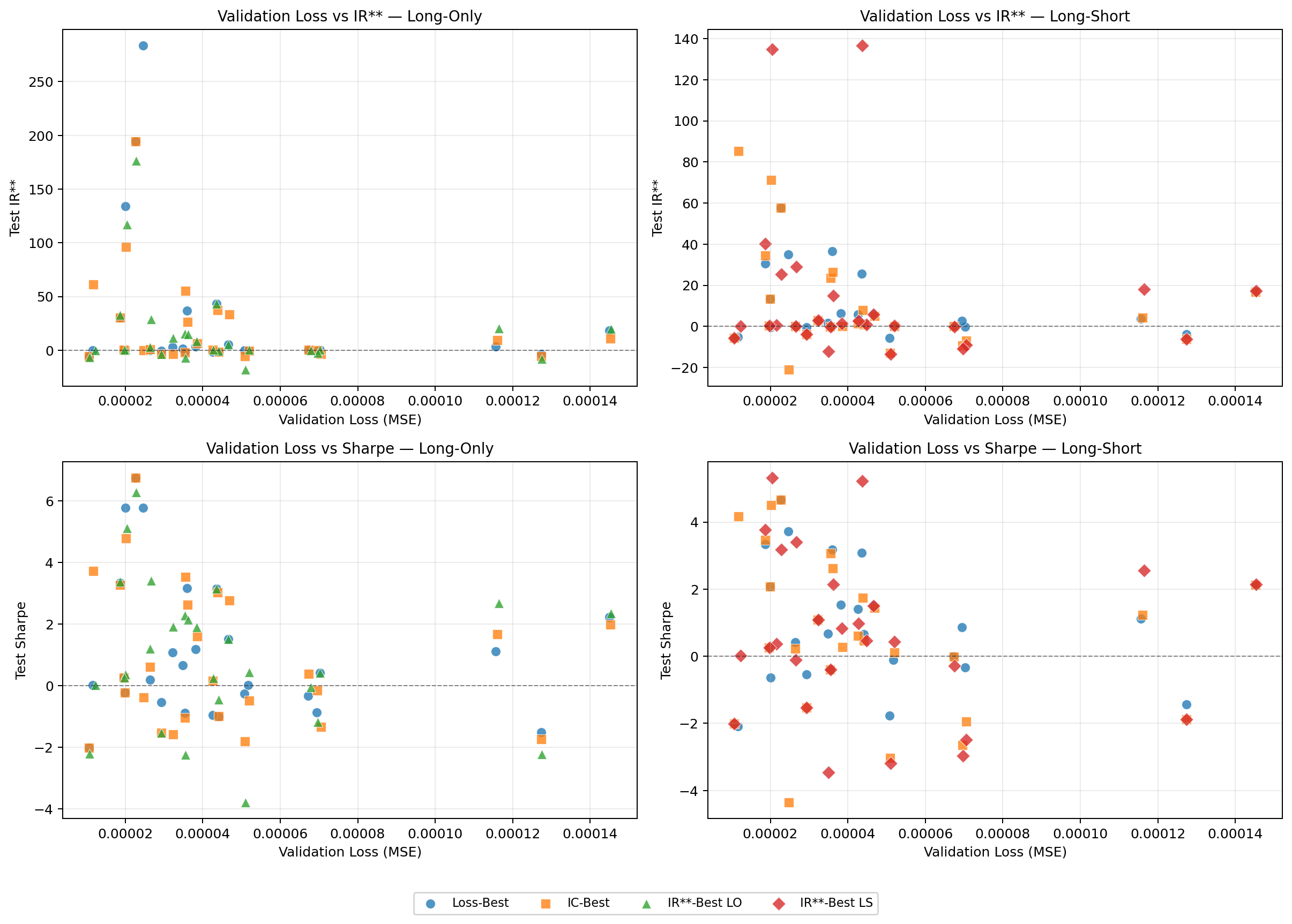}

\vspace{0.4em}
\begin{minipage}{0.95\textwidth}
\footnotesize \textit{Note:} The figure compares selected validation loss with realised out-of-sample trading metrics across fold--selector observations. Results are shown for XGBoost with OHLCV+TA+EGARCH features, MSE loss, cost-aware execution with $\lambda=2.0$, and transaction costs of 10 basis points. The diffuse point clouds illustrate that lower validation loss is not a reliable standalone proxy for stronger realised trading performance.
\end{minipage}

\label{fig:validation_loss_trading_diagnostic}
\end{figure}

Figure~\ref{fig:validation_loss_trading_diagnostic} visualises the relationship between selected validation loss and realised out-of-sample trading metrics.

\FloatBarrier

Table~\ref{tab:validation_loss_trading_diagnostic} reports the corresponding Spearman rank correlations.

\begin{table}[H]
\centering
\captionsetup{justification=centering,singlelinecheck=false}
\caption{Validation loss and realised trading performance diagnostic}
\label{tab:validation_loss_trading_diagnostic}
\small
\renewcommand{\arraystretch}{1.02}
\setlength{\tabcolsep}{7pt}

\begin{tabular}{llrrr}
\toprule
Mode & Test metric & $n$ & Spearman $\rho_s$ & $p$-value \\
\midrule

LO & IR$^{**}$ & 81 & -0.169 & 0.132 \\
LO & Sharpe    & 81 & -0.172 & 0.124 \\
LO & IR$^{*}$  & 81 & -0.190 & 0.090 \\
LO & ARC       & 81 & -0.162 & 0.148 \\

\addlinespace[1pt]

LS & IR$^{**}$ & 81 & -0.183 & 0.102 \\
LS & Sharpe    & 81 & -0.155 & 0.167 \\
LS & IR$^{*}$  & 81 & -0.172 & 0.124 \\
LS & ARC       & 81 & -0.154 & 0.171 \\

\bottomrule
\end{tabular}

\vspace{0.4em}
\begin{minipage}{0.90\textwidth}
\footnotesize \textit{Note:} LO = long-only; LS = long-short. The table reports Spearman rank correlations between selected validation loss and realised out-of-sample trading metrics across fold--selector observations. Each mode contains 81 observations, corresponding to 27 folds and three selector observations. Reported $p$-values correspond to two-sided tests of \(H_0:\rho_s=0\).
\end{minipage}

\renewcommand{\arraystretch}{1.0}
\end{table}

\FloatBarrier

A second diagnostic replaces the raw validation-loss value with the validation-loss rank within the corresponding Optuna search distribution. This transformation controls for fold-level differences in volatility and loss scale. The question is whether configurations closer to the best validation-loss trial within each fold also deliver stronger out-of-sample trading performance. Table~\ref{tab:validation_loss_rank_trading_diagnostic} reports the corresponding rank-based correlations.

\begin{table}[H]
\centering
\captionsetup{justification=centering,singlelinecheck=false}
\caption{Validation-loss rank and realised trading performance diagnostic}
\label{tab:validation_loss_rank_trading_diagnostic}
\small
\renewcommand{\arraystretch}{1.02}
\setlength{\tabcolsep}{7pt}

\begin{tabular}{llrrr}
\toprule
Mode & Test metric & $n$ & Spearman $\rho_s$ & $p$-value \\
\midrule

LO & IR$^{**}$ & 81 & 0.066 & 0.558 \\
LO & Sharpe    & 81 & 0.076 & 0.503 \\
LO & IR$^{*}$  & 81 & 0.083 & 0.460 \\
LO & ARC       & 81 & 0.050 & 0.658 \\

\addlinespace[1pt]

LS & IR$^{**}$ & 81 & 0.017 & 0.883 \\
LS & Sharpe    & 81 & 0.025 & 0.827 \\
LS & IR$^{*}$  & 81 & 0.023 & 0.838 \\
LS & ARC       & 81 & 0.040 & 0.726 \\

\bottomrule
\end{tabular}

\vspace{0.4em}
\begin{minipage}{0.90\textwidth}
\footnotesize \textit{Note:} LO = long-only; LS = long-short. Validation-loss rank is computed within each fold relative to the Optuna trial distribution, where lower ranks indicate lower validation loss. The table reports Spearman rank correlations between validation-loss rank and realised out-of-sample trading metrics. Reported $p$-values correspond to two-sided tests of \(H_0:\rho_s=0\).
\end{minipage}

\renewcommand{\arraystretch}{1.0}
\end{table}

\FloatBarrier

The diagnostic supports the interpretation that validation loss is a weak standalone proxy for realised trading value. In both trading modes, the correlations between selected validation loss and test-period IR$^{**}$, Sharpe ratio, IR$^{*}$, and ARC are small and statistically insignificant. The coefficients are directionally negative in the raw-loss diagnostic, as expected if lower validation loss were associated with stronger test performance, but the relationship is not strong enough to reject the null hypothesis of no monotonic association.

The validation-loss-rank diagnostic leads to the same conclusion. Once the comparison is normalised within each fold relative to the Optuna trial distribution, the correlations become even closer to zero and remain statistically insignificant. Configurations ranked closer to the best validation-loss trial do not systematically deliver higher realised IR$^{**}$, Sharpe ratio, IR$^{*}$, or ARC on the held-out test periods.

This finding reinforces the motivation for considering IC-based and trading-metric-based selectors in the empirical design. It does not imply that validation loss is useless: validation loss remains important for stabilising model training, early stopping, and avoiding pathological fits. However, it shows that loss minimisation alone is not sufficient as an economic selection criterion in this short-horizon BTC/USDT trading setting.

\section*{Data and code availability}
The hourly BTC/USDT dataset was collected from the Binance public REST API. The code used for data processing, model estimation, and strategy evaluation is available from the corresponding author upon request.

% ====================================================================
% REFERENCES
% ====================================================================
\bibliographystyle{elsarticle-harv}
\bibliography{references}

\end{document}